\documentclass[aps,superscriptaddress,nofootinbib]{revtex4}
\usepackage{graphicx,color}
\usepackage{amsmath}
\usepackage[normalem]{ulem}

\def \bfgr #1{ \mbox {{\boldmath $#1$}}}
\newcommand{\be}{\begin{eqnarray} &&}

 \usepackage{graphicx}
\newcommand {\br} {{\bf r}}
\newcommand {\bR} {{\bf R}}
\newcommand{\ee}{\end{eqnarray}}
\newcommand{\la}{\langle\,}
\newcommand{\ra}{\,\rangle}

\def\bp{{\bf p}}

\def\bq{{\bf q}}

\def\bk{{\bf k}}

\def \brho {{\mbox{\boldmath$\rho$}}}
\def \bkappa {{\mbox{\boldmath$\kappa$}}}
\def\nonu{\nonumber \\ &&}

\newcommand{\bm}[1] {\mbox{\boldmath{$#1$}}}

\newcommand{\Rea}{\Re e}
\newcommand{\Ima}{\Im m}

\usepackage{amsbsy,amsmath}
\newcommand{\bs}[1]{\boldsymbol{#1}}

\begin{document}

\vskip 2mm
\date{\today}
\vskip 2mm
\title{
Final state interactions and the
extraction of neutron single spin asymmetries
 from SIDIS by  a transversely polarized $^3$He target
}
\author{A. Del Dotto}
\affiliation{
Istituto Nazionale di Fisica Nucleare, Sezione di Roma,
Piazzale A. Moro 2, I-00185, Rome,  Italy
and
\\
University of South Carolina, Columbia, SC 29208, USA
}
  \author{L.P. Kaptari}
  \affiliation{Bogoliubov Lab. Theor. Phys., 141980, JINR, Dubna,
 Russia}
 \author{E. Pace}
 \affiliation{Phys. Dept. Univ. of Rome "Tor Vergata" and
    Istituto Nazionale di Fisica Nucleare, Sezione di Roma
    Tor Vergata,
  Via della Ricerca Scientifica 1, I-00133, Rome,  Italy}
 \author{G. Salm\`e}
  \affiliation{
Istituto Nazionale di Fisica Nucleare, Sezione di Roma,
Piazzale A. Moro 2, I-00185, Rome,  Italy
}

  \author{S. Scopetta}
  \affiliation{Department of Physics and Geology, University of Perugia  
and\\
     Istituto Nazionale di Fisica Nucleare, Sezione di Perugia,
     Via A. Pascoli, I-06123, Italy}
  
\begin{abstract}
The semi-inclusive deep inelastic electron scattering off 
transversely polarized $^3$He, i.e.
the process, $e + \vec{^3 {\rm He}} \to e' + h+X$,
with $h$ a detected fast hadron, is studied beyond the plane wave impulse 
approximation. To this end, 
a distorted spin-dependent spectral function of a nucleon inside an A=3 
nucleus  is  actually evaluated  through 
a generalized eikonal approximation, in order to  take into account 
the final state interactions
between the hadronizing system and the (A-1) nucleon spectator { one}.   
Our realistic description of both   nuclear target and  final state 
is  a substantial step forward  for achieving   
 a reliable 
 extraction of 
 the Sivers and Collins single spin asymmetries of the free 
neutron.
 To illustrate how and 
to what extent the model dependence due to the treatment of 
the nuclear effects
is under control,  we apply  our approach to  the extraction procedure  of the 
neutron single spin asymmetries from those measured for $^3$He 
 for  values of the kinematical variables relevant  both
for  forthcoming experiments at Jefferson Lab and, 
with an exploratory purpose,
for the future Electron Ion Collider.
\end{abstract}

\date{\today}

\pacs{13.40.-f, 21.60.-n, 24.85.+p,25.60.Gc}

\maketitle

\section{Introduction}

In recent years, special
efforts on both experimental and theoretical sides have been focused on
semi-inclusive deep inelastic scattering (SIDIS), i.e. the process
$A(l,l'h)X$ where, in the final state, a scattered lepton $l'$
and a hadron $h$ are  detected in coincidence, after  
the interaction of a lepton
$l$ with a hadronic system $A$.
Nowadays, it is clear that inclusive deep inelastic scattering (DIS), 
i.e. the process $A(l,l')X$, despite of intense experimental investigations
in the last decades, cannot answer to a few crucial questions
on hadron structure.
Indeed, at least three long standing problems
cannot be explained through DIS
measurements, namely:
(i) the fully quantitative description of the so-called EMC effect
(i.e. the modification of the nucleon partonic structure due to the nuclear
medium \cite{EMC});
(ii)  the solution of the so-called ``spin crisis'', i.e.
the fact that the nucleon spin
does not originate from
only the spins of  its valence quarks \cite{polEMC}; 
(iii) the measurement of the  chiral-odd  parton
distribution function  (PDF) called  transversity
(see, e.g., Refs. \cite{barone,barone1} and references therein quoted)
that  complements  
the leading-twist collinear description of  a
polarized  nucleon. As it is well-known, transversity is 
related to the amount of 
transversely-polarized quark  inside a transversely-polarized  nucleon 
and it is not measurable
in DIS, where a flip of the quark chirality cannot take place.
Through DIS processes, { on both proton (see, e.g., Refs.
\cite{unpol, polarized}) and nuclear targets
(see, e.g., Refs. \cite{Arneodo,Piller,emclast}), it is possible
to investigate only partonic distributions of    longitudinal momentum 
(i.e. parallel to  the direction of the incoming lepton) 
 and helicity.}
Therefore, in order to access information on the transverse structure 
of the target, either in coordinate or momentum space, 
one necessarily has  to go beyond DIS measurements (see, e.g.
Refs. \cite{markus} and \cite{3Dmio} for recent reviews on
nucleon and nuclear targets, respectively).

SIDIS processes are an important tool for increasing our knowledge on 
hadron dynamics.
Indeed, if the detected hadron is fast,
it likely originates  from the
fragmentation of the active quark,
after absorbing
the virtual photon. Hence, the  detected  hadron
opens a  valuable window on the motion of quarks
inside the parent nucleon, before  the interaction  with the photon 
occurs.
In particular  their transverse motion, not seen in the collinear case, 
represents the subject of  intense
experimental efforts 
{{in the study of}}  SIDIS reactions, 
through which one can  access  
the so-called transverse-momentum-dependent parton distributions
(TMDs) (see, e.g., Ref. \cite{barone1}). Those distributions provide
a wealth of information on the 
partonic dynamics, 
eventually shedding light on the  
challenging three issues listed above.
Beside the main topic represented by TMDs, 
one should remind that 
the detected hadron carries also
information on the hadronization mechanism itself.
The SIDIS cross sections can be parametrized, at leading twist,
by six TMDs; this number reduces to three in the collinear case
(with only two TMDs measurable in DIS \cite{barone1}) and increases to eight
once the so-called time-reversal odd TMDs  
(i.e. the Sivers
\cite{sivers} and Boer-Mulders \cite{boemu} functions) 
are considered \cite{barone1}.
%,barone1,kotzinian}.

It should be emphasized that,
in order to experimentally investigate
the  wide field of  TMDs, one should measure
cross-section asymmetries,
using different combinations of beam and  target polarizations 
(see, e.g., Ref. \cite{d'alesio}). Moreover,
for  completing the  study of TMDs, one should achieve
a sound  flavor decomposition, possible only  by collecting  a detailed 
knowledge of the neutron TMDs.
The present investigation moves from the observation that 
free neutron targets are not available
and nuclei have to be used as effective neutron targets.
In particular, the { study} of the neutron spin structure
is highly favored  by choosing a polarized $^3$He  target, as it has been
done extensively in DIS studies.
In the 90's, procedures to extract the
neutron spin-dependent structure functions from $^3$He
data in the DIS regime, taking properly into account  Fermi-motion and  
binding effects,
were proposed \cite{antico}
and  successfully applied
(see, e.g. \cite{appl}). 
Such a detailed description of the target nucleus
was obtained in plane wave impulse approximation
(PWIA) by using
the so-called spin-dependent spectral function, whose diagonal elements yield 
the probability
distribution to find a nucleon with a given momentum, missing energy and 
polarization
inside the nucleus.  
It is worth noting that, within PWIA, accurate
$^3$He spin-dependent spectral functions, based on
realistic 
calculations  of both the target nucleus and the  spectator
pair in the final state (fully interacting through the $NN$ interaction 
adopted for $^3$He),
have been built and used in the last twenty years
\cite{cda,SS,cda1,pskv,umnikovkapt,plbold}.

The question whether similar procedures 
can be extended to
SIDIS is of great relevance, due to the several experiments that exploit
a  polarized $^3$He target  (see, e.g., Ref.   \cite{He3exp}),
for  { accessing}  the transverse momentum and spin 
of the partons inside the neutron.
For instance, a wide interest has arisen about the possibility
{ to use a transversely polarized $^3$He target for measuring}
azimuthal single-spin asymmetries (SSAs) of the neutron,
which are sensitive to
time-reversal odd TMDs and to the Collins
fragmentation functions (FF) \cite{collins} generated
by leading twist final state interactions~\cite{Brodsky:2002cx}.
{{In the first measurements of SSAs, through
SIDIS off transversely polarized
proton and deuteron targets, the proton SSAs were found to be sizable
\cite{hermes}, while those of deuteron were found to be negligible 
\cite{compass}, pointing to a large cancellation between the proton and 
neutron contributions. 
{{A high luminosity environment coupled to a suitable neutron target, 
as a polarized $^3$He
(at level of 90 \% an effective neutron target),
allows one first to better assess the
flavor separation and then accurately test its sensitivity to quark 
angular momenta.}}
It became clear the need of    
increasing the  experimental knowledge on  neutron TMDs
through an independent measurement, and 
an experiment of SIDIS off transversely polarized $^3$He was soon 
proposed \cite{bro}.}} As it is well-known, some significant steps have been 
already carried out along the suggested path, since 
 azimuthal asymmetries
in the production of leading 
{ $\pi^\pm$ $(K^\pm)$}
from transversely
polarized $^3$He have been already measured at Jefferson Lab (JLab),
with a beam energy of 6 GeV
\cite{prljlab} and new experiments
will be soon performed after completing the 12
GeV upgrade \cite{jlab12}.

In view of { those} experimental  efforts, 
a realistic PWIA analysis of SIDIS
off transversely polarized $^3$He has been performed
\cite{mio}. 
A realistic spin-dependent spectral function, corresponding to the
nucleon-nucleon 
AV18 interaction \cite{av18}, has been used for the
description of nuclear dynamics and
the issue of the extraction of
the neutron information from $^3$He data
has been { addressed}. { According to Ref. \cite{mio}, 
one can safely extend to SIDIS, where both PDFs and
FFs are involved, 
the
model-independent  extraction procedure based
on the realistic evaluation
of the proton and neutron polarizations in $^3$He
and  widely used in inclusive DIS \cite{appl}. As a matter of fact, 
such an extraction procedure is
able to take into account effectively
the momentum and energy distributions
of the polarized bound nucleons in $^3$He.

%\begin{figure}[!ht]
%\centerline{
%%%%%%\includegraphics[scale=0.35,angle=0]{Fig1.eps}}
%\caption{One-photon exchange diagram for
%semi-inclusive deep inelastic scattering
%processes $\vec A(\vec l,l'\, h) X$. {  The target $A$ can be
%either longitudinally or transversely polarized}.}
%
%\label{Fig1}
%\end{figure}

In general, SIDIS off  nuclear targets can happen through
at least
two, rather different,
sets of processes:  
\begin{enumerate} 
%[i)]
\item the   {\em standard}  reaction (most familiar),
where a fast hadron is detected mainly in the
forward direction, implying
that the hadron has been  produced by the leading
quark. Therefore, this  reaction, representing 
the dominant mechanism in the kinematics 
of the { Jlab experiments of Refs. }\cite{prljlab,jlab12}, can be used  to investigate
TMDs inside the hit nucleon;
\item 
the {\em spectator}  SIDIS, where
a slow  $(A-1)$ nucleon system, acting as a spectator of the
photon-nucleon interaction, is detected, while the produced fast 
hadron is not.
\end{enumerate}

%Ref. \cite{nostro} 
%has been dedicated to 
The spectator
 SIDIS process has been proven very
useful to investigate the unpolarized DIS
functions $F_{1,2}(x)$ of a bound nucleon, and therefore
to clarify the origin of the EMC effect
(see, e.g., \cite{ourlast,scopetta,ckk,smss,veronica}).
At the same time, this process can provide {{also}} useful
information on quark hadronization in
medium, complementary to that obtained so far by the
standard SIDIS process. Noteworthy, the polarization degrees of freedom 
of the target substantially 
enrich the wealth of information one can gather, as shown 
in Ref. \cite{nostro}, where a spectator SIDIS, with a detected deuteron, 
off a polarized $^3$He target was { studied}. Through such a
polarized SIDIS, one can obtain fresh
information on the spin-dependent structure functions $g_{1,2}(x)$
for  bound nucleons and, ultimately,
on the origin of the polarized EMC effect. 

In polarized (as well as unpolarized) SIDIS processes, { the effects} of  the 
final state interaction (FSI) 
that occur among the hadronizing system (produced after the quark-photon
knock-out) and the $(A-1)$ spectator system has to be {  carefully
analyzed}. 
For the case of a polarized $^3$He, this study started in  Ref. \cite{nostro}, 
where the trinucleon {\em distorted spin-dependent spectral function}
has been introduced, but restricted to the   deuteron  spectator system.
In order to realistically take into account the above mentioned FSI, it has
been adopted 
 a generalized eikonal approximation (GEA),
i.e. a framework successfully { introduced} for describing
unpolarized SIDIS off nuclei \cite{ourlast}. To apply  such a   
{\em distorted spin-dependent spectral function} 
 to the standard  polarized SIDIS by { $^3\vec {\rm He}$}, one has to consider 
{\em all the possible states} 
  of the two-nucleon
spectator system. But  due to  FSI between the  
spectator system and the quark debris,
produced after DIS off an internal nucleon with given polarization, 
this novel distribution function is remarkably more complicated than the 
PWIA spin-dependent spectral function,
adopted
in the  description of { both DIS by unpolarized $^3$He \cite{antico} and SIDIS
\cite{mio}}. 
However,  efforts for evaluating a realistic {\em distorted spin-dependent spectral function}
are worth    attempting since  its thorough knowledge 
 represents a fundamental help for   reliably disentangling TMDs from the nuclear structure, 
  in the  experimental cross
sections.
In perspective, an experimental check of the robustness of the description 
of the nuclear effects could be in 
principle carried out { by exploiting}  the isodoublet nature of the
trinucleon bound states. In the case of a polarized $^3$H, one could extract 
(i) the proton polarized structure functions when a spectator SIDIS 
is considered 
or (ii) the relevant TMDs when a
standard SIDIS is investigated. 
The proton information extracted 
from $^3$H could be compared with { the ones} gathered using
free proton targets, shedding light on the relevance and nature
of nuclear effects.
Nowadays, the use of a polarized $^3$H target seems 
too challenging, but it is worth mentioning that important
achievements have been obtained in the 
last decade in handling such a problematic target, as demonstrated by the 
final approval (with scientific rating A), at JLab,
of an experiment dedicated to  
DIS by a $^3$H target \cite{marathon}.

As a concluding remark, it should be pointed out that, at the present stage, 
the needed relativistic description of SIDIS is restricted to the
kinematics and the elementary cross-section, as discussed in the following 
Sections. Indeed, in order to embed the very successful non relativistic
phenomenology  of the nuclear structure, developed over the past decades, 
in a fully Poincar\'e covariant approach, one could    
exploit  the Light-front framework, that originates from the seminal work 
by P.A.M. Dirac on the forms of relativistic Hamiltonian dynamics
\cite{Dirac}. A thorough formal investigation of a Light-front
spin-dependent  spectral function for a $J=1/2$ target, 
in impulse approximation,  
has been recently presented in 
Ref.  \cite{tobe} (see also \cite{lussino,Pacelc16} 
for preliminary results). Obviously, this novel distribution function
is the first step for constructing a Poincar\'e covariant
description of SIDIS reactions, since in analogy with the transition from 
the PWIA spectral function to the
distorted one, FSI effects have to be taken into account also
in the Poincar\'e covariant approach.

Aims of  the present paper are first to extend the calculation  
 of the distorted spin-dependent spectral function of
$^3$He performed in Ref. \cite{nostro}, in order to include the excited states 
of the two-nucleon spectator system (recall that in
Ref. \cite{nostro} only the deuteron state was retained). As a second
step, we apply our formalism  to the 
{\em standard} SIDIS { process, with kinematical
conditions typical of  experiments to be performed in the next years at
 JLab  and in the future (possibly near) at the electron 
ion collider (EIC)}, focusing  on the extraction of quark
TMDs inside the neutron, i.e. the needed ingredients for  making complete 
the flavor
decomposition. One can easily realize that, since in standard
SIDIS the final fast hadronic state can re-interact with a
two-nucleon scattering state, this process is much
more involved than spectator SIDIS, where FSIs occur between
the final hadronic state and the detected deuteron.

The paper is organized as follows. In  Section II we present the
basic formalism for the cross section,
valid for the standard  SIDIS process,
where a hadron $h$ is detected in coincidence with the scattered charged 
lepton.
The main quantities relevant for the calculations are presented
and the PWIA framework is reviewed, to better appreciate the
difference with { the FSI case,  discussed in the next Sections}.
In Section III, the SIDIS reaction
$\vec{^3{\rm He}}(e,e'h)X $ is
investigated in detail, introducing the distorted spin-dependent spectral
function, that represents the main ingredient of our method for implementing
FSI effects,
through  a generalized eikonal approximation.
In Section IV, the dependence of the nuclear hadronic tensor 
upon the target polarization is studied.
In Section V the expressions to be used for evaluating the nuclear SSAs, 
both in PWIA and with FSI taken into account, are presented and a 
strategy for the extraction of the neutron information is discussed. 
In Section VI, the results for the distorted spectral functions and 
light-cone momentum distributions are presented and compared  
with the corresponding PWIA calculations; furthermore the finite values of 
the momentum and energy transfers corresponding to the actually proposed 
experiments are adopted for the evaluation of the $^3$He Collins and Sivers 
asymmetries and for the extraction of neutron asymmetries with FSI effects 
taken into account and implementing the comparison with the PWIA 
calculations. Eventually, in the last Section, conclusions are drawn 
and perspectives presented. Important formal details are collected in two
appendixes.

\section{The  SIDIS cross section}

%%%%%%%%%%%%%%%%%%%%%%%%%%%%%%%%%%%%%%%%%%%%%%%%%%%%%%%%%%%%%%%%%%%%%%%%%%%%%%%%%%%%%%%%%%%%%%
%Our approach is based on the so called {\it spectator mechanism}:
The differential cross section for the  generic
SIDIS process off a polarized target $A$, i.e. $l+\vec A=l'+h+X$ 
when the final pseudoscalar  
hadron $h$ is detected, 
can be written in the laboratory frame and in one-photon exchange
approximation as follows
(cf, e.g., Refs. \cite{barone1,scopetta,nostro}),
\be
\frac{d\sigma}{d\varphi_{\ell} dx_{Bj} dy d{\bf P}_h }=
\frac{\alpha_{em}^2\ m_N}{Q^4}\,y\,~
\frac{1}{2 E_h}~
 L^{\mu\nu} W_{\mu\nu}^{s.i.}({\bf S}_A,Q^2,P_h)~,
   \label{crosa-1}
   \ee
where, for  incoming and outgoing charged leptons with 4-momentum
$k^\mu=({\cal E}, \vec k)$ and
$k'^{\mu}=({\cal E'}, \vec k')$, one has 
$Q^2 =-q^2= -(k-k')^2 = \vec q^{\,\,2} - \nu^2=4 {\cal E}
{\cal E}' \sin^2(\theta_{\ell}/ 2)$, i.e. the square 4-momentum transfer  in
 ultrarelativistic approximation (with
$\vec q = \vec k - \vec k'$, $\nu= {\cal E} - {\cal E}' $ and
$\theta_{\ell}\equiv \theta_{\widehat{\vec k  \vec k'}}$). 
Moreover, 
$x_{Bj} = Q^2/(2m_N\nu) $ is   the Bjorken scaling variable, $y=\nu/{\cal E}$,
$m_N$ the nucleon mass,  $\alpha_{em}$  the electromagnetic
fine structure constant,
$\varphi_{\ell}$ the azimuthal angle of the detected 
charged lepton, $P_h=(E_h,{\bf{P_h}})$  the 
4-momentum of the detected-hadron $h$, with mass $m_h$
and ${\bf{S}}_A$ the polarization vector of the target nucleus. 

The unpolarized leptonic tensor $L_{\mu\nu}$ is an exactly calculable quantity
in QED. In the ultrarelativistic limit it gets the form
\be
L_{\mu\nu} = 2 \left [k_\mu k_\nu' +k_\mu' k_\nu -(k\cdot k') 
g_{\mu\nu} \right]~.
\label{lmunu}
\ee
The semi inclusive (s.i.) hadronic tensor of the
target with polarization   four-vector $S_A$ { and mass $M^2_A=P^2_A$ } 
is defined as
\begin{eqnarray}
W_{\mu\nu}^{s.i.}({\bf{S}}_A,Q^2,P_h) 
& = & 
\frac{1}{2 M_A} {\sum\limits_{X}}
 \la  S_A, P_A|\hat J_\mu|P_{h},X\ra  \la P_{h},X|\hat J_\nu| S_A, P_A 
\ra \nonumber \\
& \times & 
\delta^4\left(P_A+q-P_X-P_{h}\right)~d\tau_X ~,
%\frac{d{\bf P}_h}
%{2E_h(2\pi)^3}~,
\label{wmunuA}
\end{eqnarray}
where the covariant normalization
$\la p| p'\ra =2E(2\pi)^3\delta\left( {\bf p-p}'\right)$ has been assumed and
$d\tau_X$ is the suitable phase-space factor for the undetected
state $X$, given in turn by a state $X'$  with baryon number 1
and an $A-1$ recoiling nuclear system.
One should notice that, in Eq. (\ref{wmunuA}), 
the integration over the phase-space volume of the detected hadron, 
$h$, does not have
to  be performed.
%In what follows, in order to keep the notation for the nuclear part
%as close as possible to the non relativistic
%case,  the factor  $2M_A$ in  the nuclear
%matrix elements are included  into the
%normalization of the  nuclear states, i.e. { for them} 
%we changed the covariant normalization
%to the
%non-covariant one, according to
%$\la p| p'\ra =2E(2\pi)^3\delta\left( {\bf p-p}'\right)\longrightarrow
%\la {\bf p}| {\bf p'}\ra =(2 \pi)^3 \delta\left({\bf p-p}'\right)$~.
%\begin{eqnarray}\!\!\!\!\!\!
%W_{\mu\nu}^{\hat {\bf S}_A}=\frac{1}{(2\pi)^4}\overline{\sum\limits_{X}}
% \la  S_A, P_A|J_\mu|P_{h},X\ra   \la P_{h},X|J_\nu| S_A, P_A \ra
%(2\pi)^4\delta^4\left(P_A+q-P_X-P_{h}\right) d\tau_X
%\label{wmunuA}
%\end{eqnarray}

In the following, the cross section for SIDIS off transversely polarized
$^3$He will be worked out, taking into account final state
interaction effects. To this aim, it is necessary first to recall
the results obtained in PWIA.

Within PWIA, the nuclear tensor Eq. (\ref{wmunuA}) is approximated
using the following assumptions:
(i) the nuclear current operator is written as the sum 
of single nucleon operators $\hat j_\mu^N$; 
(ii) the FSI between the debris originating by the struck nucleon
and the fully interacting (A-1) nuclear system is disregarded,
as suggested by the kinematics of the process under investigation;
(iii) the coupling of the 
virtual photon with the 
$(A-1)$ system is disregarded, due to the { large} 4-momentum transferred 
in the process; (iv) the effect of boosts is not considered
(they will be properly taken into account in a Light-front 
framework elsewhere, following the procedure addressed in Refs. 
\cite{tobe, lussino,Pacelc16}).
In this way,
the complicated final baryon states
$|P_{h},X\rangle$ are approximated by a
tensor product of hadronic states,  viz
\be
|P_{h},X\rangle^{PWIA}=  |P_{A-1}\ra \otimes|P_h\ra \otimes |X'\ra~,
\label{tensprod}
\ee
where  $|P_{A-1}\ra$ indicates the state (properly antisymmetrized)
of the fully-interacting 
$(A-1)$-nucleon  system,
which acts merely as a spectator, $|X'\ra$ describes
the baryonic state, that originates
together with $|P_h\ra$ from the hadronization of both { the quark which
has absorbed the virtual photon and } the other colored remnants.
The nuclear
tensor $W_{\mu\nu}^{s.i.}({\bf{S}}_A,Q^2,P_h)$ can be related
therefore to the one of a single nucleon.
%$w_{\mu\nu}^{s.i.}(S_N,Q^2,P_h)$. 
This is obtained inserting
in Eq. (\ref{wmunuA}) complete sets of nucleon plane waves and
$(A-1)$-nucleon   interacting states, given by
\be
\sum_\lambda \int {d {\bf p}_N\over 2 {E}_N (2\pi)^3} ~ |\lambda,p_N \ra
 \la  \lambda, p_N|  ~= 1~,
\label{comp}
\ee
\be
\sum_{f_{A-1}}
\sum \! \!\! \!\! \!\! \!\int_{~\epsilon^*_{A-1}}
\rho\left(\epsilon^*_{A-1}\right)
~\int {d {\bf  P}_{A-1}\over 2 E_{A-1} (2\pi)^3} \, 
| \Phi_{\epsilon^*_{A-1}}^{f_{A-1}},{\bf P}_{A-1}\ra
 \la  \Phi_{\epsilon^*_{A-1}}^{f_{A-1}} ,{\bf P}_{A-1}|  ~= 1~,
\label{comp2}
\ee
where $p_N\equiv \{{{E}}_N=\sqrt{m^2_N+|{\bf p}_N|^2}, {\bf p}_N\}$
is the on-shell four-momentum of a nucleon,
$\Phi_{\epsilon^*_{A-1}}^{f_{A-1}}$ is the intrinsic part of the  
$(A-1)$-nucleon
state with quantum numbers 
${f_{A-1}}$ and energy
eigenvalue $\epsilon^*_{A-1}$. Moreover,
$E_{A-1}=\sqrt{(M^*_{A-1})^2+|{\bf P}_{A-1}|^2}$ with $M^*_{A-1}=Z_{ A-1} m_p+
(A-1-Z_{ A-1})m_n
+\epsilon^*_{A-1}$. The symbol with the sum overlapping the
integral
indicates that the $(A-1)$ system has both discrete and 
continuum energy spectra: this
corresponds to negative and positive values of  the eigenvalue
 $\epsilon^*_{A-1}$.
In Eq. (\ref{comp2}), $\rho\left(
\epsilon^*_{A-1}\right)$ is the proper state density, 
that  for  $A=3$ reads
\be
\rho_{2bbu}= {1 \over (2 \pi)^3}~~,
\quad \quad \quad\quad \quad
\rho_{3bbu}= {1 \over (2 \pi)^6}~{m_N\sqrt{m_N\epsilon^*_2} \over 2}~,
\ee
with the labels $2bbu$ and $3bbu$ indicating  the
two-body and three-body break-up  channels, respectively.
Furthermore, recalling that Eq. (\ref{tensprod}) implies
\be
\sum_X~d\tau_{X} \to ~\sum_{X'}~d\tau_{X'} \sum_{f_{A-1}}
\sum \! \!\! \!\! \!\! \!\int_{~\epsilon^*_{A-1}}
\rho\left(\epsilon^*_{A-1}\right)
~\int{d{\bf P}_{A-1} \over 2 E_{A-1} (2\pi)^3}~,
\ee
one obtains the following expression for the nuclear tensor in PWIA
\be
W_{\mu\nu}^{s.i.{;IA}}({\bf{S}}_A,Q^2,P_h) 
=
{\sum\limits_{X',\lambda\lambda'}}
\sum_N\int dE~ { { P}^{N \,{\bf S}_A}_{\lambda\lambda'}}
(E,{\bf p}_N)
 {\frac{1}{2E_N }}
\la \lambda',{ p}_N | 
\hat j_\mu^N
|P_h,X' \ra 
\la P_h,X'| 
\hat j_\nu^N 
|\lambda,  { p}_N\ra
\nonu
 \times 
\delta^4\left( P_A+q-P_{A-1}-P_h-P_{X'}\right)d\tau_{X'} \,
{d{\bf P}_{A-1} \over (2 \pi)^3}~,
\label{munuN}
\ee
where, w.r.t. Eq. (\ref{wmunuA}),  $P_{X'}+P_{A-1}$ is in place of $P_X$,
and the nucleon three-momentum,
% $ { \tilde p}_N\equiv \{E_N,{\bf p}_N \}$ with
${\bf p}_{N}={\bf P}_A-{\bf P}_{A-1}$, is fixed
by the translation invariance of the
initial nuclear vertex, viz
\be
\la  \Phi_{\epsilon_{A-1}}^{* \, f_{A-1}},{\bf P}_{A-1}; 
\lambda, p_N|  {\bf{S}}_A,  P_A\ra=
\sqrt{2E_N~2E_{A-1}~2M_A}~(2 \pi)^3~\times \nonu
\delta\left ( {\bf P}_{A}-{\bf P}_{A-1}-{\bf p}_{N}\right)~
\la\Phi_{\epsilon_{A-1}}^{* \, f_{A-1}}; \lambda, {\bf p}_N|  {\bf{S}}_A,\Phi_A\ra
~~.\label{overl}\ee
In Eq. (\ref{overl}),   $\Phi_A$ is the intrinsic wave function of the 
target nucleus, with mass $M_A$  and $(P_A- P_{A-1})^2\ne m^2_N$. 

{ The matrix elements ${ P}^{N  \, 
{\bf S}_A}_{\lambda\lambda'}(E,{\bf p}_N)$ in Eq. (\ref{munuN})
contain the description of the 
nuclear structure  and are given in PWIA by 
\be 
{ P}^{N \, {\bf S}_A}_{\lambda\lambda'}(E,{\bf  p}_N) = ~
%{{ 1/2M_A}}
\sum_{f_{A-1}} 
\sum \! \!\! \!\! \!\! \!\int_{~\epsilon^*_{A-1}}\rho\left(
\epsilon^*_{A-1}\right)~
 {\cal O}^{N \, {\bf S}_A}_{\lambda\lambda'}(\epsilon^*_{A-1},{\bf p}_N)
   ~
{  \delta\left( { E+ M_A-m_N-M^*_{A-1}}\right)}~,
\label{overlaps}
\ee}
where  $E$ is 
the usual missing { or removal energy, 
$E=M^*_{A-1} +m_N-M_A=\epsilon^*_{A-1}+B_A$, }
with $B_A$ the binding energy of the target nucleus. The quantity 
$m_N-E$ is the off-shell mass of a nucleon inside the target nucleus, 
when the $(A-1)$ system acts
as a spectator. { In Eq. \eqref{overlaps}, 
${\cal O}^{N \, {\bf S}_A}_{\lambda\lambda'}(\epsilon^*_{A-1},{\bf p}_N)$ is the following product of PWIA overlaps
\be
{\cal O}^{N \, {\bf S}_A}_{\lambda\lambda'}(\epsilon^*_{A-1},{\bf p}_N)=\la  \Phi_{\epsilon^*_{A-1}}^{f_{A-1}},
\lambda,{\bf p}_N|  {\bf{S}}_A, \Phi_A\ra
 \la   {\bf{S}}_A, \Phi_A| \Phi_{\epsilon^*_{A-1}}^{f_{A-1}},\lambda',
 {\bf p}_N \ra~.
\ee}
The quantities
{ ${ P}^{N \, {\bf S}_A}_{\lambda\lambda'}(E,{\bf p}_N)$}, 
Eq. (\ref{overlaps}),
are the matrix elements of the $2\otimes 2$ spin-dependent
spectral function of a nucleon inside the nucleus $A$,
with polarization ${\bf S}_A$ \cite{cda1}.
 The trace of the spectral function yields
the probability distribution to find a nucleon in the
nucleus $A$ with three-momentum ${\bf p}_N$, removal  energy $E$ and
spin projection equal to $\lambda$. The suitable normalization is 
\be
{1 \over 2} \sum_{\lambda N} \int dE~ \int  d{\bf  p}_N~
{ { P}^{N \, {\bf S}_A}_{\lambda\lambda}( E,{\bf p}_N)}~=1~.
\label{compl1}\ee

Assuming the polarized target in a pure state, 
the nuclear wave function has
definite spin projections on the spin quantization axis, chosen as usual
along  the polarization vector ${\bf S}_A$. 
In agreement with the definition of the spin-dependent spectral 
function given in Refs. \cite{SS,cda1},  in the 
complete set of the nucleon plane waves, the spin projections 
$\lambda$ and  $\lambda'$ 
are defined with respect to the $z$ axis.

As for the Cartesian coordinates, 
we  adopt the  DIS convention, i.e. the  $z$ axis
is directed along the three-momentum transfer
${\bf q}$ and   the plane $(x,z)$ is  the scattering plane. 
Notice that, in the DIS limit, the
direction of the three-momentum transfer coincides with that of the lepton
beam, i.e. ${\bf q}\ ||\ {\bf k}_e$.

The nuclear tensor Eq. (\ref{munuN}) can be written 
\be
W_{\mu\nu}^{s.i.{;IA}}({\bf{S}}_A,Q^2,P_h) =
\sum\limits_{\lambda\lambda'}\sum_N \int d{\bf p}_N \int
dE~
{m_N \over E_N }
w_{\mu\nu}^{N \,s.i.}(\tilde p_N,P_h,\lambda'\lambda)
{ { P}^{N \, {\bf S}_A}_{\lambda\lambda'}(E,{\bf p}_N)}~,
\label{A_si}
\ee
where the integration over ${\bf P}_{A-1}$ has been changed to
the one over ${\bf p}_N={\bf P}_A
-{\bf P}_{A-1}$, and 
the  semi-inclusive  nucleon  tensor  (cf. Eq. (\ref{wmunuA})) is given by
\begin{eqnarray}
w_{\mu\nu}^{N \, s.i.}(\tilde p_N,P_h,\lambda'\lambda)= 
\frac{1}{2m_N}
\sum\limits_{X'}
\la  { p}_N,\lambda'|\hat j_\mu^N | P_h,X'\ra \la P_h,X'|\hat j_\nu^N|
{p}_N,\lambda\ra
 \delta^4\left ( \tilde p_N+q-P_h-P_{X'}\right)d\tau_{X'}~, 
 \label{wm}
\end{eqnarray}
where $ \tilde{p}_N =P_A -P_{A-1}$ is such that 
$\tilde{p}^2_N\ne m^2_N={p}^2_N$.

%
%, the cross section Eq. (\ref{crosa-1}),
%the cross section for standard SIDIS, when the hadron $h$ is detected,
%is obtained as follows
%\be
%\frac{d\sigma(h_l, {\bf S}^A)}{d\varphi_e dx_{Bj} dyd{\bf P}_h }
%=\frac{\alpha_{em}^2\ m_N y}{Q^4}
%\frac{1}{2E_h}
%  \ L^{\mu\nu}(h_l) 
%W_{\mu\nu}^{s.i.}(q,S_A,P_h)
%\label{crosa-2}
%\ee
%where the nuclear hadronic tensor is given by

Eventually, for the nuclear cross section given in Eq. (\ref{crosa-1}),
$\sigma_A({\bf S}_A)
\equiv 
~\displaystyle\frac{d\sigma({\bf S}_A)}
{d\varphi_{\ell} dx_{Bj} dyd{\bf P}_h }$,
one {gets the following expression in PWIA}
\be
\sigma^{A ; IA}({\bf S}_A)
= \sum\limits_{\lambda\lambda'}\sum_N \int d{\bf p}_N \int
  dE~{~\tilde\alpha ~m_N\over E_N}
   \sigma^N_{\lambda\lambda'}
 ~{ { P}^{N \, {\bf S}_A}_{\lambda\lambda'}(E,{\bf p}_N)}~,
  \nonu  \label{crosa-3}
   \ee
where 
\be
\sigma^N_{\lambda\lambda'} 
\equiv
\frac{d\sigma^N_{\lambda \lambda'}}{d\varphi_{\ell} dx_{Bj} dyd{\bf P}_h }
=
\frac{\alpha_{em}^2\ m_N}{Q^4}
\frac{m_N \nu}{{(p_N \cdot k)}}
\frac{1}{2 E_h}
\ L^{\mu\nu}
w_{\mu\nu}^{N \, s.i.}(\tilde p_N,P_h,\lambda'\lambda)
\label{crosa-2}
\ee
represents the corresponding cross
section for the scattering of a
{charged lepton}  from a polarized moving
nucleon. { In Eq. (\ref{crosa-3}), $\tilde\alpha$ is given by 
\be
\tilde\alpha\equiv \frac{{(p_N \cdot  k)} } {{\cal E} m_N}
\ee 
and it is usually called
 the "flux factor". When 
 energies are close to the
Bjorken limit $\tilde\alpha$ coincides with the light-cone momentum 
fraction of the nucleon inside the nucleus, i.e.
\be
\lim_{\to Bj}\tilde \alpha=
\frac {A(p_N\cdot q) } {(P_A\cdot  q)}~~.
\ee}

\section{The distorted spin-dependent spectral function}

In order to go beyond PWIA (cf. Eq. \eqref{tensprod}), it is necessary 
to deal with the FSI 
between the debris, originating from the struck nucleon, and 
the fully interacting (A-1) nuclear system. In view of this, 
the dependence upon the space coordinates in the current operator is kept, 
since we will focus on the action of
the  current onto the final state in coordinate space.

The starting point is the hadronic tensor written as follows
\be
W_{\mu\nu}^{s.i.}({\bf{S}}_A,Q^2,P_h) 
= 
\frac{1}{2 M_A} {\sum\limits_{X}}
 \la  S_A, P_A|\hat J_\mu(\hat{\bf r}_i)|P_{h},X\ra  \la P_{h},X|\hat 
 J_\nu(\hat{\bf r}_i)| S_A, P_A 
\ra  ~ 
\delta\left(M_A+\nu-E_X-E_{h}\right)~d\tau_X~,
%\frac{d{\bf P}_h}
%{2E_h(2\pi)^3}
\label{wmunuB}
\ee

\begin{figure} [t]
\includegraphics[width=0.5\textwidth]{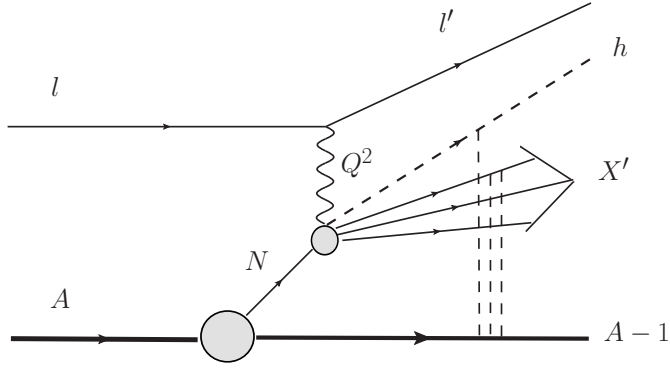}
\caption{
The SIDIS process $A(e,e'h)X$, with final state interactions
taken into account
} 
\label{GEA}
\end{figure}

For a $^3$He target, the matrix element 
 of the  current operator
$\hat J_\mu(\hat{\bf r}_1,\hat{\bf r}_2,\hat{\bf r}_3)$
between the nuclear ground state,
%$|\vec\Psi_3^{{\bf S}_A}(\br_1,\br_2,\br_3)\ra$
$|\Psi_3^{{\bf S}_A}(1,2,3)\ra$,
and a generic final state,
$
| 
\Psi^{f}(1,2,3)\ra
$,  is evaluated by introducing the following approximation
\be
\hat J_\mu(\hat{\bf r}_1,\hat {\bf r}_2,\hat{\bf r}_3)\approx\sum\limits_i \hat j_\mu(\bf{r}_i)~,
\ee
where $\hat j_\mu(\bf{r}_i)$ is the one-body 
transition current operator, that describes the
electromagnetic response of the single nucleon inside the target. 
In this way the matrix element becomes
\be
\la 
P_{h},X| \hat J_\mu (\hat{\bf r}_1,\hat{\bf r}_2,\hat{\bf r}_3)| {\bf{S}}_3, 
P_3
\ra
=
\la  
\Psi^{f}(1,2,3)|
\hat J_\mu(\hat{\bf r}_1,\hat{\bf r}_2,\hat{\bf r}_3)
|\Psi_3^{{\bf S}_3}(1,2,3)
\ra \approx 
%\la  
%\Psi^{f}(1,2,3)|
%\sum\limits_i \hat J_\mu(i)
%|\vec\Psi_3^{{\bf S}_A}(1,2,3)\ra
%= 
 \sum\limits_i 
\la  
\Psi^{f}(1,2,3)|
\hat j_\mu(\hat{\bf r}_i)
|\Psi_3^{{\bf S}_3}(1,2,3)\ra=
\nonu=
3 \la  
\Psi^{f}(1,2,3)|
\hat j_\mu(\hat{\bf r}_1)
|\Psi_3^{{\bf S}_3}(1,2,3)\ra~.
\label{r1}
\ee
In what follows, for the sake of concreteness, 
the active nucleon is labeled "i=1" and the spectator indexes
are "23". 

For constructing a realistic approximation of FSI, it is 
useful to consider that, in  SIDIS  processes, we aim at investigating,
the momentum transfer ${\bf q}$ is rather large, and therefore
$h$, the leading pseudoscalar meson to be detected, and $X'$, that has  baryon number equal to 
$1$ (cf Eq. \eqref{tensprod}),  move throughout the $A-1$ 
remnants 
with high
velocity.  This observation motivates the introduction of the  
 generalized eikonal approximation (see, e.g., Refs. \cite{nostro,ourlast} and references
 quoted therein) for  estimating
the rest of FSI not taken into account through  PWIA (cf Eq. \eqref{tensprod}).
Then,  the final state  can be  approximated in coordinate space as
\be
\la\br_1 \br_2 \br_3 | \Psi^{f}(1,2,3)\ra
\approx {{\cal A} \over 
% {\sout{(2\pi)^{3/2}}}
\sqrt{V}~\sqrt{3}}
\Psi_{23}^{f}(\br_2,\br_3)  
\chi_{\lambda_Y} \phi(\xi_Y)~{\sqrt{2 E_Y}} ~e^{i{\bf p}_Y\br_1}
~ {\cal G}(\br_1,\br_2,\br_3)~,
\label{fsi}
\ee
where   ${\cal A}$ is the antisymmetrization operator that 
 acts on the
final state,  given by a recoiling two-nucleon system and
a debris { $Y$ originated by the struck nucleon (see below)},
 $\Psi_{23}^{f}(\br_2,\br_3)$ is the
properly antisymmetrized 
wave function of the recoiling two-nucleon  system,
$V$ is the normalization volume of the global motion
of the final state, and  the amplitude  
${\cal G}(\br_1,\br_2,\br_3)$,   
identically equal to 1
in PWIA,  is the non singular part of the matrix elements of the Glauber
operator, i.e.
\be
\langle {\bf r}'_1, {\bf r}'_2, {\bf r}'_3|{\cal {\hat{G}}}|{\bf r}_1, 
{\bf r}_2, {\bf r}_3\rangle= ~\delta({\bf r}_1'-{\bf
r}_1)~\delta({\bf r}_2'-{\bf
r}_2)~\delta({\bf r}_3'-{\bf
r}_3)~{\cal G}({\bf r}_1, {\bf r}_2, {\bf r}_3)~.   
\ee
The Glauber amplitude depends only upon
intrinsic coordinates, 
$\br, \brho$,
related to ${\bf r}_i$  through
\be
\br_1=\frac23 \brho +\bR~, \nonu
\br_2=-\frac13 \brho+\frac12 \br+\bR~, \nonu
\br_3=-\frac13 \brho- \frac12 \br+\bR~, \label{jac}
\ee
and therefore
\be
{\cal G}({\bf r}_1, {\bf r}_2, {\bf r}_3)~ \to~{\cal G}(\br, \brho)~.
\ee
In Eq. (\ref{fsi}), $Y$ is the final debris produced
by the nucleon  after the absorption
of the virtual photon. In the process under consideration,
it coincides with a leading pseudoscalar meson  
to be detected and a baryonic remnant
$X'$  (cf. Fig. \ref{GEA}).  
The function $\phi(\xi_Y)$ characterizes the internal structure of the debris  
that will hadronize in $h+X'$,
$\chi_{\lambda_Y}$ its spin state, 
while $e^{i{\bf p}_Y\br_1}$ is the plane wave describing
the propagation of the c.m. of the debris.

By using intrinsic coordinates,  the final state in Eq. \eqref{fsi}  becomes
\be
\la\br_1 \br_2 \br_3 | \Psi^{f}(1,2,3)\ra
\approx {{\cal A} \over 
%{\sout{(2\pi)^{3/2}}}
\sqrt{V}~\sqrt{3}} ~ 
{{\sqrt{2 E_Y}}}~e^{i{\bf p}_Y (2\brho/3+\bR)}~\chi_{\lambda_Y} 
\phi(\xi_Y) ~ \sqrt{2 E_{23}}
e^{i{\bf {P}}_{23}(\bR-\brho/3)}~\phi_{\epsilon^*_{23}}^{f_{23}}(\br)  
~{\cal G}(\br, \brho)~,
\label{psint}\ee
where ${\bf{P}}_{23}$ is the total momentum of the $(2,3)$ system and 
the intrinsic part of the  
two-nucleon
state, $\phi_{\epsilon^*_{23}}^{f_{23}}(\br)$, has quantum numbers 
${f_{23}}$ and energy
eigenvalue $\epsilon^*_{23}$.

Disregarding the photon coupling to the spectator pair, one can apply the
familiar approximation 
\be
\la  
\Psi^{f}(1,2,3)|
\hat j_\mu(\hat{\bf r}_1)
|\Psi_3^{{\bf S}_3}(1,2,3)\ra  
\approx{ 1 \over \sqrt{3} 
%{\sout{(2\pi)^{3/2}}}
\sqrt{V}}~{ \sqrt{2 E_Y}}
\nonu \times ~\int d\br_1 d\br_2 d \br_3 \Psi_{23}^{*f}
(\br_2,\br_3)
e^{-i{\bf p}_Y\br_1} \chi^+_{\lambda_Y}\phi^*(\xi_Y)
~{\cal G}(\br_1,\br_2,\br_3) \hat j_\mu(\br_1)
\Psi_3^{{\bf S}_3}(\br_1,\br_2,\br_3)~, \nonu
\label{r2}
\ee
with 
\be
{{
\Psi_3^{{\bf S}_3}(\br_1,\br_2,\br_3)
=\sqrt{2E_3}
e^{i {\bf{P}}_3 \cdot {\bf{R}} } \psi_3^{{\bf S}_3}({\bf{r}},{\bf{\rho}})
=\sqrt{2M_3}
\psi_3^{{\bf{S}}_3}({\bf{r}},{\bf{\rho}})~,
}}
\ee
where $\psi_3^{{\bf{S}}_3} ({\bf{r}},{\bf{\rho}})$ is the intrinsic
nuclear wave function and the total momentum of the nucleus 
is ${\bf{P}}_3=0$.

Moreover, if  
${\cal G}(\br_1,\br_2,\br_3)$ is such that:
(i) it does not depend upon spins 
and (ii) it commutes  with $\hat j_\mu(\br_1)$
(as it does in PWIA, since ${\cal G}(\br_1,\br_2,\br_3)\equiv 1$ ), 
one can write
\be
\la  
\Psi^{f}(1,2,3)|
\hat j_\mu(\hat{\bf r}_1)
|\Psi_3^{{\bf S}_3}(1,2,3) \ra 
\approx{1 \over \sqrt{3} \sqrt{V}} ~
{ \sqrt{2 E_Y }}
\nonu \times ~\int d\br_1 d\br_2 d \br_3 \Psi_{23}^{*f}(\br_2,\br_3)
~
e^{-i{\bf p}_Y\br_1} \chi^+_{\lambda_Y}\phi^*(\xi_Y)
\hat j_\mu(\br_1)
~ {\cal G}(\br_1,\br_2,\br_3)\Psi_3^{{\bf S}_3}(\br_1,\br_2,\br_3)~.
\label{r3}
\ee
This is the main  assumption of our approach, that is exact when 
the one-body operator 
$\hat j_\mu$ does not contain the momentum $\hat {\bf p}$. Otherwise 
one can have a non-zero commutator 
$[\hat j_\mu, {\cal G}]$. In the present SIDIS case,
the explicit expression of { the transition current operator}
$\hat j_\mu$ is unknown and we cannot compute the
commutator, but  we assume 
a vanishing result, namely $[\hat {\bf p}, {\cal G}(1,2,3)] \sim
\partial/\partial {\brho} ~{\cal G}({\bf r},{\brho})\sim 0$.
It is worth noting that if only the longitudinal part of the current 
operator is relevant
and the dependence on the coordinates in the Glauber operator 
is mainly given by the
transverse components,  
one can largely justify our assumption. 
{ As a matter of fact,} we adopt
in the following the same approach used in 
Ref. \cite{nostro}, where the distorted spectral function
was evaluated only in the 2bbu channel.
This amounts to consider GEA
(see, e.g., Ref. \cite{ourlast} and references therein).
In this scheme, { the Glauber amplitude reads 
\be
{\cal G}( \br_1, \br_2, \br_3)=\prod\limits_{i=2,3}
\left[ 1-
\theta( \br_{i||}- \br_{1||})
\Gamma \left(  
\br_{i\perp}- \br_{1\perp}
, 
\br_{i||}- \br_{1||}
\right) \right ]~,
\label{gl}
\ee}
where the parallel and perpendicular components of the vectors $\br_i$ 
are determined with respect to 
${\bf{p}}_Y$, i.e. to the
direction of
propagation of the debris. In DIS, when $|\bq|^2 \gg  |{\bf p}_N|^2$, 
this direction coincides with the direction of
$\bq$. 
The profile function 
$
\Gamma \left( \br_{i\perp}-\br_{1\perp},\br_{i||}-\br_{1||}
\right)
$
in Eq. (\ref{gl}), unlike in the standard Glauber approach,
depends not only upon the  transverse relative separation
 %${\bf b}=
%{\bf b}_1-{\bf b}_2$
but also upon the longitudinal one.
The Heavyside function $\theta(\br_{i||}-\br_{1||})$ 
assures causality in the re-scattering process.
In the following we adopt for
$
\Gamma \left( \br_{i\perp}-\br_{1\perp},\br_{i||}-\br_{1||}
\right)
$
the expression already used in Refs.
\cite{nostro,ourlast}, based on the hadronization model
of Ref. \cite{ckk}
to evaluate the total cross section of the debris-nucleon
interaction, depending
on the kinematics of the process, viz
\be
\Gamma \left( 
\br_{i\perp}-\br_{1\perp}
,
\br_{i||}-\br_{1||}
\right)={(1-i\eta)~\sigma_{eff}(
\br_{i||}-\br_{1||}
)\over 4\pi b_0}~ exp\left[-{
(
\br_{i\perp}- \br_{1\perp}
)^2\over 2 b^2_0}\right]
\label{profile}\ee
%In 
%Refs. \cite{nostro,ourlast}, details on the model and on the 
%values of the effective hadronic cross-section $\sigma_{eff}(\br_{1||})$, as
%well as  the
%parameters $\eta$ and $b_0$ can be found}.
{{ 
In this approach, the resulting Glauber operator
turns out to be mildly dependent on the longitudinal
distance, so that 
the assumption of a vanishing commutator between the operator
and the current is qualitatively
justified in the present scheme. Details on the model and on
the corresponding parameters can be found
in Refs. \cite{nostro,ourlast}.

An important issue has now to be addressed.
The effective cross section, $\sigma_{eff}$, in Eq. (32),  
models the hadronization of the debris interacting with the recoiling 
nuclear system. The debris consists of one nucleon and radiated mesons 
and gluons. The number of radiated gluons depends on the momentum scale 
of the process, given by $Q^2$. Besides, the emission of mesons and 
gluons will stop when a maximum longitudinal distance is reached, 
which increases with the invariant mass, $W_Y$, 
of the debris. As a consequence, $\sigma_{eff}$ 
depends also on $W_Y$. 
Therefore, in Eq. (\ref{profile}) one should write
$\sigma_{eff}(\br_{i||}-\br_{1||},Q^2,W_Y)$ 
and not simply  
$\sigma_{eff}(\br_{i||}-\br_{1||})$.
Nevertheless, in the kinematics we are going to discuss 
in this paper it occurs that: i) for a given value of ${\cal{E}}$, 
the range of variation of $Q^2$ is not wide enough to produce 
important changes in the gluon radiation rate; ii) 
$\sigma_{eff}$  depends weakly on the maximum longitudinal distance. 
In other words, in the kinematics we are going to analyze, 
for a given ${\cal {E}}$, the dependence 
of $\sigma_{eff}$ on
$Q^2$ and $W_Y$ is weak. 
As a matter of facts, in Refs. \cite{ourlast,nostro},
$\sigma_{eff}(\br_{i||}-\br_{1||},Q^2,W_Y) \simeq  
\sigma_{eff}(\br_{i||}-\br_{1||})$ was assumed in actual
calculations.
In Ref. \cite{ourlast},
the model of $\sigma_{eff}$ with this assumption was proven to be able
to reasonably describe data of Ref. \cite{sigmaeff}
for unpolarized spectator
SIDIS processes, in a kinematics which is close to the
one we are discussing. Therefore, to avoid a too heavy notation, 
throughout the paper we drop the 
dependence of $\sigma_{eff}$ on
$Q^2$ and $W_Y$ in the relevant expressions.}}

For completeness we mention that, in the actual form for 
${\cal G}({\bf r},{\brho})$,  Eq. \eqref{gl}, there is 
a theta-function that generates a contribution to the commutator  
proportional to $\delta^3(\brho)$. Obviously, such a contribution  
is  vanishing if  not too much severe singularities are  present
in both target and spectator wave functions.
It is worth noticing that in the quasi-elastic case, where an explicit 
form of the current operator is commonly accepted,
the above assumption, called the {\em factorized form} of FSI
has been discussed against the  unfactorized one  in  Ref. \cite{unfact}.

Coming back to  Eq. (\ref{r3}) and
following the spirit of the standard procedure adopted in PWIA,
one can { insert the one-nucleon }completeness (cf Eq. \eqref{comp})
{ \be
%\hat I 
%\delta(\br_1-\br_1')
%=
%\int d \br_1 d \br_2 d \br_3
\sum_{\lambda} \int\frac{d\bf k}{{2 E_k}(2\pi)^3 } 
 |k, \lambda \ra \la k, \lambda | 
= I,
\ee 
%\be
%\sum_\beta \Rightarrow \sum_{S_\beta}\int\frac{d{\bf k}_{2}}{(2\pi)^3} 
%\frac{d{\bf k}_3}{(2\pi)^3}
%\ee
where $I$ is the identity, and the free { nucleon }states} $| k , \lambda \ra$ are normalized
according to $ \la k , \lambda |  k' , \lambda' \ra = 2 E_k \delta_{\lambda
\lambda'}~{ (2\pi)^3 \delta({\bf k}-{\bf k}')}$.
Then, one can obtain from 
Eq. (\ref{r3}) the  following expression
\be
\la  
\Psi^{f}(1,2,3)|
\hat j_\mu(\hat{\bf r}_1)
|\Psi_3^{{\bf S}_3}(1,2,3)\ra  
\approx{1 \over \sqrt{3} 
%{\sout{(2\pi)^{3/2}}}
\sqrt{V}}~
 \int { \frac{d\bf k}{(2\pi)^3 2E_k}}\sum_{\lambda}  
{  \la p_Y, \lambda_Y; \phi(\xi_Y)| \hat j_\mu(\hat\br_1')| k, 
\lambda \ra }
 \nonu \times~
  \left[{ \sqrt{2E_k}}
\int d\br_1 d\br_2 d \br_3 \chi_\lambda^\dagger
e^{-i{\bf k} \br_1 }
\Psi_{23}^{*f}(\br_2,\br_3) \,
{\cal G}(\br_1,\br_2,\br_3)
\Psi_3^{{\bf S}_3}(\br_1,\br_2,\br_3)
 \right ]~.
 \label{w1}
 \ee
By  changing  coordinates, (see Eq. \eqref{psint}), and exploiting the 
translation invariance of the initial vertex in Eq. \eqref{overl}, 
one gets  for Eq. (\ref{w1}) 
\be
\la  
\Psi^{f}(1,2,3)|
\hat j_\mu(\hat{\bf r}_1)
|\Psi_3^{{\bf S}_3}(1,2,3)\ra  
\approx{1 \over \sqrt{3} 
%{\sout{(2\pi)^{3/2}}} 
\sqrt{V}}
\sum_{\lambda } 
\int
 \frac{d {\bf k}}{(2\pi)^3 2 E_k} \, 
\la p_Y, \lambda_Y\phi(\xi_Y) |\hat j_\mu(0)|  k ,\lambda\ra 
~
(2\pi)^3 \delta( {\bf q } + {\bf k}-{\bf p}_Y)\nonu \times ~
(2\pi)^3 \sqrt{2 E_k 2E_{23} 2 M_3}~\delta(\bk + {\bf P}_{23})~
\int d\br d\brho\left[  
\chi_\lambda^\dagger 
e^{-i2{\bf k}\brho/3}~e^{i{\bf P}_{23}\brho/3}~
\phi_{\epsilon_{23}^*}^{f_{23}*}(\br)
{\cal G}(\br,\brho) 
\Psi_3^{{\bf S}_3}(\br,\brho)
\right]=
\nonu
=
{(2\pi)^{
%\sout{{3/2}} 
3 }
\over \sqrt{3} \sqrt{2E_{mis}V}}~\sqrt{ 2E_{23} 2 M_3}
~\delta( {\bf q } - {\bf p}_{mis}-{\bf p}_Y)\nonu \times ~
\sum_{\lambda } 
 \,
\la p_Y, \lambda_Y\phi(\xi_Y) |\hat j_\mu(0)|  p_{mis} ,\lambda\ra 
 ~
\int d\br d\brho\left[  
\chi_\lambda^\dagger 
e^{i{\bf p}_{mis}\brho}
\phi_{\epsilon_{23}^*}^{f_{23}*}(\br)
{\cal G}(\br,\brho) 
\Psi_3^{{\bf S}_3}(\br,\brho)
\right]
~~,
\label{r5}
\ee
where    { $\la p_Y, \lambda_Y\phi(\xi_Y) |\hat j_\mu(0)|  k ,\lambda\ra$ 
is the
matrix element of the unknown { transition current}  operator involved in 
the quark-photon vertex
(notice that the factor $\sqrt{2E_Y 2 E_k}$ is put inside 
the matrix element),} and
$ p_{mis}\equiv \{E_ {mis}=\sqrt{M^2_N+|{\bf p}_{mis}|^2},~ -{\bf
p}_{mis}\}$, 
{ with}  ${\bf p}_{mis}={\bf q } -{\bf p}_Y={\bf P}_{23}$
the three-momentum of the  system "23" in the final state, 
 $\Psi_{23}^{f}(\br_2,\br_3)$. Indeed, ${\bf p}_{mis}$ is also the 
three-momentum 
 of the initial spectator system and
 eventually of the nucleon (with opposite sign) before absorbing the virtual photon.
This is a consequence of the assumed commutativity between  the one-body current
and  the Glauber amplitude.
 It should be pointed out  that the matrix element
 $\la p_Y, \lambda_Y\phi(\xi_Y) |\hat j_\mu(0)| p_{mis}\lambda\ra$ 
describes SIDIS off a free nucleon, within our approach.

Summarizing the above results and recalling that $X\to (A-1) \otimes Y 
\to (A-1) \otimes h\otimes X'$, one can write the hadron tensor for a polarized
$^3$He target as follows 
\be
W_{\mu\nu}^{s.i.}({\bf{S_3}},Q^2,P_h) 
= 
\frac{1}{2 M_3} {\sum\limits_{X}}
 \la  S_3, P_3|\hat J_\mu|P_{h},X\ra  \la P_{h},X|\hat J_\nu| S_3, P_3 
\ra  ~
\delta\left(M_3+\nu-E_{X'}-E_{h}-E_{23}\right)~d\tau_X =
\nonu
\approx
\frac{3}{ V} (2 \pi)^3 {\sum\limits_{X'}} d\tau_X'\sum_{f_{23}}
\sum \! \!\! \!\! \!\! \!\int_{~\epsilon^*_{23}}
\rho\left(\epsilon^*_{23}\right)
~\int {d {\bf  p}_{mis}\over 2 E_{mis}}
\delta\left(M_3+\nu-E_{X'}-E_{h}-E_{23}\right)\nonu
 \delta( {\bf q } - {\bf p}_{mis}-{\bf p}_h -{\bf p}_{X'})
\sum_{\lambda } 
 \,
\la p_Y, \lambda_Y\phi(\xi_Y) |\hat j_\mu(0)|  p_{mis} ,\lambda\ra 
 ~
\int d\br d\brho\left[  
\chi_\lambda^\dagger 
e^{i{\bf p}_{mis}\brho}
\phi_{\epsilon_{23}^*}^{f_{23}*}(\br)
{\cal G}(\br,\brho) 
\Psi_3^{{\bf S}_3}(\br,\brho)
\right]
\nonu
 \delta( {\bf q } - {\bf p}_{mis}-{\bf p}_h-{\bf p}_{X'})
\sum_{\lambda' } 
 \,
\la p_{mis}, \lambda'
|\hat j_\mu(0)|\phi(\xi_Y)  p_Y, \lambda_Y \ra 
 ~
\int d\br d\brho\left[  
\chi_{\lambda'}^{ \dagger} 
e^{i{\bf p}_{mis}\brho}
\phi_{\epsilon_{23}^*}^{f_{23}*}(\br)
{\cal G}(\br,\brho) 
\Psi_3^{{\bf S}_3}(\br,\brho)
\right]^*=
\nonu
=3 (2 \pi)^3 {\sum\limits_{X'}} d\tau_X'\sum_{f_{23}}
\sum \! \!\! \!\! \!\! \!\int_{~\epsilon^*_{23}}
\rho\left(\epsilon^*_{23}\right)
~\int {d {\bf  p}_{mis}\over  2 E_{mis}}
~\delta\left(M_3+\nu-E_{X'}-E_{h}-E_{23}\right)
 \delta( {\bf q } - {\bf p}_{mis}-{\bf p}_h -{\bf p}_{X'} )\nonu 
\sum_{\lambda } 
 \,
\la p_Y, \lambda_Y\phi(\xi_Y) |\hat j_\mu(0)|  p_{mis} ,\lambda\ra 
 ~
\int d\br d\brho\left[  
\chi_\lambda^\dagger 
e^{i{\bf p}_{mis}\brho}
\phi_{\epsilon_{23}^*}^{f_{23}*}(\br)
{\cal G}(\br,\brho) 
\Psi_3^{{\bf S}_3}(\br,\brho)
\right]
\nonu
\sum_{\lambda' } 
 \,
\la \lambda', p_{mis} |\hat j_\mu(0)|\phi(\xi_Y)  p_Y, \lambda_Y\ra 
 ~
\int d\br d\brho\left[  
\chi_{\lambda'}^{ \dagger} 
e^{i{\bf p}_{mis}\brho}
\phi_{\epsilon_{23}^*}^{f_{23}*}(\br)
{\cal G}(\br,\brho) 
\Psi_3^{{\bf S}_3}(\br,\brho)
\right]^*
\ee
where ${\bf p}_Y={\bf p}_h +{\bf X'}$ has been inserted and the following phase
space of the spectator system has been adopted
\be
\sum_{f_{23}}
\sum \! \!\! \!\! \!\! \!\int_{~\epsilon^*_{23}}
\rho\left(\epsilon^*_{23}\right)
~\int {d {\bf  P}_{23}\over  (2\pi)^3 2 E_{23}} = \sum_{f_{23}}
\sum \! \!\! \!\! \!\! \!\int_{~\epsilon^*_{23}}
\rho\left(\epsilon^*_{23}\right)
~\int {d {\bf  p}_{mis}\over  (2\pi)^3 2E_{23}}~.
\ee
 
In conclusion, the nuclear hadronic tensor reads 
\be
W_{\mu\nu}^{s.i.}({\bf{S_3}},Q^2,P_h) =
\sum\limits_{\lambda\lambda'}\sum_N \int d{\bf p}_{mis} \int
dE~
{m_N \over E_{mis}}
w_{\mu\nu}^{N \, s.i.}(\tilde p_{mis},P_h,\lambda'\lambda)
{\cal P}^{ N \,{\bf S}_{3}}_{\lambda\lambda'}(E,{\bf p}_{mis})~,
\label{A_fsi}
\ee
%where ${\bf p_N} = - {\bf p}_{mis}$,
where
the  semi-inclusive  nucleon  tensor  (cf. Eq. (\ref{wm})) is given by
\begin{eqnarray}
w_{\mu\nu}^{N \,s.i.}(-\tilde p_{mis},P_h,\lambda'\lambda)= 
\frac{1}{2m_N}
\sum\limits_{X'}
\la  {   p}_{mis},\lambda'|\hat j_\mu^N | P_h,X'\ra \la P_h,X'|\hat j_\nu^N|
{  
p}_{mis},\lambda\ra
\delta^4
\left (  q - \tilde p_{mis} - P_h - P_{X'} \right)
d\tau_{X'}\, ,
\label{wmfsi}
\end{eqnarray}
with $\tilde p_{mis}\equiv \{E-m_N, {\bf p}_{mis}\}$ and the isospin formalism
has been released (i.e. $3\to \sum_N$).
In Eq. \eqref{A_fsi}, it has been introduced the {\it distorted} spin-dependent 
spectral function given by the following expression for a polarized $^3$He target
\be
{ {\cal P}}^{N \,{\bf S}_3}_{\lambda\lambda'}(E,{\bf p}_{mis})=
\sum_{f_{23}} 
\sum \! \!\! \!\! \!\! \!\int_{~\epsilon^*_{23}}\rho\left(
\epsilon^*_{23}\right)\,
{ \tilde {\cal O}}_{\lambda\lambda'}^{N \, {\bf S}_3 \, f_{23}}
(~\epsilon^*_{23},{\bf p}_{mis})
\,
{ \delta\left( { E+ M_3-m_N-M^*_{23}}\right)}~,
\label{spectr}
\ee
 with
 the  product of distorted overlaps defined by 
\be
{ \tilde {\cal O}}_{\lambda\lambda'}^{N \, {\bf S}_3 \, f_{23}}
(\epsilon^*_{23},{\bf p}_{mis})=
\nonu =
\la \lambda, 
e^{-i{\bf p}_{mis}\brho} 
\phi_{\epsilon_{23}^*}^{f_{23}}(\br) 
{\cal G}(\br,\brho)
| 
\Psi_3^{{\bf S}_3}(\br,\brho)\ra
\la \Psi_3^{{\bf S}_3}(\br',\brho')| {\cal G}(\br',\brho')
\phi_{\epsilon_{23}^*}
^{f_{23}}
(\br') 
e^{-i{\bf p}_{mis}\brho'}, \lambda' \ra.
\label{overfsi}
\ee
{ with an obvious meaning of the adopted notation (see the { Appendix
\ref{overls}
for} the
detailed expression of the overlaps)}. 

{{One shoud notice that the distorted spectral function depends,
through the profile function Eq. (\ref{profile}), on the 
effective cross section $\sigma_{eff}(\br_{i||}-\br_{1||})$.
As discussed above, below Eq. (\ref{profile}),
this quantity depends, in principle, also on $Q^2$
and $W_Y$. 
As a consequence, the distorted spectral function is a 
process dependent quantity, at variance with the spectral function 
evaluated in PWIA. In principle, at any kinematical point 
(given by ${\cal E}, \theta_e, x_{Bj}$, and $\theta_{p_{mis}q}$)  
one should evaluate a different distorted spectral function. 
Nevertheless, for the reasons discussed below Eq. (32), 
in the kinematics we are going to study, 
for a fixed initial electron energy 
${\cal E}$
and scattering angle $\theta_e$ the 
dependence of $\sigma_{eff}$ on $Q^2$ and 
$W_Y$ is rather mild and can be disregarded. 
As a consequence, also the spectral function, for fixed
${\cal E}$
and $\theta_e$, can be considered independent on
$x_{Bj}$ and $\theta_{p_{mis}q}$. To avoid a too heavy notation,
this dependence is not shown throughout the paper.
}}

The generalization of the above formalism  to a polarized nuclear target 
with $A$ nucleon is
straightforward. In particular,
for the nuclear cross section
$\sigma^A({\bf S}_A)\equiv ~\displaystyle\frac{d\sigma({\bf S}_A)}
{d\varphi_e dx_{Bj} dyd{\bf P}_h }$
one has
\be
   \sigma^A({\bf S}_A)
= \sum\limits_{\lambda\lambda'}\sum_N \int d{\bf p}_{mis} \int
  dE~{~\tilde\alpha m_N\over E_N}
   \sigma^N_{\lambda\lambda'}
 ~{\cal P}^{N \, {\bf S}_A}_{\lambda\lambda'}(E,{\bf p}_{mis})~.
   \label{crosa-b}
   \ee
%The quantity
% $\tilde\alpha\equiv{(p_{mis} k) }/{{\cal E} m_N}$, usually referred to as  
%the "flux factor",
%at high energies  coincides with the light-cone fraction of the nucleon 
%momentum inside the nucleus, $\alpha\equiv{A(p_{mis} \cdot q) }/{(P_A q)}$.

One should notice that, formally, Eq. (\ref{A_fsi}) coincides with
Eq. (\ref{A_si}), relative to the PWIA case, 
if the distorted spectral function is substituted by the
PWIA one.
This is a consequence of the assumption made between
Eqs. (\ref{r2}) and (\ref{r3}), concerning the commutation
property of the Glauber operator with the nucleon current.
The FSI described in this manner, called factorized FSI in the literature
(see, e.g. Ref. \cite{unfact}
and references therein),
lead to convolution-like formulas, as the ones
obtained in the PWIA case, where the distorted spectral function appears
instead of the PWIA one. The latter can be recovered just
putting the Glauber operator identically equal to 1. 
This observation has crucial consequences in the
following sections
of the present paper.
\section{The dependence of the nuclear hadronic tensor
upon the target nucleus polarization}
As a matter of facts, { the whole } formalism developed in the PWIA
case in Ref. \cite{mio} can be exploited now in the present
scenario, once the distorted overlaps are properly evaluated
and inserted in the relevant equations.

Notice that, in PWIA,
the spectral function $ P^{{\bf S}_3}_{\lambda\lambda}(E,{\bf p}_{mis})$ 
in (\ref{spectr}) defines
the probability to remove from a polarized $^3$He with polarization   
${\bf S}_3$
a polarized nucleon with momentum $-{\bf p}_{mis}$  and polarization
${\bf s}_N$ (characterized by spin projection $\lambda$ 
on the quantization axis)
leaving the remnant $(A-1)$ system with { removal } energy $E$.
Once the full FSI is taken into account, even through GEA, 
the probabilistic interpretation
of the distorted spectral function is somehow lost.

A further issue is represented by the fact that the direction  
of the target polarization-axis, 
$ {\bf S}_3$, may  not always be parallel to the direction 
which determines the eikonal ${\cal G}$-matrix, i.e.
the direction of $\bp_Y$ (or, in DIS, the direction of \bq).  
In particular, in the SIDIS process of interest here, 
the target nucleus is transversely polarized,
i.e. $ {\bf S}_3\perp \bq$.
To reconcile the polarization axis and the eikonal approximation, 
one needs to rotate the quantization axis of the target wave function
from the direction of ${\bf q}$ to the direction of the polarization
${\bf  S}_3$,  namely
\be
\la \theta,\phi|\Psi_{^3He}{\ra_{\hat{\bf  S}_3}} = \la
\theta',\phi'|D^{1/2}(0,\beta,0)|\Psi_{^3He}\ra_{\hat{\bf  q}}=\nonu
=\cos(\beta/2)\
\la \theta',\phi'|\Psi^{{  M}=1/2}_{^3He}\ra_{\hat{\bf  q}}
+ \sin(\beta/2)\ \la \theta',\phi' |\Psi^{{  M}=-1/2}_{^3He}
\ra_{\hat{\bf  q}}~,
\label{he}
\ee
where the subscript indicates the direction of the quantization axis,
$\cos \beta= {\hat{\bf  S}_3 }\cdot \hat{\bf  q}$ and 
the polarization vector ${\bf S}_3$ is
supposed to be in the $(x,z)$ plane. In Eq. \eqref{he},
{ 
$D^{1/2}_{\sigma' \sigma}$ are  the suitable  Wigner D-functions} \cite{varsha}.
{ Therefore in the general case}, the nuclear tensor in Eq. (\ref{A_fsi})
is modified and reads 
\begin{eqnarray}
{W_{\mu\nu}^{s.i.}({\bf S}_3,Q^2,P_h)}
=\cos^2(\beta/2)\ W_{\mu\nu}^{\frac12\frac12}+\
\sin^2(\beta/2)W_{\mu\nu}^{-\frac12-\frac12}+ \sin\beta\ \left[\frac12\left(
W_{\mu\nu}^{\frac12-\frac12}+W_{\mu\nu}^{-\frac12\frac12}\right)\right]
\label{roty}
\end{eqnarray}
\begin{eqnarray}
{W_{\mu\nu}^{s.i.}(-{\bf S}_3,Q^2,P_h)}=
\sin^2(\beta/2)\ W_{\mu\nu}^{\frac12\frac12}+\
\cos^2(\beta/2)W_{\mu\nu}^{-\frac12-\frac12}- \sin\beta\ \left[\frac12\left(
W_{\mu\nu}^{\frac12-\frac12}+W_{\mu\nu}^{-\frac12\frac12}\right)\right]~.
\label{roty1}
\end{eqnarray}
In the above equations, we have defined
\be
W_{\mu\nu}^{MM'} =
\sum\limits_{\lambda\lambda'}\sum_N \int d{\bf p}_{mis} \int
dE~
{m_N \over E_N}
w_{\mu\nu}^{N \, s.i.}{(-\tilde p_{mis}},P_h,\lambda'\lambda)
~{\cal P}^{N \, MM'}_{\lambda\lambda'}(E,{\bf p}_{mis})~,
\label{A_fsinew}
\ee
where the third components $M$ and $M'$ are defined with
respect to the direction $\hat
\bq$. 
In Eq. (\ref{A_fsinew})
one has (cf Eq. (\ref{spectr}))
\be
{\cal P}^{N \,MM'}_{\lambda\lambda'}(E,{\bf p}_{mis})=
\sum_{f_{23}} 
\sum \! \!\! \!\! \!\! \!\int_{~\epsilon^*_{23}}\rho\left(
\epsilon^*_{23}\right)\,
{ \tilde {\cal O}}_{\lambda\lambda'}^{N \, MM' \, f_{23}}
(~\epsilon^*_{23},{\bf p}_{mis})
\,
{ \delta\left(  E+ M_3-m_N-M^*_{23}\right)}~
~,
\label{spectrg}
\ee
with $ \tilde {\cal O}_{\lambda\lambda'}^{N \,M \, M' \, 
f_{23}}$,
a natural non-diagonal generalization of Eq. (\ref{overfsi}), viz
\be
{ \tilde {\cal O}}_{\lambda\lambda'}^{N \,M \, M' \, f_{23}}
(\epsilon^*_{23},{\bf p}_{mis})=
\nonu =
\la \lambda, 
e^{-i{\bf p}_{mis}\brho}
\phi_{\epsilon_{23}^*}^{f_{23}}(\br) 
{\cal G}(\br,\brho)
| 
\Psi_3^{M}(\br,\brho)\ra
\la \Psi_3^{M'}(\br',\brho')| {\cal G}(\br',\brho')
\phi_{\epsilon_{23}^*}^{f_{23}}(\br') 
e^{-i{\bf p}_{mis}\brho'} \lambda' \ra.
\label{overfsig}
\ee
It is worth noticing that, in Eq. (\ref{roty}),
the upper scripts $\frac12 \frac12$  $\left(-\frac12 -\frac12\right )$  
denote a nucleus polarized
along (opposite) the quantization-axis, while  
$\pm\frac12 \, \mp\frac12$ indicate a nucleus polarized in
the perpendicular (wrt the quantization-axis) plane, i.e., in our case,
along the $x$-axis.

Let us consider first a longitudinally polarized nucleus; in this case,
we have to consider in Eq. (\ref{roty}) only the terms with
$ M=M'=\pm 1/2$.
One gets the following longitudinal contribution to the hadronic tensor
\be
W_{\mu\nu}^{||}({\bf S}_3,Q^2,P_h) = \sum_{\lambda\lambda'}
\sum_N \int d{\bf p}_{mis} \int
dE~
{m_N \over E_N}
\left[
 \cos^2 \frac\beta2\, {\cal P}^{N \,\frac12 \frac12}_{\lambda\lambda'}\ 
w_{\mu\nu}^{N \, \lambda\lambda'} +
 \sin^2 \frac\beta2\, {\cal P}^{N \, -\frac12 -\frac12}_{\lambda\lambda'}\ 
w_{\mu\nu}^{N \, \lambda\lambda'}\right]~. 
\label{paral}
\ee
In Eq. (\ref{paral}), $w_{\mu \nu}^{N \, \lambda \lambda'}$
is a short-hand notation for 
$w_{\mu \nu}^{N \, s.i.}(p_{mis},P_h,\lambda' \lambda)$, previously used.
In the SIDIS process under investigation, 
since leptons are unpolarized,
the leptonic tensor is symmetric and, as a consequence, only
the symmetric part of
the hadronic spin-dependent tensor, 
$w_{\mu\nu}^{sN \lambda\lambda'}$, is involved. 
For the diagonal terms of the symmetric part of the nucleon tensor 
(see, e.g., Ref. \cite{barone} for its general structure),
one gets
\be
\la\frac12|\hat w_{\mu\nu}^{sN}|\frac12\ra
= -\la-\frac12|\hat w_{\mu\nu}^{sN}|-\frac12\ra~,
\label{inc}
\ee
while for the off-diagonal terms one has 
\be
\la-\frac12|\hat w_{\mu\nu}^{sN}|\frac12\ra=~
\la \frac12|\hat w_{\mu\nu}^{sN}|-\frac12\ra^* \quad .
\label{cc}
\ee
Then, making use of the properties under complex conjugation
of the quantities (\ref{overfsig}), 
defined with respect to the quantization
axis, namely
\be
 {\tilde{\cal O}}^{{ N \, M M}'\, f_{23}}_{\lambda\lambda'}(E,{\bf p}_{mis})=
 \left(-1\right)^{{  M}+{M }'+\lambda+\lambda'}\left(
{\tilde{\cal O}}^{N \, -{  M}-{  M}'\, f_{23}}_{-\lambda-\lambda'}
(E,{\bf p}_{mis})\right)^*~,    
\label{conj1}\\&&
 {\tilde{\cal O}}^{{  N \, M M' }\, f_{23}}_{\lambda\lambda'}(E,{\bf p}_{mis})=
\left( {\tilde{\cal O}}^{N \,{  M'}{ M}\, f_{23}}_{\lambda'\lambda}
(E,{\bf p}_{mis})
\right)^*~,
\label{prop}
\ee
one obtains 
\be
W_{\mu\nu}^{||}({\bf S}_3,Q^2,P_h)  
=
\cos\beta\, 
\sum_N \int d{\bf p}_{mis} \int
dE~
{m_N \over E_N}
\left\{\left[ {\cal P}^{N \, \frac12 \frac12}_{\frac12 \frac12}- 
{\cal P}^{N \, \frac12 \frac12}_{-\frac12 -\frac12}\right]
w_{\mu\nu}^{s N \, \frac12 \frac12}
\right. \nonu
+ \left.
\left[ {\cal P}^{N \, \frac12 \frac12}_{\frac12 -\frac12}\ 
w_{\mu\nu}^{s N \, \frac12 -\frac12}+
{\cal P}^{N \, \frac12 \frac12}_{-\frac12 \frac12}\ 
w_{\mu\nu}^{s N \, -\frac12 \frac12}\right]
\right\}~.
\label{spectrlong}
\ee

In Eq. (\ref{spectrlong}) the first term in square brackets represents 
the {\it parallel} spin-dependent
spectral function. 

We are interested in single spin asymmetries measured
with transversely polarized targets.
The relevant hadronic tensor is therefore
\be
\Delta W^{s.i.}_{\mu\nu}({\bf  S}_\perp,Q^2,P_h) 
= W_{\mu\nu}^{s.i.}({\bf  S}_3={\bf S}_\perp, Q^2,P_h)
 -W_{\mu\nu}^{s.i.}({\bf  S}_3=-{\bf S}_\perp, Q^2,P_h)\quad, 
\label{f}
\ee
where we choose ${\bf S}_\perp $ along the $x$ axis, i.e. 
$\beta=90^o$. Then,
using Eqs. (\ref{roty}) and (\ref{roty1}), the quantity
relevant to describe the JLAB experiments turns out to be
\be
\Delta W^{s.i}_{\mu\nu} ({\bf S}_\perp,Q^2,P_h)=
W_{\mu\nu}^{\frac12-\frac12}+W_{\mu\nu}^{-\frac12\frac12}~.
\label{w}
\ee
Therefore, we have to evaluate 
\be
  \Delta W^{s.i}_{\mu\nu}({\bf S}_3,Q^2,P_h)= 
\sum_{\lambda\lambda'}
\sum_N \int \hspace{-1mm} d{\bf p}_{mis} \int
\hspace{-1mm} dE~
{m_N \over E_N}
\left[ {\cal P}^{N \, \frac12 -\frac12}_{\lambda\lambda'}(E,{\bf p}_{mis})\ 
w_{\mu\nu}^{N s \lambda\lambda'}
+ {\cal P}^{N \, -\frac12 \frac12}_{\lambda\lambda'}(E,{\bf p}_{mis})\ 
w_{\mu\nu}^{N s \lambda\lambda'}\right]
=
\nonu
= 
\sum_N \int d{\bf p}_{mis} \int
dE~
{m_N \over E_N}
\left\{
\sum_\lambda
\left[ {\cal P}^{N \, \frac12 -\frac12}_{\lambda\lambda}(E,{\bf p}_{mis}) 
+ {\cal P}^{N \, -\frac12 \frac12}_{\lambda\lambda}(E,{\bf p}_{mis})\right] 
w_{\mu\nu}^{sN \lambda\lambda} \right. 
\nonu \left. +
\sum_\lambda
\left[ {\cal P}^{N \, \frac12 -\frac12}_{\lambda-\lambda}(E,{\bf p}_{mis}) 
+ {\cal P}^{N \, -\frac12 \frac12}_{\lambda-\lambda}(E,{\bf p}_{mis})\right] 
w_{\mu\nu}^{sN \lambda-\lambda}
\right\} \quad . 
\label{DW}
\ee
Therefore one obtains, for the term in the last line
of Eq. (\ref{DW}),
\be
\sum_\lambda
\left[ {\cal P}^{N \, \frac12 -\frac12}_{\lambda-\lambda}(E,{\bf p}_{mis}) 
+ {\cal P}^{N \, -\frac12 \frac12}_{\lambda-\lambda}(E,{\bf p}_{mis})\right] 
w_{\mu\nu}^{sN \lambda-\lambda}=
\nonu
= 2 \Rea \left\{\left[ {\cal P}^{N \, \frac12 -\frac12}_{\frac12-\frac12}(E,{\bf p}_{mis}) 
+ {\cal P}^{N \, -\frac12 \frac12}_{\frac12-\frac12}(E,{\bf p}_{mis})\right] 
w_{\mu\nu}^{sN \frac12-\frac12}\right\}=
\nonu
=2 \Rea \left[ {\cal P}^{N \, \frac12 -\frac12}_{\frac12-\frac12}(E,{\bf p}_{mis}) 
+ {\cal P}^{N \, -\frac12 \frac12}_{\frac12-\frac12}(E,{\bf p}_{mis})\right] 
 \Rea \Bigl[w_{\mu\nu}^{sN \frac12-\frac12}\Bigr]
\nonu
- 2 \Ima  \left[ {\cal P}^{N \, \frac12 -\frac12}_{\frac12-\frac12}(E,{\bf p}_{mis}) 
+ {\cal P}^{N \, -\frac12 \frac12}_{\frac12-\frac12}(E,{\bf p}_{mis})\right] 
 \Ima \Bigl[w_{\mu\nu}^{sN \frac12-\frac12}\Bigr]
 \label{pr}
\ee
where the relations (\ref{cc}) and (\ref{prop}) have been used.  

In Appendix \ref{GSF} it is shown that the contribution 
of the last line { in Eq. (\ref{pr})
 can be safely neglected, being of higher order in ${\bf p}_\perp/m_N$, where
 ${\bf p}_\perp$ is the nucleon transverse-momentum inside the target, with
 ${\bf p}=-{\bf
 p}_{mis}$}.
Besides, in the remaining expression, only the zero order term
in { ${\bf p}_\perp/m_N$  yields a sizable contribution.
Hence, ${\bf p}_\perp $ does not give} relevant
contributions to 
the hadronic tensor,  and  the 
expression of the nucleon hadronic tensor obtained
in a collinear frame, where $\bf p_\perp = 0 $,
for example
the one given in Ref. \cite{barone} 
for the Collins process
{ (cf section 6.5)}, can be safely used.
As a consequence, the final expression for the nuclear hadronic tensor,
suitable for calculations of SSAs, reads:

\be
\Delta W^{s.i}_{\mu\nu}({\bf S}_3,Q^2,P_h)
= ~\sum_N \int d{\bf p}_{mis} \int
dE~
{m_N \over E_N}
\left \{
{\cal P}^{N \, \perp}(E,{\bf p}_{mis}) w_{\mu\nu}^{N \perp} + 
%\right. 
%\nonu
%\left.+ i \Ima
%\left[ {\cal P}^{N \, \frac12 -\frac12}_{\frac12 -\frac12}(E,{\bf p}_{mis})\ 
%+ ~ {\cal P}^{N \, -\frac12 \frac12}_{\frac12 -\frac12}(E,{\bf p}_{mis})\ 
%\right]
%\left [w_{\mu\nu}^{N\frac12 -\frac12} - w_{\mu\nu}^{N -\frac12 \frac12} \right]
%+ \right. 
%\nonu
%\left.+ \
2 
\Rea \left[
{\cal P}^{N \, (\perp-||)}(E,{\bf p}_{mis})
\right]
w_{\mu\nu}^{s N \, \frac12 \frac12}\right\}~,
\label{transv}
\ee
where { Eqs. (\ref{inc}) and (\ref{conj1})} have been used
to obtain the last term.

In Eq. (\ref{transv}), the transverse spectral function has been introduced
\be
{\cal P}^{N \, \perp}(E,{\bf p}_{mis}) =
 \Rea \left [ {\cal P}^{N \, \frac12 -\frac12}_{\frac12 -\frac12}(E,{\bf p}_{mis})\  + ~
 {\cal P}^{N \, -\frac12 +\frac12}_{\frac12 -\frac12}(E,{\bf p}_{mis})\ 
\right ]~;
\ee
and the quantity
\be
w_{\mu\nu}^{N \perp}
\equiv
 \left[ w_{\mu\nu}^{s N \, \frac12 -\frac12}
+w_{\mu\nu}^{s N \, -\frac12 \frac12}
      \right]~
\ee
has been defined.
Furthermore, in Eq. (\ref{transv}), the 
transverse-longitudinal spectral function, { 
\be
{\cal P}^{N \, (\perp-||)}(E,{\bf p}_{mis})={\cal P}^{N \, \frac12 -\frac12}_{\frac12 \frac12}(E,{\bf p}_{mis})+
 {\cal P}^{N \, -\frac12 \frac12}_{\frac12 \frac12}(E,{\bf p}_{mis})~
\ee 
is a real} quantity which represents, in PWIA, the
probability to find a longitudinally polarized nucleon in a transversely 
polarized nucleus.
%is the hadronic tensor of a transversely polarized nucleon with 
%polarization ${\bf S}^N$ along 
% the  $x$-axis, i.e.  along ${\bf S}_3$. 
It should be pointed out that, {in PWIA},
the transverse spectral function ${\cal P}^{N \, \perp}(E,{\bf p}_{mis})$
yields the probability to find a transversely polarized 
nucleon in a transversely polarized nucleus with
a polarization vector  ${\bf S}_{3}$ along the $x$-axis.

For the nuclear cross section Eq. (\ref{crosa-b})
one gets
\be
\sigma^3({\bf S}_{3})
= \sum_N \int d{\bf p}_{mis} \int
  dE~{~ \tilde \alpha  m_N\over E_N}
  \left[
    \sigma^{N \, \perp}
 ~{\cal P}^{N \, \perp}(E,{\bf p}_{mis})+
  \sigma^{N \, ||}
 ~{\cal P}^{N \, (\perp-||)}(E,{\bf p}_{mis})~\right]~,
 \label{crosa-4}
   \ee
where  $\sigma^{N \, \perp}$
and  $\sigma^{N \, ||}$
are the cross sections Eq. (\ref{crosa-2})
for transversely and longitudinally polarized nucleons,
respectively.

Note also that, in PWIA, one has
\be
{P}^{N \, \perp}(E,p,\cos\theta_{pq}) = 
{P}^{N \, ||}(E,p,\cos\theta_{p{S}_3})=
{P}^{N \, ||}(E,p,\sin\theta_{pq})~,
\ee
where ${P}^{N \, ||}(E,p,\theta)$ is the spin-dependent
spectral function considered, for example, in Ref. \cite{mio}. 
It has to be pointed out that 
${P}^{N \, \perp}\ne {P}^{N \, ||}$ in the relativistic case
(see, e.g., Ref. \cite{Pacelc16}).

\section{The Collins and Sivers asymmetries for $^3$He}

As discussed in the Introduction,
a series of SIDIS experiments are planned at JLab,
using a transversely polarized $^3$He target and an unpolarized
electron beam, detecting a fast pion (kaon) 
in the final state. The Sivers and Collins
SSAs of $^3$He will be therefore
measured, with the aim of 
extracting the
corresponding neutron quantities.
The formal results of the present approach
for the $^3$He SSAs, and for the extraction
of the neutron information, are presented in this Section. 

The Sivers and Collins asymmetries are defined through
proper moments of the experimental SIDIS cross sections, viz
 \begin{eqnarray}
A_{3}^{Col(Siv)} \equiv \frac
{
\int d \phi_{S_3} d \phi_h \sin ( \phi_h \pm \phi_{S_3})
\left[\sigma^3({\bf S}_{3}, \phi_h,\phi_{S_3},z) - 
\sigma^3({\bf S}_{3}, \phi_h,\phi_{S_3}+\pi,z)\right]}
{\int   d \phi_h
\sigma^{3}_{unpol}(x_{Bj},Q^2,{\bf P}_h)}~,
\label{acolsiv}
\end{eqnarray}
where  $\phi_h$ is the azimuthal angle between
the hadron and the lepton planes,  $\phi_{S_3}$ is the azimuthal angle between
the target polarization and the lepton plane, according to 
the conventions  fixed in Ref.
\cite{trento}; $z=E_h/\nu$ is the fraction of energy transfer
carried by the detected meson.
Inserting 
the cross section Eq. (\ref{crosa-4}) in 
the above equation,
one gets
\begin{eqnarray}
A_{3}^{Col(Siv)} =
\frac
{
\int_{x_{Bj}}^3 
d\alpha
\left[
\Delta \sigma_{Col(Siv)}^n\left (x_{Bj}/\alpha ,Q^2, z
\right  ) 
f^{\perp,i}_n(\alpha ,Q^2,{\cal E})+
2\Delta 
\sigma_{Col(Siv)}^p\left (x_{Bj}/\alpha ,Q^2, z \right ) 
f^{\perp,i}_p(\alpha ,Q^2,{\cal E}) \right]
}
{\int d\alpha\left[
 \sigma^n\left (x_{Bj}/\alpha ,Q^2, z \right ) f_n^i(\alpha ,Q^2,{\cal E})+
2\sigma^p\left (x_{Bj}/\alpha ,Q^2, z \right ) f_p^i(\alpha ,Q^2,{\cal E}) \right]}~,
\label{A3}
\end{eqnarray}
where ${\cal E}$ is the energy of the incoming lepton (see below Eq.
\eqref{crosa-1}) and  $f^{\perp,i}_{p(n)}(\alpha ,Q^2,{\cal E})$ are
the light-cone momentum distributions
of transversely polarized nucleons in a transversely polarized
nucleus for $i=$ PWIA or FSI. One defines
\be
f^{\perp,i}_{N}(\alpha ,Q^2,{\cal E})=
\int_{E_{min}}^{E_{max}} dE
f^{\perp,i}_{N}(\alpha,Q^2,{\cal E},E)~, 
\label{falpha_perp}
\ee
where
\be
f^{\perp,i}_{N}(\alpha ,Q^2,{\cal E},E) =
%\int\limits_{{p}_{min}( \alpha,Q^2\ldots)}^{p_{max}
%( \alpha,Q^2\ldots)}
\int d {\bf p}_{mis}~~
\frac{m_N}{E_N}
%S^{\perp,i}_{p(n)}(E,{\bf p}_{mis})
{\cal P}^{N \, \perp, \, i} \, (E,{\bf p}_{mis})
\delta\left( \alpha+\frac{\tilde p_{mis} \cdot q}{m_N \nu}\right)
\theta \left ( W_Y^2 - (m_N+m_\pi)^2\right) ~,
\label{integranda1}
\ee
with $W_Y$ the invariant mass of the debris $Y$, that hadronizes in a
nucleon and, at least, one
pseudoscalar meson. 
For the sake of definiteness, in Eq. \eqref{integranda1} and in what 
follows we consider a $\pi^-$ in the final state.
Let us recall that in the unpolarized case,
the light-cone momentum distributions read
\be
f^{i}_{N}(\alpha,Q^2,{\cal E})=
\int_{E_{min}}^{E_{max}} dE 
f^{i}_{N}(\alpha,Q^2,{\cal E},E)~,
\label{falpha}
\ee
with
\be
f^{i}_{N}(\alpha, Q^2, {\cal E}, E)=
\int d {\bf p}_{mis}~
%\limits_{{p}_{min}( \alpha,Q^2\ldots)}^{p_{max}
%( \alpha,Q^2\ldots)}
\frac{m_N}{E_N}~ { {\cal P}^{Ni}(E,{\bf p}_{mis})}
\delta\left( \alpha+\frac{\tilde p_{mis} \cdot q}{m_N \nu}\right)
\theta \left ( W_Y^2 - (m_N+m_\pi)^2\right) ~,
\label{integranda2}
\ee
where ${ {\cal P}^{Ni}(E,{\bf p}_{mis})}= 
\sum_{\lambda} {\cal P}_{\lambda \lambda}^{Ni}$.
In Eqs. (\ref{integranda1}) and (\ref{integranda2}), 
the delta function can be eliminated by integrating over
the angle between  ${\bf {p}}_{mis}$ and ${\bf q}$;
the limits of integration on 
$|{\bf p}_{mis}|$, i.e.
$|{\bf p}_{min}|$ and $|{\bf p}_{max}|$,
and on $E$, $E_{min}$ and $E_{max}$,  
are determined from the 
condition $|\cos \theta_{pq}|\le 1$ and,
 from 
the requirement $W_Y^2 \ge (m_N+m_\pi)^2$, { since we consider SIDIS with at least one pion in the 
final state}.
As a consequence, $|{\bf p}_{min}|$ and $|{\bf p}_{max}|$ 
are functions of
$\alpha,E,Q^2,{\cal E}$. One should notice that,
in the Bjorken limit, they would be functions of $\alpha$
and $E$ only.  In Eqs. \eqref{falpha_perp} and \eqref{falpha}, one has
$E_{min}=B_3-B_2\sim 5.5~ MeV$.

Moreover, as explained in the previous section, one can obtain
the distributions for the two cases, $i=$PWIA, FSI, just substituting,
in the same equations, the corresponding spectral functions
${\cal P}^{N \, i}(E,{\bf p}_{mis})$ 
and ${\cal P}^{N\perp \, i}(E,{\bf p}_{mis})$.
The evaluation of   ${\cal P}^{N \, i}(E,{\bf p}_{mis})$, when 
 both the nuclear structure and the effects of FSI are included, 
is the main technical achievement
of this paper. 
Actual numerical results, based on (i) two and three nucleon wave functions
\cite{pisa}
evaluated with the nucleon-nucleon AV18 interaction \cite{av18},
and (ii) the GEA mechanism,  are discussed in detail
in the following Section. {In what follows, when 
the distorted spectral functions will be
considered in Eqs. \eqref{integranda1} and \eqref{integranda2},   we will call
  the
distribution functions in Eqs. \eqref{falpha_perp} and \eqref{falpha} 
{\em  distorted
light-cone momentum distributions} { (see Appendix \ref{GSF})}. 

 In Eq. \eqref{acolsiv}, one should notice that, 
after multiplying the nuclear hadronic tensor by
$\sin(\phi_{{S}_3} \pm \phi_h)$
and integrating over $\phi_{{S}_3}$, the transverse-longitudinal 
term in Eq. (\ref{crosa-4}) 
does not contribute to the numerators
in the asymmetries above defined, due to the properties of the 
spin-dependent SIDIS nucleon tensor
\cite{barone1}~.

In Eq. (\ref{A3}), the quantities $\Delta \sigma_{Col(Siv)}^{N}$
and $\sigma^{N}$,
related to the structure of the bound nucleon,
are defined as follows { (see, e.g., \cite{barone1})}
\begin{eqnarray}
\Delta \sigma_{Col}^N\left (x_{Bj},Q^2, z \right  )
& = &
{ {1 - y \over 1 - y + y^2/2}}
\nonumber \\
& \times &
\sum_q e_q^2
\int  d^2 {\bkappa_T}
d^2 {\bf k}_T \delta^2 ( {\bf k}_T + {\bf q}_T  - \bkappa_T )
 {{\bf \hat{P}}_{h\,\perp} \cdot {\bkappa_T} \over m_h}
 h_1^{q,N} (x_{Bj}, {\bf k}_T^2 )
 H_1^{\perp q,h} (z, (z {\bkappa_T})^2 )~,
\label{dcoll}
\end{eqnarray}

\be
\Delta \sigma_{Siv}^N\left (x_{Bj},Q^2, z \right  )
=
 \sum_q e_q^2
\int d^2 {\bkappa_T}
d^2 {\bf k}_T
\delta^2 ( {\bf k}_T + {\bf q}_T  - {\bkappa_T} )
{ {\bf \hat{P}}_{h\,\perp} \cdot {\bf{k}_T} \over m_N}
f_{1T}^{\perp q,N} (x_{Bj}, {\bf{k}}_T^2 )
  D_1^{q,h} (z, (z \bkappa_T)^2 )~,
\label{dsiv}
\ee

\be
\sigma^N\left (x_{Bj},Q^2 , z\right  )
=
\sum_q e_q^2
\int d^2 {\bkappa_T} d^2 {\bf k}_T
\delta^2 ( {\bf k}_T + {\bf q}_T  - {\bkappa_T} )
f_1^{q,N} (x_{Bj},{\bf k}_T^2 )
D_1^{q,h}  (z, (z {\bkappa_T})^2 )~.
\label{unpol}
\ee

In the last three equations, 
the quantities ${\bf k}_T$ and  ${\bkappa_T}$ are the intrinsic 
transverse momenta of the parton in the bound nucleon
and in the produced hadron, respectively; 
following the notation of SIDIS, a subscript $T$ means
transverse with respect to ${\bf{P}}_h$ 
(the three-momentum of the final pion or kaon), while the
subscript $\perp$ means transverse with respect to ${\bf{q}}$.
The transverse momentum dependent parton distributions, 
$h_1^{q,N} (x_{Bj}, {\bf k}_T^2 )$, 
$f_{1T}^{\perp q,N} (x_{Bj}, {\bf k}_T^2 )$,
$f_{1}^{q,N} (x_{Bj}, {\bf k}_T^2 )$,
and the transverse momentum dependent fragmentation functions,
$D_1^{q,h}  (z, (z {\bkappa_T)^2} )$,
$H_1^{\perp q,h} (z, (z {\bkappa_T)^2} )$,
appearing in Eqs. (\ref{dcoll}), (\ref{dsiv}) and (\ref{unpol}),
have been evaluated
using experimental data
whenever possible, or using proper model estimates. 
One should realize that the main goal of the present study is the estimate
of nuclear effects in the extraction of the neutron information, rather
than obtaining absolute predictions on the SSAs of $^3$He, which would be 
affected anyhow by the poor present knowledge of some of the
distributions necessary to perform the actual calculation.
Any reasonable choice of the distribution functions of the
nucleon is therefore suitable for our study.
In particular, in the actual calculations we have made use
of the same functions adopted in Ref. \cite{mio}, namely:

\begin{enumerate}

\item{}
for the unpolarized parton
distribution, $f_1^{q,N}(x_{Bj})$, it has been used 
the parametrization of Ref. \cite{gluu},
with a gaussian ansatz for the ${\bf k}_T$ dependence;

\item{}
for the transversity distribution,
$h_1^{q,N}$, it has been exploited the ansatz $h_1 = g_1$, i.e.,
the transversity distribution has been taken to be
equal to the helicity distribution.
This gives certainly the correct order of magnitude.
In particular, the parametrization of Ref. \cite{glup}
has been used;

\item{}
for the Sivers function,
$f_{1T}^{\perp q} (x_{Bj}, {\bf k}_T^2)$ in Eq. (\ref{dsiv}),
it has been adopted the fit 
proposed in Ref. \cite{ans};

\item{}
for the unpolarized fragmentation function
$D_1^{q,h}(z)$, different models are used for
evaluating the Sivers and Collins asymmetries.
In particular, for the Sivers asymmetry, 
the parametrization in Ref. \cite{kret} has been used while, 
for  the Collins one,  
the model calculation of Ref. \cite{amr} has been adopted
(see \cite{mio} for details);

\item{}
for the basically unknown Collins fragmentation function,
$H_1^{\perp q}(z, (z {\bkappa}_T)^2 )$, appearing
in Eq. (\ref{dcoll}), the model calculation of Ref. \cite{amr}
has been used.

\end{enumerate}

 Equation (\ref{A3}) has been presented  in Ref. \cite{mio} within  PWIA. 
{ As already
noticed, within GEA the theoretical expression of the nuclear asymmetries does not formally
change in presence of FSI.
 Therefore 
Eq. (\ref{A3}) can be 
exploited also in this case, but using
the suitable ingredient, i.e. the distorted spin-dependent spectral function, 
and eventually  evaluating 
 the distorted light-cone momentum
distributions.}

Let us discuss now the crucial issue of the extraction of the neutron
information from $^3$He data.
A strategy for extracting the neutron Sivers and Collins asymmetries
from $^3$He data, developed
in Ref. \cite{mio}, is summarized and applied in the following.

If the results of the calculation were able to simulate $^3$He data, the
problem would { amount to} unfolding the convolution formula.
This can be done taking into account that
the light-cone momentum distributions 
$f_N(\alpha,Q^2,{\cal E})$  and $f_N^\perp(\alpha,Q^2,{\cal E})$ 
exhibit sharp maxima at
$\alpha\sim 1 $, i.e. $f_N(\alpha,Q^2,{\cal E})\sim \delta(\alpha-1)$
even in presence of FSI, as we will show
in the next Section. Let us remind that this peak is expected since 
$\alpha= -(\tilde p_{mis}\cdot q)/m_N\nu$  plays the
role of the Bjorken variable for a bound nucleon.
Assuming that the delta-like  behavior for the light-cone distributions  
is a reliable approximation (as shown in what follows), then $\Delta
\sigma_{Col(Siv)}^n\left (x_{Bj}/\alpha ,Q^2,{\bf S}_\perp^n,z \right )\sim 
\Delta \sigma_{Col(Siv)}^n\left (x_{Bj} ,Q^2,{\bf S}_\perp^n,z \right )$, 
and the calculated asymmetries    
$A_3$ can be written as { (notably, the dependence on ${\cal E}$ becomes
milder and milder, approaching the Bjorken limit)}
\be
A_3^{Col(Siv)}  \simeq 
\displaystyle\frac{\Delta\sigma_3^{Col(Siv)}}{\sigma}
\nonu
 \simeq  
\displaystyle
\frac
{
\Delta \sigma_{Col(Siv)}^n\left (x_{Bj} ,Q^2,{\bf S}_\perp^n,z \right ) 
\int d\alpha f_n^\perp(\alpha,Q^2)
+2\Delta \sigma_{Col(Siv)}^p\left (x_{Bj},Q^2,{\bf S}_\perp^n,z \right ) 
\int d\alpha f_p^\perp(\alpha,Q^2)
}
{ \sigma^n(x_{Bj} ,Q^2,z)  \int d\alpha f_n(\alpha,Q^2)  
+ 2\sigma^p(x_{Bj} ,Q^2,z)  \int d\alpha f_p(\alpha,Q^2) }~.
\label{inter}
\ee

Let us introduce the so-called ``dilution" factors as
\begin{eqnarray}
d_{p(n)}(x_{Bj},z)=
\displaystyle\frac{\sigma^{p(n)} (x_{Bj} ,Q^2,z)}
{\langle N_n\rangle  \sigma^{n} (x_{Bj} ,Q^2,z)+2\langle 
N_p\rangle\sigma^{p}(x_{Bj} ,Q^2,z)},
\label{dilf}
\end{eqnarray}
where
{
\begin{eqnarray}
\label{neff}
\langle N_{p(n)}\rangle=
%\int\limits_{{p}_{min}( \alpha,Q^2\ldots)}^{p_{max}
%( \alpha,Q^2\ldots)}
\int_{E_{min}}^{E_{max}} dE
\int d {\bf p}_{mis}~~
%\frac{m_{p(n)}}{E_{p(n)}}
%S^{\perp,i}_{p(n)}(E,{\bf p}_{mis})
{\cal P}^{p(n)} \, (E,{\bf p}_{mis})
%\delta\left( \alpha+\frac{\tilde p_{mis} \cdot q}{m_N \nu}\right)
\theta \left ( W_Y^2 - (m_{p(n)}+m_\pi)^2\right) ~,
\end{eqnarray}
Notice that, within PWIA and in the Bjorken 
limit, when 
%{ no kinematical restrictions affect $f(\alpha)$
$W_Y \rightarrow \infty$}, 
%if one takes
%${\alpha m_N}/{E_N}=1$ in Eq. (\ref{integranda2})},
then $\langle N_{p(n)}\rangle$ 
must strictly be 1, providing an obvious physical 
meaning. In presence of FSI there is a
depletion that spoils the above interpretation in terms of number of 
nucleons involved in the
elementary process.

{ By  using }the dilution factors, 
% the approximated expression
Eq. (\ref{inter}) can be approximated as follows 
\begin{eqnarray}
 A_{^3He}^{Col(Siv)} \simeq p^\perp_n \, d_n A_{n}^{Col(Siv)}+
 2\, p_p^\perp\, d_p A_{p}^{Col(Siv)}~,
\label{appr}
 \end{eqnarray}
where  $A^{Col(Siv)}_{n(p)} $ are the free nucleon asymmetries and
$ p^\perp_{n (p)} $ are the average, or effective, 
transverse polarizations of the 
neutron (proton) in a transversely polarized $^3$He nucleus, given by
{
\begin{eqnarray}
p_{p(n)}^\perp =
\int_{E_{min}}^{E_{max}} dE
\int d {\bf p}_{mis}~~
%\frac{m_{p(n)}}{E_{p(n)}}
%S^{\perp,i}_{p(n)}(E,{\bf p}_{mis})
{\cal P}^{p(n)\,\perp} \, (E,{\bf p}_{mis})
%\delta\left( \alpha+\frac{\tilde p_{mis} \cdot q}{m_N \nu}\right)
\theta \left ( W_Y^2 - (m_{p(n)}+m_\pi)^2\right) ~.
\label{effpol} 
\end{eqnarray}
In the Bjorken limit, they are $Q^2$-independent
and
%, if one takes, in Eq. (\ref{integranda1}),
%$\alpha m_N/{E_N}=1$, 
%they 
can be obtained directly from the nuclear wave function,
without evaluating the complicated final states
entering the spectral function.}
{ In such a limit,} by adopting the nucleon-nucleon AV18 
interaction and disregarding relativistic corrections
(see Ref. \cite{tobe}) one gets  that the effective 
longitudinal and transverse polarizations
coincide and are equal to
$$p^\perp_n=p^{||}_n=p_n\simeq 0.878~~, \quad p^\perp_p=p^{||}_p=p_p\simeq -0.024~~.$$
It is important to stress that, using another realistic potential,
these values  change by a few percent at most \cite{pskv}.
We also note that, to obtain Eq. (\ref{appr}),
the term $m_N/E_N$ in the definition
of the light cone momentum distibutions $f_N^\perp$,
Eq. (\ref{falpha_perp}), and
$f_N$, Eq. (\ref{falpha}), has been neglected in Eq.
(\ref{inter}). We checked that this procedure introduces
a change in the nuclear asymmetries of the order of a few parts
in one thousand, not relevant phenomenologically.
  
The free nucleon asymmetries $A^{Col(Siv)}_N $ can be calculated
in terms of the quark distributions and fragmentation functions
previously described, using their
leading twist definitions \cite{barone1}
\begin{eqnarray}
A_N^{Col} & = &
{ 1 - y \over 1 - y + y^2/2}
{
\sum_q e_q^2
\int  d^2 {\bkappa_T}
d^2 {\bf{k}_T} \delta^2 ( {\bf k}_T + {\bf q}_T  - {\bkappa_T} )
\Bigl({ {\bf {\hat P}_{ h\,\perp}} \cdot {\bkappa_T} / m_h}\Bigr)~
 h_1^{q,N} (x_{Bj}, {\bf k}_T^2 )
 H_1^{\perp q,h} (z, (z {\bkappa_T)^2} )
\over \sum_q e_q^2 \int  d^2 {\bkappa_T} d^2 {\bf k}_T
\delta^2 ( {\bf k}_T + {\bf q}_T  - {\bkappa_T} ) 
f_1^{q,N} (x_{Bj},{\bf k}_T^2 )
D_1^{q,h}  (z, (z {\bkappa_T)^2} )
}~,
\label{coll}
\end{eqnarray}
and
\begin{eqnarray}
A_N^{Siv} =
{ \sum_q e_q^2
\int d^2 {\bkappa_T}
d^2 {\bf k}_T
\delta^2 ( {\bf k}_T + {\bf q}_T  - {\bkappa_T} )
\Bigl({{\bf \hat{P}}_{h\,\perp} \cdot {\bf k}_T / m_N}\Bigr)~
f_{1T}^{\perp q, N} (x_{Bj}, {\bf k}_T^2 )
  D_1^{q,h} (z, (z {\bkappa_T)^2} )
\over
\sum_q e_q^2
\int d^2 {\bkappa_T} d^2 {\bf k}_T
\delta^2 ( {\bf k}_T + {\bf q}_T  - {\bkappa_T} )
f_1^{q,N} (x_{Bj},{\bf k}_T^2 )
D_1^{q,h}  (z, (z {\bkappa_T)^2} )
}~.
\label{siv}
\end{eqnarray}

If Eq. (\ref{appr}) were a good approximation of reality,
it would be possible to use it
to extract the neutron asymmetry according to the following
recipe, suggested in Ref. \cite{antico} for the polarized DIS case,
and in Ref. \cite{mio} for polarized SIDIS in PWIA and in the
Bjorken limit (for $j$=Collins, Sivers):
\begin{equation}
A^j_n \simeq {1 \over p_n d_n} \left ( A^{exp,j}_3 - 2 p_p d_p
A^{exp,j}_p \right )~.
\label{extr}
\end{equation}

A theoretical check of Eq. (\ref{extr}) can be performed if a 
realistic calculation of the $^3$He single spin asymmetries, 
{$A^{theo,j}_3$,}
is introduced in Eq. (\ref{extr})
in place of the forthcoming
experimental data
$A^{exp,j}_3$, 
and models for  $A^{exp,j}_p$ 
and $A^{j}_n$
are used
in the theoretical calculation
of {$A^{theo,j}_3$,} 
and in the r.h.s. of the above equation.
If nuclear effects were
safely taken care of by Eq. (\ref{appr}), one should be able
to extract, according to Eq. (\ref{extr}), the neutron asymmetry
used as an input { for calculating $A^{theo,j}_3$}. 
Namely a self-consistency check can be carried out, in preparation
of the future extraction from the experimental $A^{exp,j}_3$.
It has to be noticed that a more stringent test 
of Eq. (\ref{extr})
could be attained if 
SSAs of $^3$H will become available at some time in the future 
(let us remind that some steps forward in 
the actual use of 
unpolarized $^3$H target in DIS
experiments have been accomplished \cite{marathon}). 

\section{Results and discussion}

{ Now we} are ready  to present the results of our calculation.

\begin{figure}[t]
\includegraphics{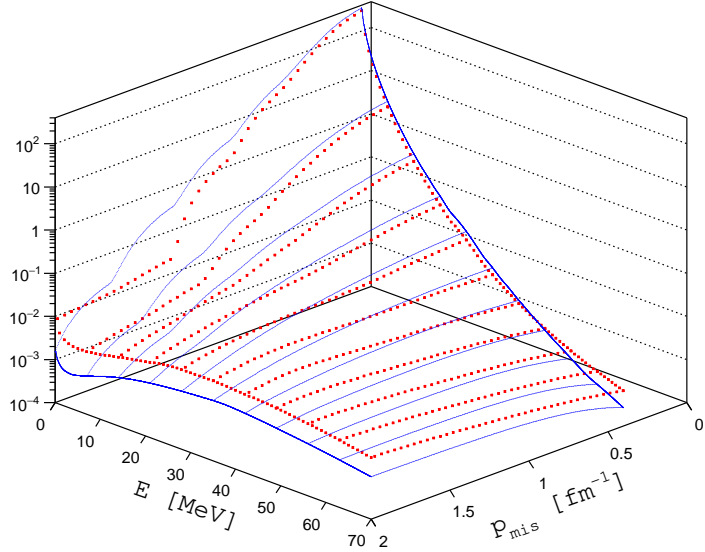}
\vskip 10.5cm
\caption{(color online) 
The $^3$He spectral function, for the neutron,
in the unpolarized case, as a function of $p_{mis}=|{\bf p}_{mis}|$ and
of the { removal energy $E$}, in PWIA (full lines) and with
 FSI taken into account within GEA framework (dotted lines), 
The kinematical ranges of $p_{mis}$ and $E$ 
correspond to the ones relevant for
the calculation of the unpolarized light-cone distribution
{{
for $\alpha=1$,
${\cal E}=$ 11 GeV (cf Eq. \eqref{integranda2}),
$x_{Bj}=0.48$, and $Q^2=7.6$ GeV$^2$.
}}
} 
\label{3d}
\end{figure}

%\begin{figure}[h]

\begin{figure}[t]
\includegraphics{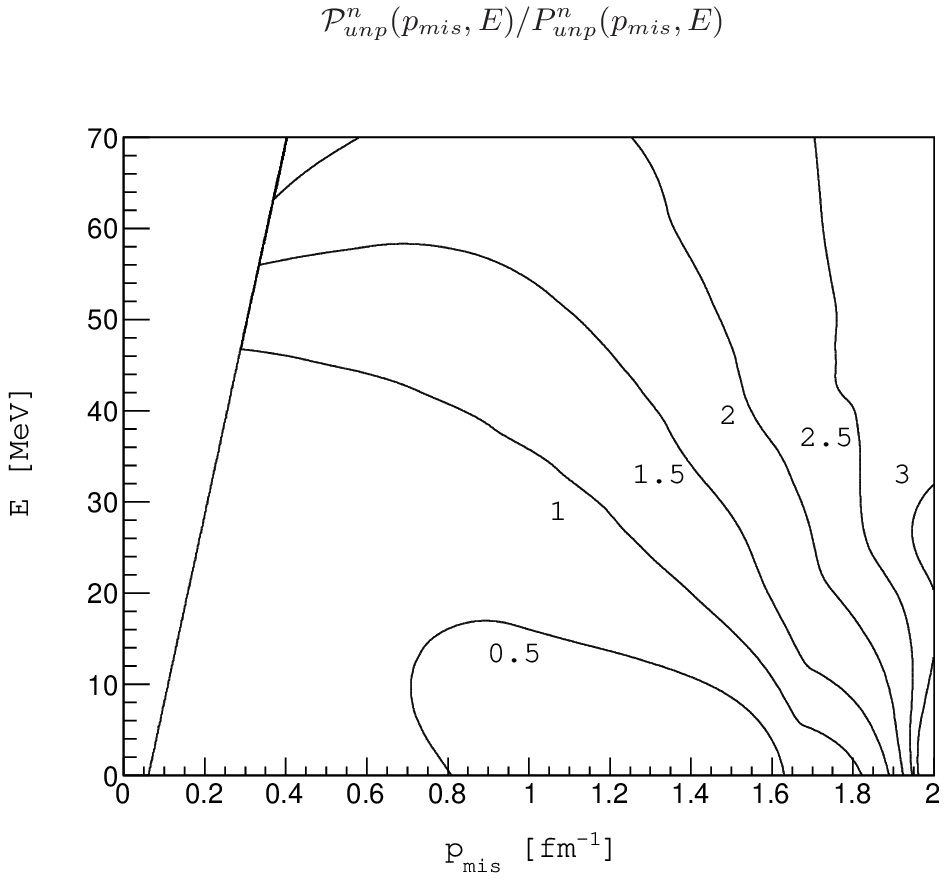}
\vskip 10.5cm
\caption{ 
The ratio between the unpolarized neutron spectral function
with FSI interactions and the corresponding
quantity in PWIA, { that are shown in Fig. \ref{3d}.}}
\label{contour}
\end{figure}

\begin{figure} [t]
\includegraphics[width=0.7\textwidth]{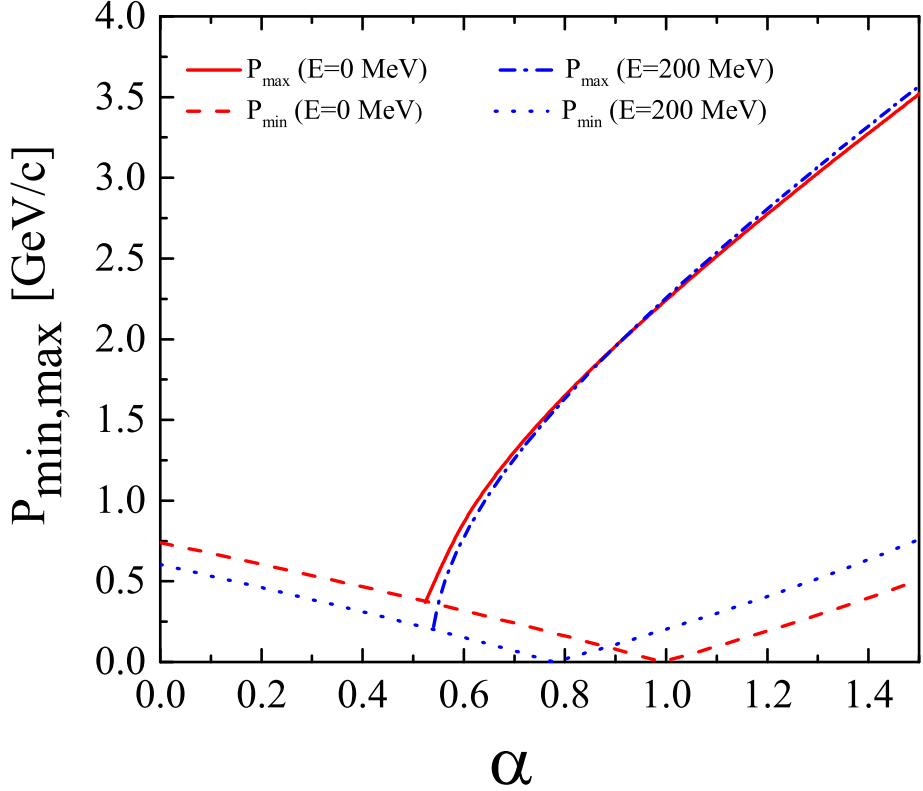}
\caption{Dependence on $\alpha$ of the integration limits 
$p_{min,max}=|{\bf p}_{min,max}|$ in Eqs. 
(\ref{integranda1}) 
and
(\ref{integranda2}) 
for the 3bbu channel and two
values of the  removal energy $E$, in the kinematics of the forthcoming
JLab experiments (corresponding to an 
initial electron energy ${\cal E}$=8.8 GeV)}.
\label{figKin}
\end{figure}

\begin{figure} [t]
\includegraphics[width=1.0\textwidth]{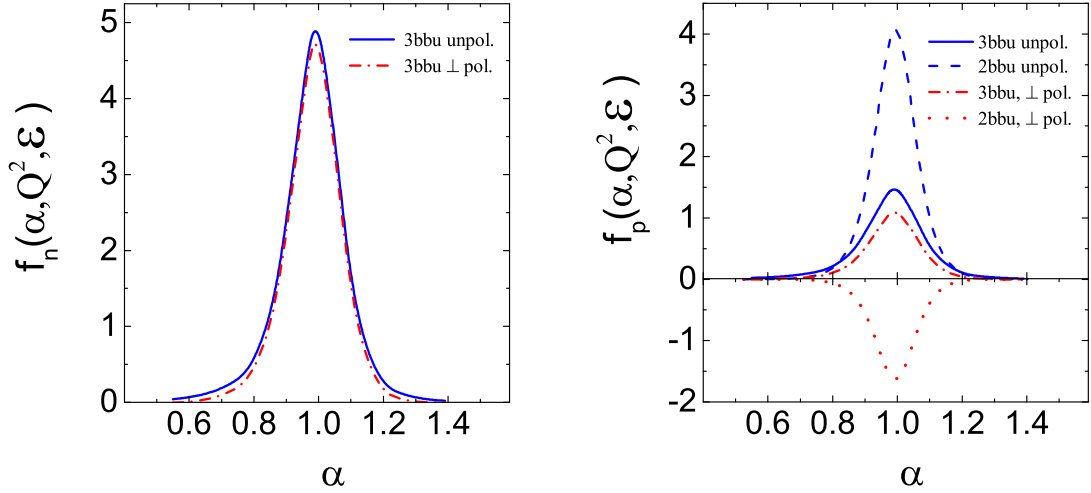}
\vskip -3.cm
\caption{(color online) The PWIA distribution functions 
 $f^\perp_N(\alpha,Q^2,{\cal E})$, Eq. \eqref{falpha_perp}, 
and $f_N(\alpha,Q^2,{\cal E})$, 
Eq. (\ref{falpha}), for the neutron
(left panel) and the proton
(right panel) at   ${\cal E}$=8.8 GeV,   and   
$Q^2= 5.73~ {\rm (GeV/c)^2}$. 
For the polarized proton, in the 2bbu and 3bbu channels,
these distributions are almost equal and opposite in sign, resulting
in a very small total distribution.}
%{\Blue in ordinata va aggiunta la dipendenza da $Q^2$ and ${\cal E}$
%KAPTARI LA STA MODIFICANDO}} 
\label{falphaUnpol}
\end{figure}

\begin{figure} [t]
\includegraphics[width=0.48\textwidth]{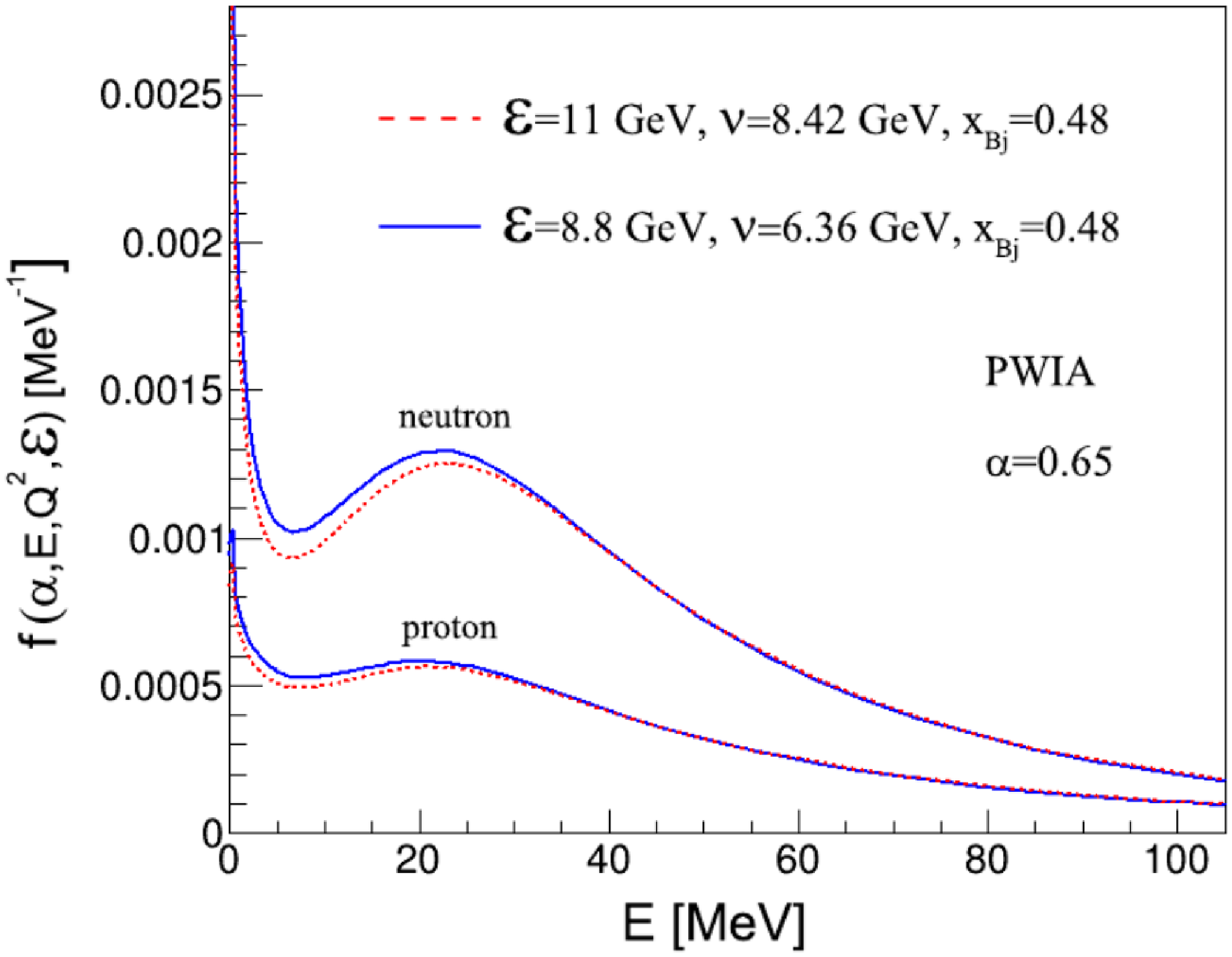}
\includegraphics[width=0.48\textwidth]{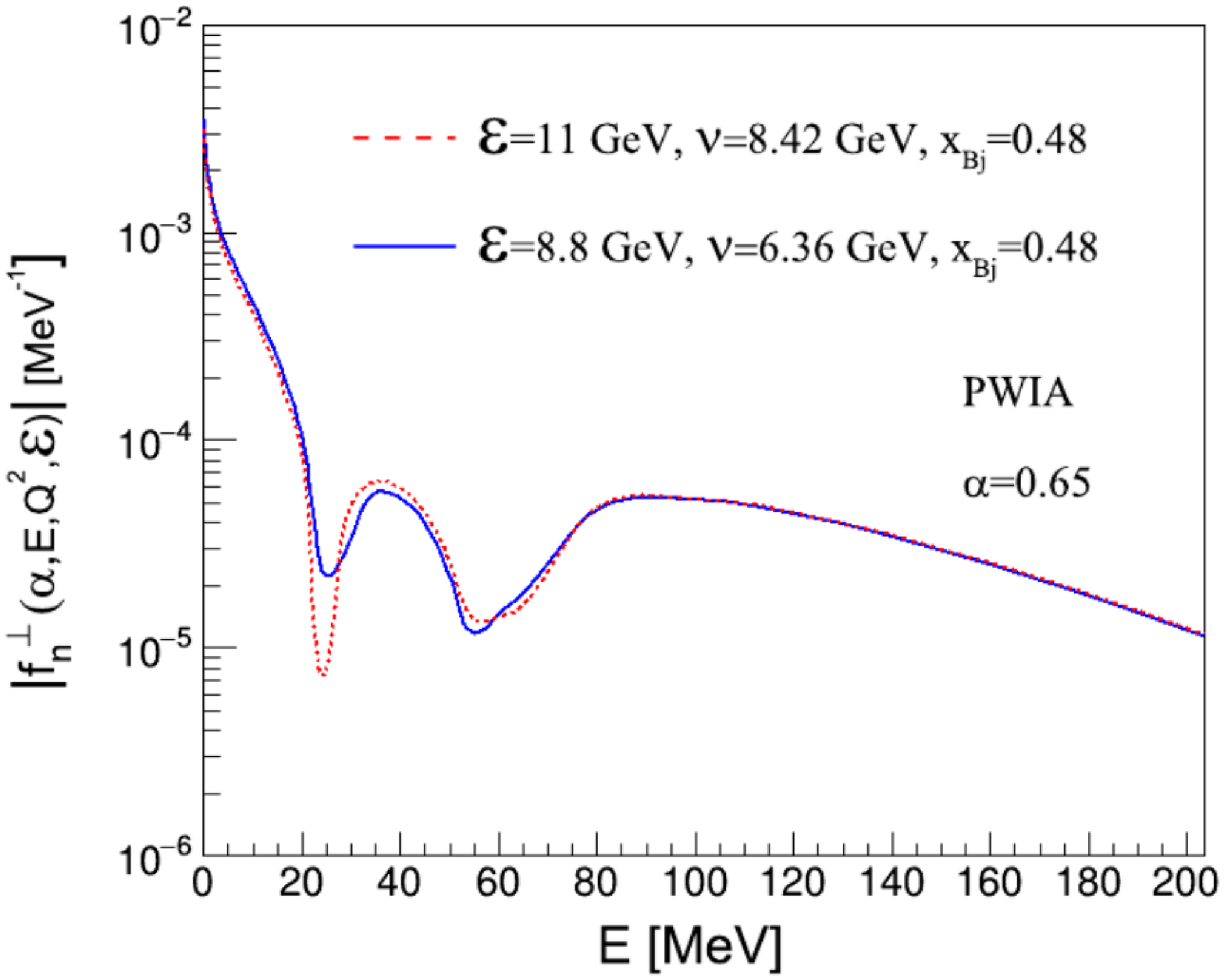}
\caption{(color online)
The  functions $f_N(\alpha=0.65,Q^2,{\cal E},E)$  
and $f^\perp_N(\alpha=0.65,Q^2,{\cal E},E)$, 
Eqs. 
(\ref{integranda2}) and (\ref{integranda1}) respectively, 
evaluated in PWIA, for the following kinematics:    
a) ${\cal E }=$ 11 GeV   
and $Q^2=7.58~{\rm(GeV/c)^2}$ (dashed lines);
b) ${\cal E }=$ 8.8 GeV,  
and $Q^2=5.73~{\rm(GeV/c)^2}$ (solid lines). 
Left panel: the proton and neutron functions 
$f_N(\alpha=0.65,Q^2,{\cal E},E)$ in an unpolarized $^3$He. Right panel: 
the functions $f^\perp_N(\alpha=0.65,Q^2,{\cal E},E)$ for a 
transversely polarized neutron in a
transversely polarized $^3$He. For a transversely polarized proton, 
the corresponding function, very small, is not shown.}
%{\Blue{KAPTARI LA MODIFICA} {C'e' anche $E_0$ invece di ${\cal E}$}
\label{figKin1}
\end{figure}

Let us start providing a pictorial view of the main quantity
of interest, {i.e.} the distorted spectral function, 
evaluated using
$^3$He and $\Psi_{23}$ wave functions computed 
within the AV18 potential \cite{av18}. 
As an example, the
neutron spectral function,
in the unpolarized case,
is shown in Fig. \ref{3d}, 
in PWIA and with FSI between debris and spectator taken into account, 
within GEA framework.
It is clearly seen that, as found in previous studies
dedicated to {quasi-elastic} scattering \cite{unfact}, the effect
of FSI increases with ${\bf p}_{mis}$, 
as it is easily understood by thinking that,
when ${\bf p}_{mis}={\bf p}_Y -{\bf q}$ is low,
the final debris $Y$ has to be very fast.
{{The low impact of FSI for small values of $|{\bf p}_{mis}|$   
is  illustrated in more detail
in Fig. \ref{contour}, where it is shown 
the ratio  of the  unpolarized distorted spectral function of the neutron, 
evaluated for $\alpha=1$,  to the PWIA one. 
Also the increase of the relevance of FSI when,
at fixed $|{\bf p}_{mis}|$,  
the  removal energy $E=M^*_{23}+m_N-M_3$ 
increases, is
physically expected. As a matter of fact from the energy conservation 
\be
M_3+\nu =  \sqrt{M^{*2}_{23}+|{\bf p}_{mis}|^2} +\sqrt{M^2_{Y}+|{\bf p}_{Y}|^2}
\ee
with $M_{Y} \ge m_\pi+m_N$, one can realize that
the momentum $|{\bf p}_{Y}|$ has to decrease { (i)
for any}
$|{\bf p}_{mis}|$, 
when the removal energy
increases, 
and { (ii) for any} $\epsilon_{23}^*$, when
$|{\bf p}_{mis}|$ increases.
Then, the
debris gets slower and FSI sizably affects the distorted spectral function. 
This is indeed what can be seen in Fig. \ref{contour}}.}

{The results for
the spin-independent} and spin-dependent light-cone momentum
distributions have already been evaluated and shown
in Ref. \cite{mio}, in PWIA,
using the AV18 interaction \cite{av18}, but assuming the Bjorken limit
($|\vec q| \simeq \nu$).
{Let us perform a first step forward, 
by illustrating in Figs. \ref{figKin}
and \ref{falphaUnpol} the effect of JLab kinematics,
at finite values of $\nu$ and $Q^2$,
on the light-cone momentum distributions 
(\ref{falpha_perp}) and (\ref{falpha}), using the  PWIA spectral function 
already exploited in Ref. \cite{mio}.}
As already mentioned, in the kinematics under scrutiny, the
distribution functions $f_{n(p)}(\alpha, Q^2, {\cal E})$ and 
$f^\perp_{n(p)}(\alpha, Q^2, {\cal E})$ 
depend on both the energy $\nu$ 
and  the momentum ${\bf q}$  through the   
limits of integration { $|{\bf p}_{min(max)}|$} and the 
invariant mass of the debris.  Figure \ref{figKin} shows
$|{\bf p}_{min}|$ and $|{\bf p}_{max}|$ as a function of the 
light-cone variable $\alpha$, 
for   {two values of the removal energy $E$, i.e.} 
$E=0$ and $200$ MeV, given { the electron} beam energy, ${\cal E}$=8.8 GeV, 
and $Q^2= 5.73~{\rm (GeV/c)^2}$. {For this kinematical
choice},
 it is seen that  one can explore only the region where  $\alpha \ge 0.55$ 
{ (i.e. when $|{\bf p}_{max}|>|{\bf
p}_{min}|$)}. 
{By changing
the kinematics} one can investigate a 
wider interval of $\alpha$. 
Figure~\ref{falphaUnpol}, where  the 
PWIA distribution function
$f_{N}(\alpha, Q^2, {\cal E})$ and $f^\perp_{N}(\alpha, Q^2, {\cal E})$
are presented for the above kinematical conditions,
shows that, as it happens in the Bjorken limit, 
the polarization of the  $^3$He nucleus 
is almost entirely determined by the neutron
one, while the contribution of the proton  
polarization is very small.
It is worth mentioning that the { existence of a kinematically forbidden} 
region  $\alpha\ <0.55 $  
can lead to slight modifications in the 
normalization conditions for both the unpolarized and the polarized
light-cone momentum distributions.  

In Fig.~\ref{figKin1}, the investigation on the PWIA light-cone distributions 
becomes more detailed.  
The functions $f^\perp_{n(p)}(\alpha,Q^2,{\cal E},E)$  
and $f_{n(p)}(\alpha,Q^2,{\cal E},E)$
of Eqs. (\ref{integranda1}) 
and (\ref{integranda2}), respectively, 
are  shown for two different choices of 
kinematics, corresponding to the planned experiments at JLab, 
and $\alpha=0.65$. Such a value of $\alpha$ belongs
to the region where  the neutron light-cone momentum distributions 
(unpolarized and transversely polarized) have shown the
biggest differences in PWIA. 
In correspondence with the different kinematical choices,
the calculated curves are hardly distinguishable and one
can conclude that
the dependence upon kinematics is rather mild in PWIA.

\begin{figure} [htb]\begin{center}
\includegraphics[width=0.9\textwidth]{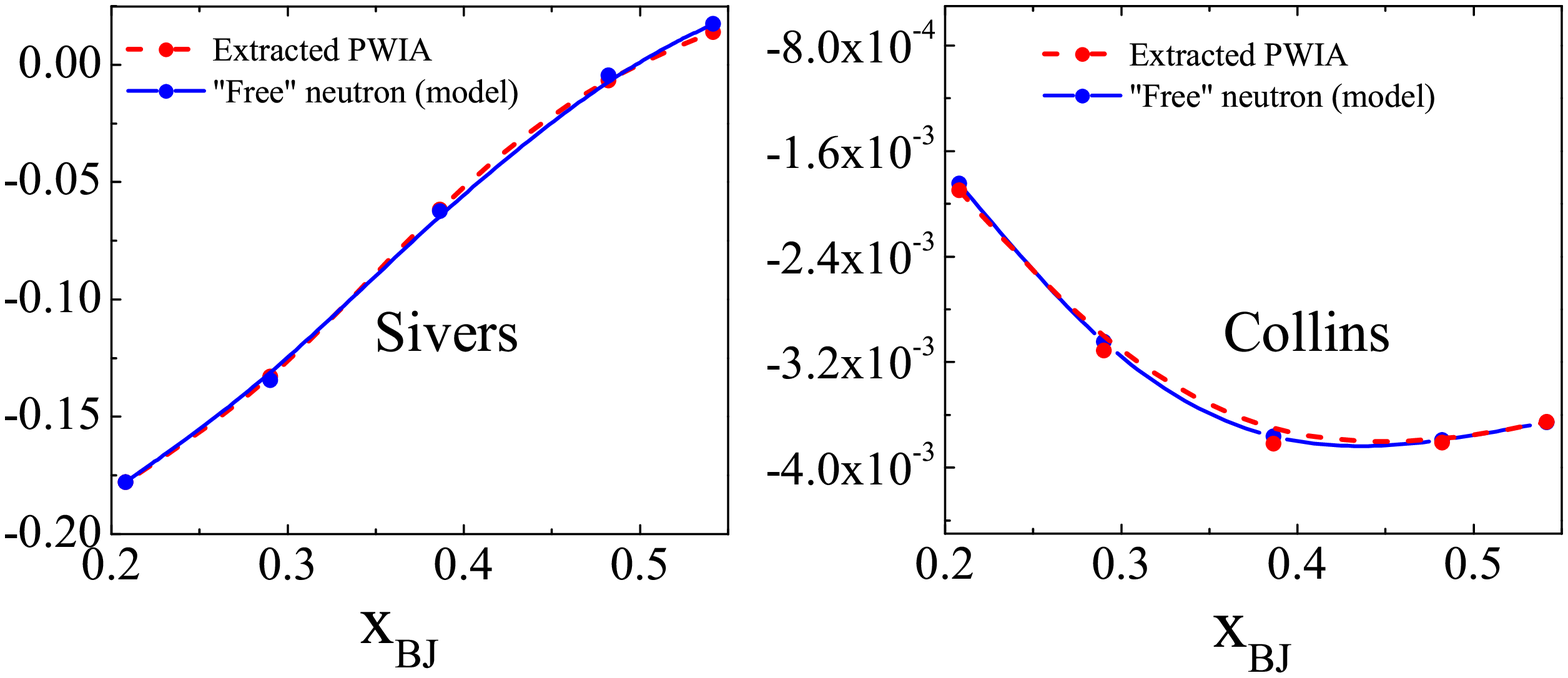}
\end{center} 
\caption{(color online)
The neutron Sivers (left panel)
and Collins (right panel) asymmetries 
for the JLAB kinematics at an initial electron energy of
$\cal E$=8.8 GeV.
Full line: the model for the neutron asymmetry used in the calculation;
dashed line: the neutron asymmetry extracted from the PWIA calculation using
Eq. (\ref{extr}). 
{{Calculations have been performed at $Q^2 = 5.73$ (GeV/c)$^2$, {i.e.
the central $Q^2$ value for an energy beam}  $\cal E$=8.8 GeV (see text).}} }
\label{asym}
\end{figure}

The extraction procedure {shown in } Eq. (\ref{extr}) and 
proposed in Ref. \cite{mio} for SIDIS adopting PWIA and  Bjorken limit,
works  very well and 
it has  been already applied in the experimental analysis
of the JLab data collected at 6 GeV \cite{prljlab}.
In the actual JLab kinematics, 
a non trivial $Q^2$ dependence is introduced in the integration
limits of the convolution formula (cf. Fig. \ref{figKin}). This amounts
{to a deviation} of the quantities  
$\langle N_n \rangle, \, \langle N_p \rangle, \,
p_n, \,  p_p $
from their values 
obtained in the Bjorken limit,
namely $1,\,1,\,0.878,\,-0.024$, respectively.
In the kinematics of JLab@12 GeV
\cite{He3exp}, this deviation is found to be a few parts
in one thousand.
In Fig. \ref{asym}, it is shown that the excellent
performance of the extraction procedure of Eq. (\ref{extr})
does not change appreciably when we move  from the Bjorken limit 
to the experimental kinematics
of JLab@12 GeV \cite{He3exp}, corresponding to finite values
of $Q^2$ and $\nu \neq | \vec q |$. Hence, the Sivers (left panel)
and the Collins (right panel) asymmetries are well determined when our
theoretical check of Eq. (\ref{extr}) is carried out.

Now it comes the basic issue
of understanding to what extent FSI effects between debris and remnants
can modify the outcomes  obtained through  Eq. (\ref{extr}) and 
shown in Fig. \ref{asym}. This is a  crucial step for a reliable
extraction of the neutron information.
As pointed out in Sections III, IV and V, the formal expressions
for the Collins and Sivers asymmetries
obtained within  PWIA, Eq. (\ref{A3}),
still work when FSI are considered within GEA.

\begin{figure}[h]
\includegraphics[width=0.45\textwidth]{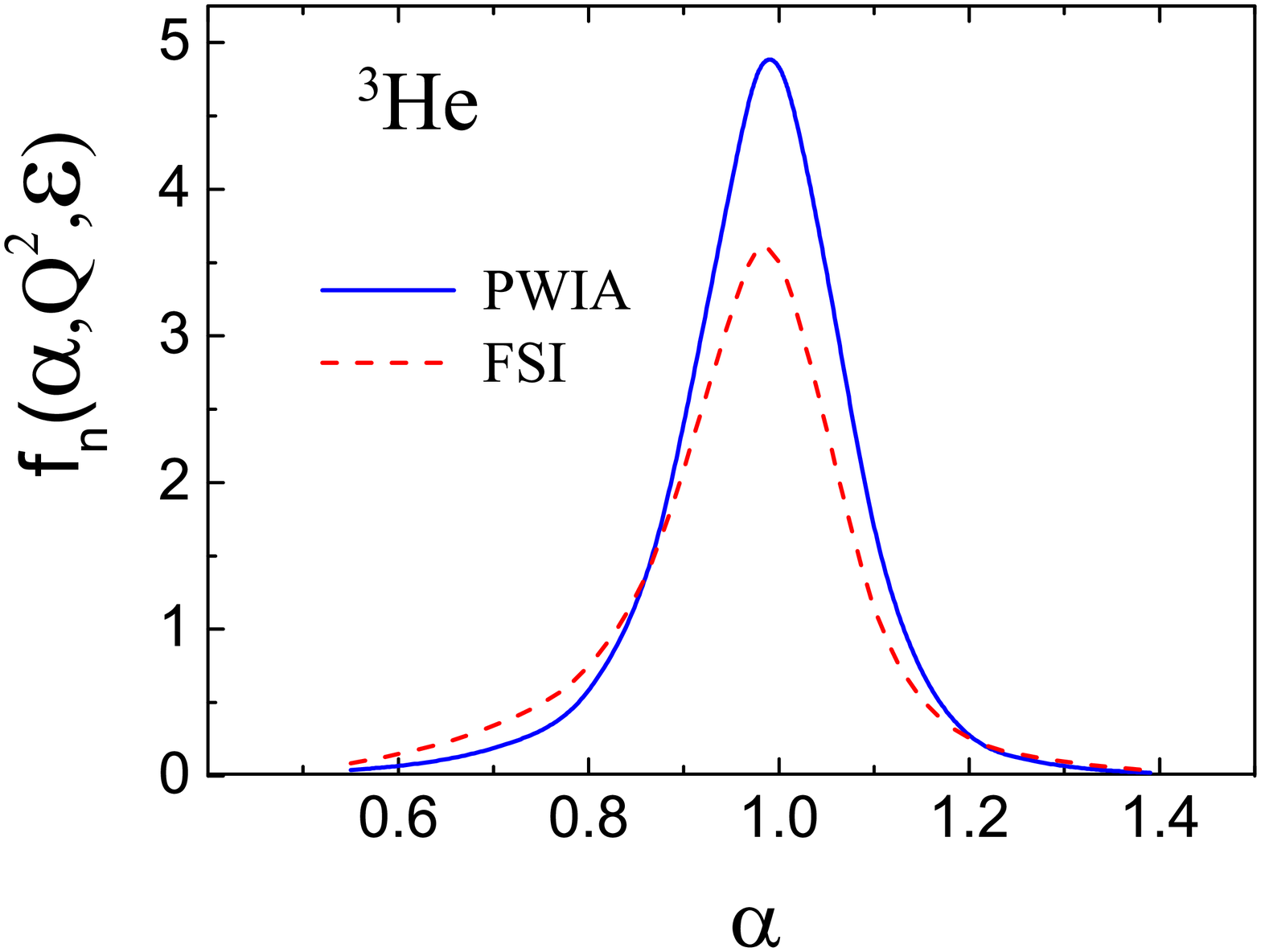}
\includegraphics[width=0.45\textwidth]{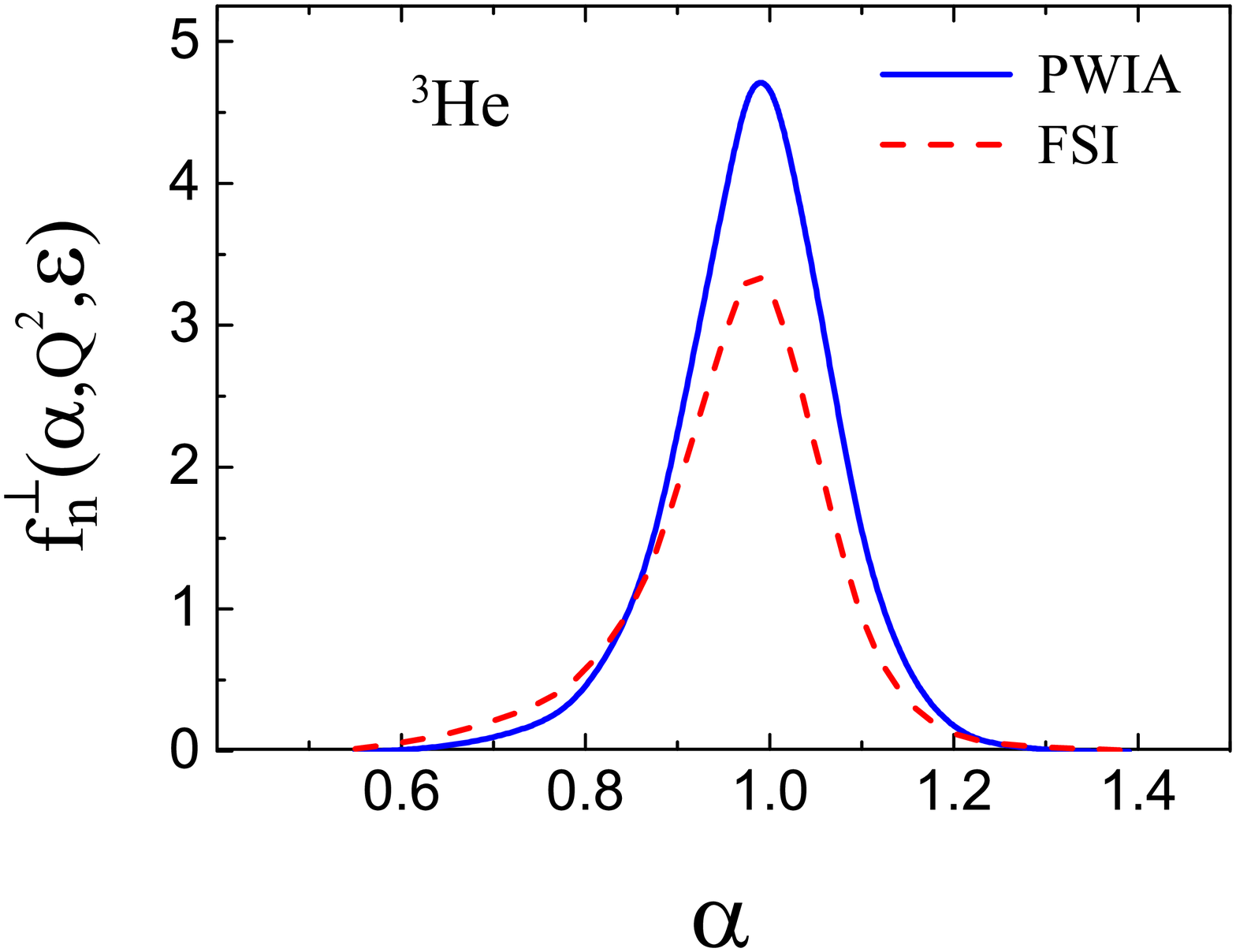}
\caption{
(color online)
The neutron unpolarized and transversely polarized distributions,
Eqs. (\ref{falpha}) and (\ref{falpha_perp}). 
Solid lines are PWIA results,
while dashed lines include effects of FSI. 
JLab kinematics has been assumed, i.e.,
the initial electron energy is $\cal E$=8.8 GeV 
and $Q^2 = 5.73$ (GeV/c)$^2$,
which is the central $Q^2$ value for an energy beam $\cal E$=8.8 GeV
according to JLab kinematics (see text).}
\label{FsiNtrUnpol}
\end{figure}

\begin{figure}[t]
\includegraphics[width=0.40\textwidth]{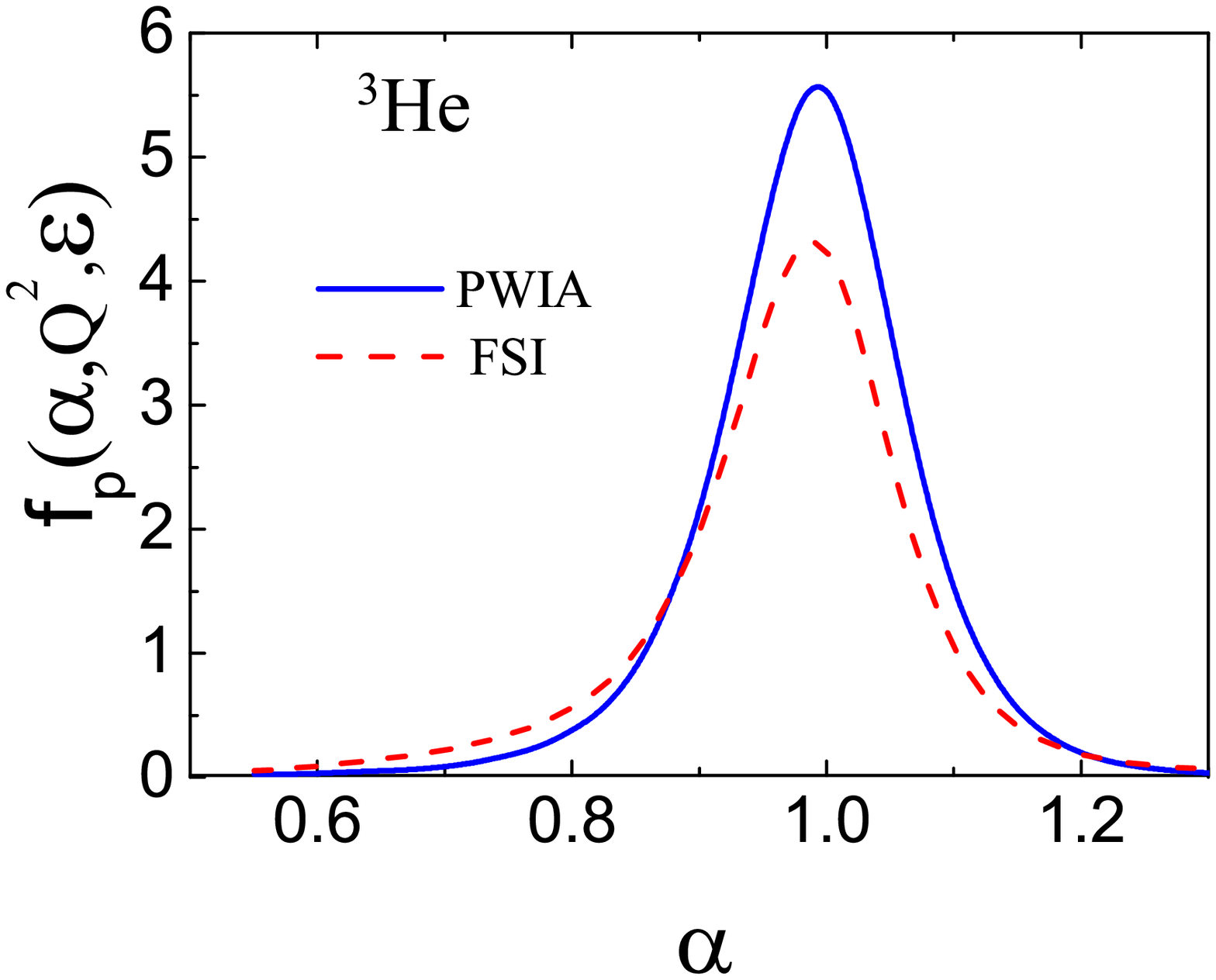}
\includegraphics[width=0.43\textwidth]{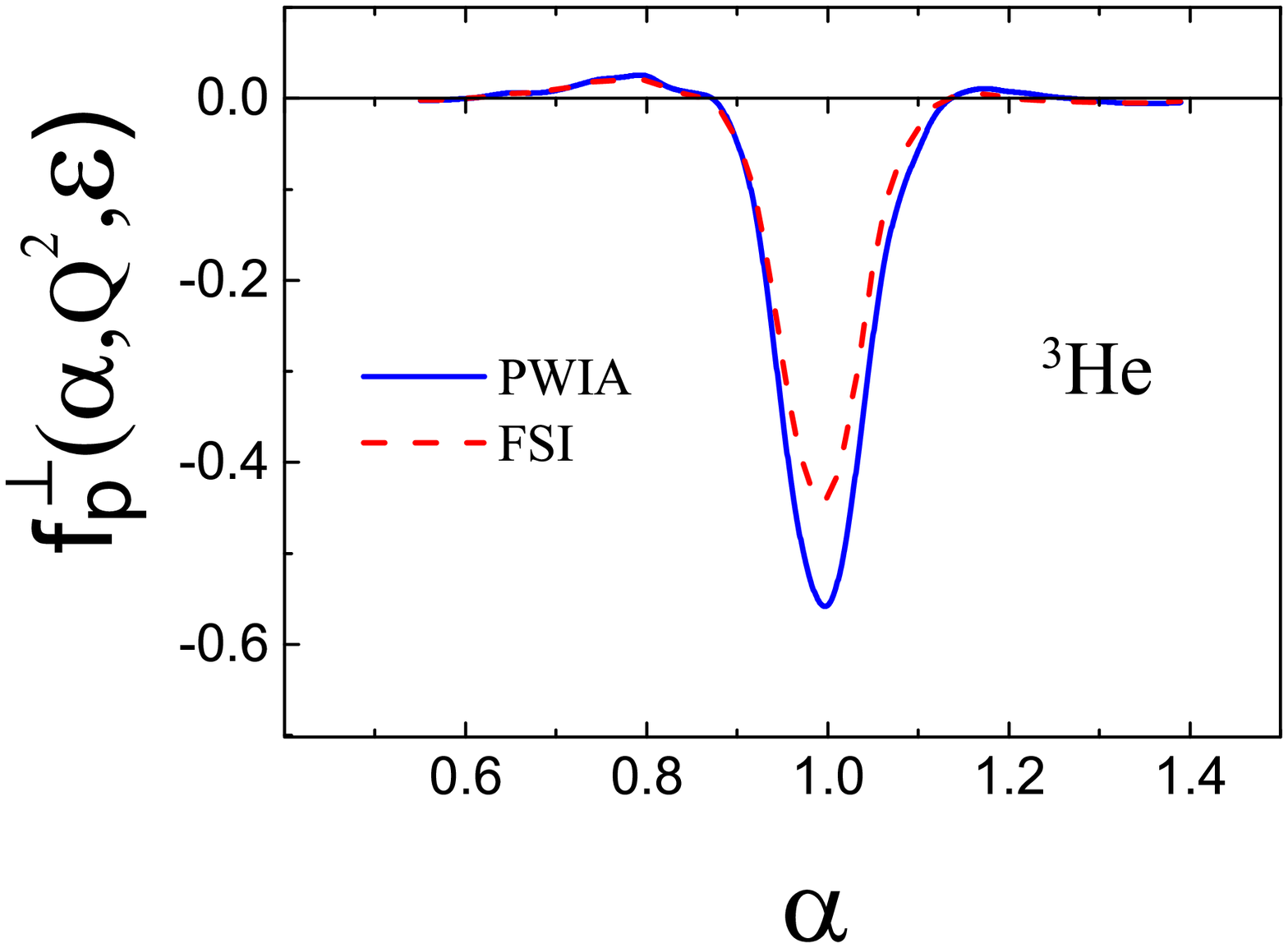}
\caption{
(color online) The same as in Fig. \ref{FsiNtrUnpol},
but for the proton distributions.
} 
\vskip 2cm
\label{FsiPrtUnpol}
\end{figure}

\begin{figure}[h]
\includegraphics[width=0.6\textwidth]{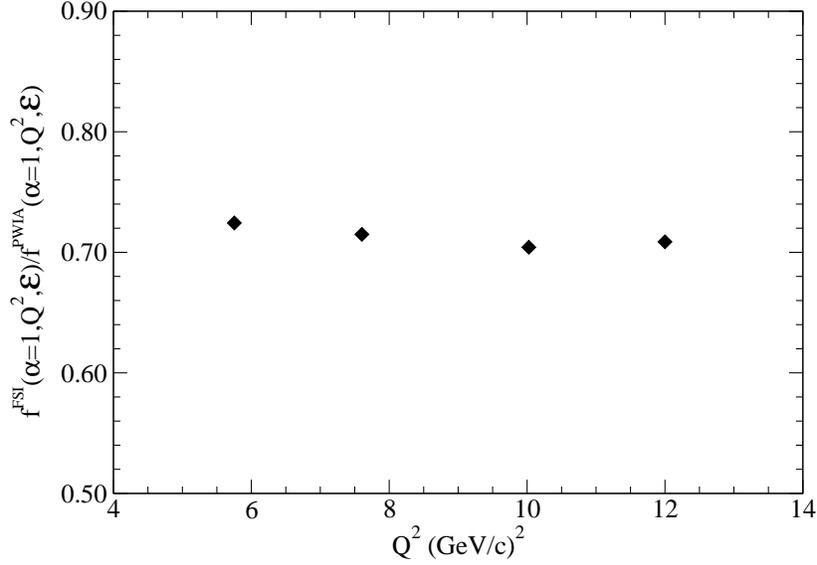}
\caption{
The ratio of the light-cone spin-independent momentum
distribution evaluated taking into account FSI to the 
corresponding quantity obtained in PWIA. The ratio is shown
in the neutron case, for $\alpha=1$, namely the value
where the distributions reach their maximum value,
as a function of the momentum transfer, $Q^2$,
corresponding to four different kinematical conditions: {the ones with $Q^2<
9~(GeV/c)^2$ have been evaluated by using  Jlab kinematical conditions, 
while the rightmost diamonds are appropriate for EIC kinematics (see text)}.
} 
\label{figeic}
\end{figure}

\begin{figure} [t]
\includegraphics{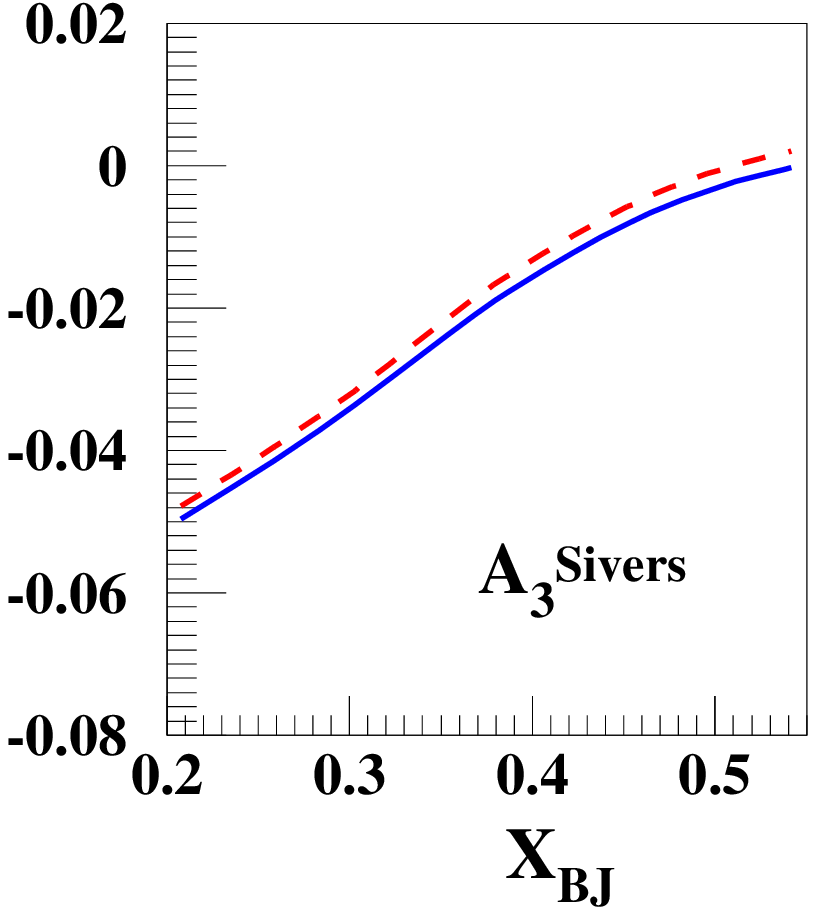}
\includegraphics{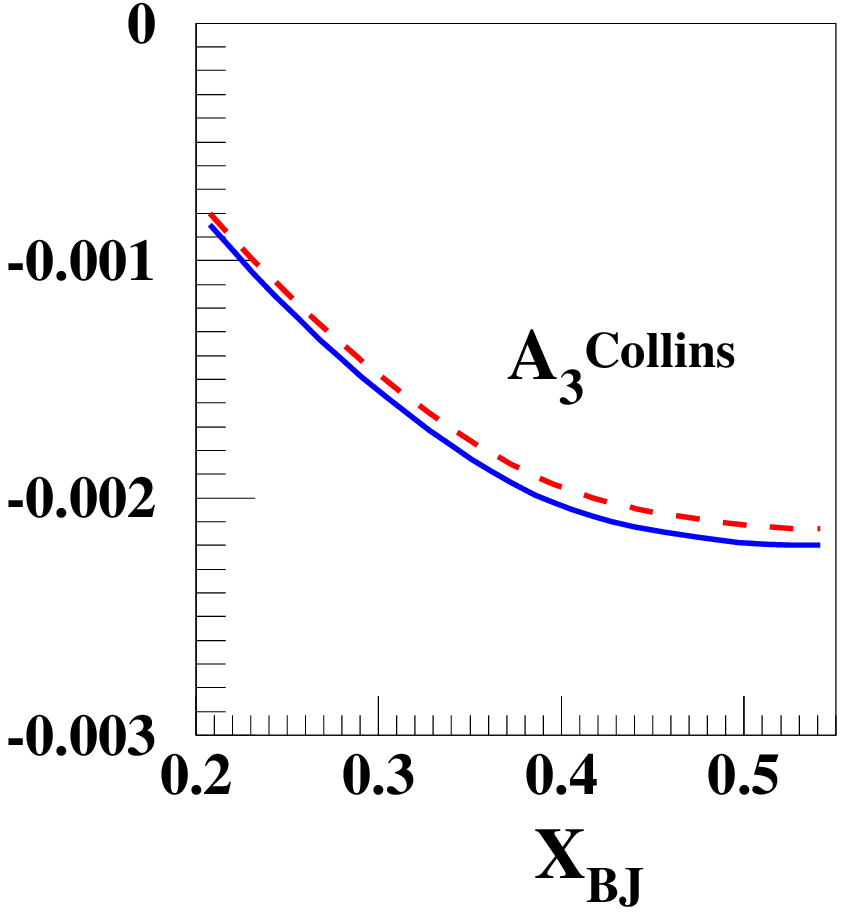}
\vskip 8.cm
\caption{(color online)
Left panel:  The $^3$He Sivers asymmetry (see Eqs. (\ref{A3})
and \eqref{dsiv}), evaluated 
taking into account FSI
effects (full line) and in PWIA (dashed line). Right Panel: the same,
but for  the $^3$He Collins asymmetry (see Eqs. (\ref{A3}) 
and \eqref{dcoll}).
Calculations have been performed at $Q^2 = 5.73$ (GeV/c)$^2$,
{i.e. the central $Q^2$ value for an energy beam} 
 $\cal E$=8.8 GeV (see text).}
\label{figtot}
\end{figure}

\begin{figure}[h]
\includegraphics[width=0.9\textwidth]{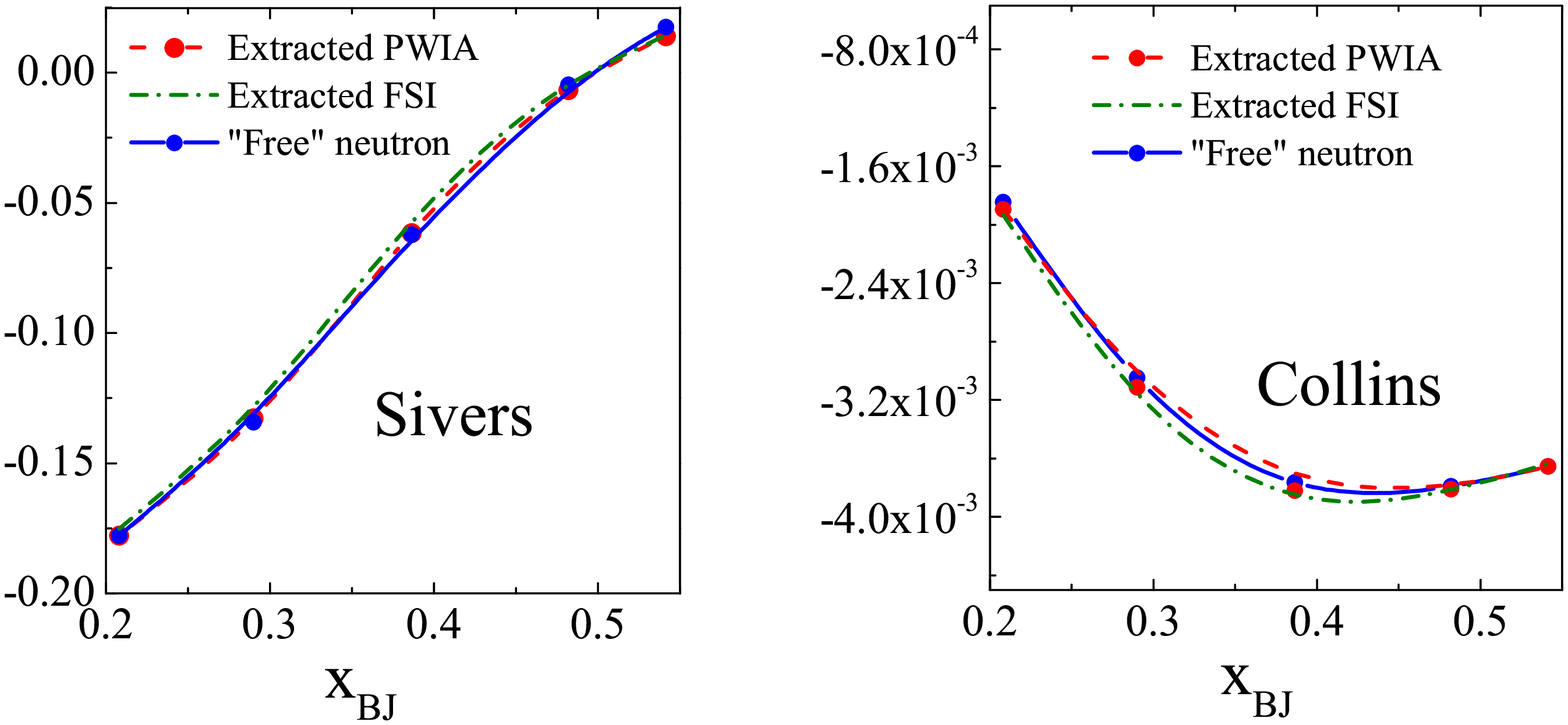}
\caption{(color online)
The neutron Sivers (left panel)
and Collins (right panel) asymmetries 
for the JLab kinematics at an initial electron energy of
$\cal E$=8.8 GeV.
Full line: the model for the neutron asymmetry
used in the calculation;
dot-dashed line: the neutron asymmetry
extracted from the full calculation of $A^j_3$
with FSI taken into account, using the
extraction formula Eq. (\ref{extr_final});
dashed line: the result obtained using Eq. (\ref{extr_final}) to extract
the neutron asymmetries from PWIA results.  
Calculations have been performed at $Q^2 = 5.73$ (GeV/c)$^2$, {i.e.
the central $Q^2$ value for an energy beam }  $\cal E$=8.8 GeV (see text).}
\label{fig8}
\end{figure}

Also Eqs. (\ref{falpha_perp}) - (\ref{falpha}) remain
formally unchanged if FSI
are included: the only difference
amounts to use there the distorted
spectral function for 
 obtaining the distorted light-cone momentum distributions,
instead of adopting the corresponding PWIA expressions.
{ In Figs. \ref{FsiNtrUnpol} and  \ref{FsiPrtUnpol}, 
 neutron and proton light-cone momentum distributions, obtained within GEA
for the 
unpolarized 
and the
transversely polarized cases,  are shown  
for ${\cal E}$=8.8 GeV, a value of the beam energy  typical for the 
 planned JLab@12 experiments,
and for $Q^2$= 5.73 (GeV/c)$^2$ (i.e. one of the  values which will be 
tested at  $\cal E$=8.8 GeV).
Moreover, they are}
 compared with the corresponding quantities 
calculated within PWIA. 
 The differences between the results with and without FSI are quite  
sizable and therefore the quantities defined in
Eqs. \eqref{dcoll}, \eqref{dsiv} and \eqref{unpol},
necessary to calculate Collins and Sivers asymmetries,
are largely affected by FSI effects, that have
to be carefully taken into account.
In particular,
$\langle N_n \rangle, \, \langle N_p \rangle, \,
p_n , \,  p_p 
$,
defined according to Eqs. 
\eqref{neff} and \eqref{effpol}, respectively,
and calculated at the actual JLab 
kinematics corresponding to 
$Q^2\sim 3 \div 7$ (GeV/c)$^2$,
are affected by FSI and exhibit deviations from 
their  values 
in the Bjorken limit (given above)  as large as 20 \%.

The $Q^2$ dependence of the above results is quite important, in view of the
possible construction of the EIC 
(see, e.g. Ref. \cite{accardi} for
the presentation of the physics case),  that could open unprecedented
possibilities in the studies of the nucleon TMDs. In order to give a first idea
of the impact on the future measurements, in Fig. \ref{figeic}, 
it is shown  the ratio of the light-cone spin-independent momentum
distribution, evaluated taking into account FSI, to the 
corresponding quantity obtained in PWIA, 
for different values of $Q^2$, at the peak, i.e.
$\alpha=1$.  Four different kinematical conditions have been chosen,
two of them, namely
(i) $\cal E$=8.8 GeV, $Q^2 \simeq 5.7$ (GeV/c)$^2$, { $x_{Bj} \simeq
0.48$}
and (ii)  $\cal E$=11 GeV, $Q^2 \simeq 7.7$ (GeV/c)$^2$, {$x_{Bj} \simeq
0.48$,
are typical for  JLab@12 }.
The third and the fourth ones are { kinematics occurring}
at the planned EIC, namely
at ${\cal E}_{collider}=11$ GeV, 
${\cal E}_{collider}^{^3He}$ = 40 GeV,
$Q^2=$ 10 and 12 (GeV/c)$^2$, { $x_{Bj} \simeq 0.48$
(noteworthy, this value of $x_{Bj}$ could be achieved by a beam energy ${\cal E}=293$ GeV
 for a fixed target experiment).
It is important to recall that a single point
in Fig. \ref{figeic} represents the outcome of  a one-week run
on the
ZEFIRO INFN-facility in 
 Pisa, Italy.}
What is found is that the effects of FSI, evaluated  within GEA framework, is 
 almost $Q^2$ independent, 
but rather sizable
at JLab and EIC
energies.
Could  one think that the extraction procedure shown in
Eq. (\ref{extr}),
{had to } be abandoned in favor of more involved and
model dependent techniques?
Actually, a crucial observation is now in order.
It is clearly seen in Figs. \ref{falphaUnpol}, 
\ref{FsiNtrUnpol} and \ref{FsiPrtUnpol}
that the spin-independent and spin-dependent light-cone
momentum distributions are strongly peaked around $\alpha=1$,
both in PWIA and with FSI effects taken into account.
This means that the approximation given in Eq. (\ref{inter})
{for  the nuclear Sivers and Collins asymmetries (cf Eq.
(\ref{A3}))},
should basically hold. Moreover, looking at the same figures,
it is also rather apparent that FSI produces a decrease of all 
the distributions in a similar { way}, both
qualitatively and quantitatively.
From Eq. (\ref{inter}), it is easy to see 
that the results
for the nuclear asymmetries obtained in PWIA,
$A^{PWIA,j}_3$,
or taking into account FSI, $A^{FSI,j}_3$ (recall that 
$j=$ Sivers or Collins),
 should not sizably differ from each other, due to a cancellation
of effects present in both the numerator and the denominator.
The realization of this fact in the actual calculation
of Eq. (\ref{A3}) is shown in Fig.
\ref{figtot}.
In principle, in this figure and in the two
following ones, at any $x_{Bj}$ should correspond
a slightly different value of $Q^2$. 
Nevertheless, in the $x_{Bj}$ range explored
at fixed ${\cal{E}}$, the dependence on $Q^2$
of the light-cone 
momentum distributions
$f_{p(n)}(\alpha,Q^2,{\cal E})$
and $f_{p(n)}^\perp(\alpha,Q^2,{\cal E})$
is rather
mild and therefore we will
show the results for the nuclear asymmetries,
Eq. (\ref{A3}),
at a fixed value of $Q^2$, namely
5.73 (GeV/c)$^2$.

Our full evaluations of the $^3$He Collins and Sivers
asymmetries, presented in  Fig. \ref{figtot}, {strongly encourage} the
investigation  of 
the extraction formula, 
Eq. (\ref{extr}), that relies on the validity of the approximation
Eq. (\ref{appr}), where effective polarization {\it and}
dilution factors are multiplied by each other. In particular, we want to assess
if Eq. (\ref{extr}) can be safely (or better with a low
degree of uncertainty) applied to the experimental data, 
{where FSI is certainly acting.}
Noteworthy, the relevant product of effective polarizations and
dilution factors is found to have
a very little dependence on FSI, as one can straightforwardly realize
by inspecting  Tables 1 and 2, where the dilution
factors, the effective polarizations and their products
are presented with or without FSI effects taken into account,  
by adopting the kinematics
of the forthcoming JLab experiments. 

Considering that (i) $A^{PWIA,j}_3 \simeq A^{FSI,j}_3$ 
(see Fig. \eqref{figtot}), and (ii) the products
of effective polarizations and dilution factors are almost
the same in PWIA and including FSI,
one has
\begin{eqnarray}
A^j_n 
%& 
\simeq 
%& 
{1 \over p_n^{PWIA} d_n^{PWIA}} \left ( A^{PWIA,j}_3 
- 2 p_p^{PWIA} d_p^{PWIA}
A^{exp,j}_p \right )
\simeq 
{1 \over p_n^{FSI} d_n^{FSI}} \left ( A^{FSI,j}_3 
- 2 p_p^{FSI} d_p^{FSI}
A^{exp,j}_p \right )~.
%\nonumber 
%\\
%& \simeq & 
%{1 \over p_n^{PWIA} d_n^{PWIA}} \left ( A^{FSI,i}_3 
%- 2 p_p^{PWIA} d_p^{PWIA}
%A^{exp,i}_p \right )~.
\label{extr_final}
\end{eqnarray}

In Fig. \ref{fig8}, the reliability of the above relations in the 
extraction of $A^j_n$ is illustrated
through our theoretical test, where the experimental $A_3^{exp,j}$ 
is replaced
by our full calculation.
Indeed,
in Fig. \ref{fig8},  
the model Collins and Sivers asymmetries for the neutron
used in the full calculations of $^3$He asymmetries  
are hardly distinguishable from the neutron asymmetries
extracted through Eq. (\ref{extr_final}) {by  using PWIA
effective polarizations and dilution factors, or by
considering the corresponding quantities calculated within GEA
(a preliminary version of this figure was presented
in Ref. \cite{dani}).
It should be pointed out that these quantities can be evaluated 
in any kinematical configuration
using our model of FSI, which is rather well constrained
phenomenologically, and could be improved checking our predictions
against the spin-dependent cross sections which will be soon available.

In addition to the above extraction procedure, 
one could adopt the following one where 
the experimental inputs are 
$A_3^{exp,j}$ and $A_p^{exp,j}$, while the theoretical
quantities reduce to  the PWIA effective polarization  in the Bjorken limit. 
In this case, one has a nice possibility 
to extract the neutron information through
another extraction scheme,  
independent
of the FSI model.
The procedure is based  on the following expression
\begin{eqnarray}
A^j_n 
%& 
\simeq 
{1 \over p_n^{PWIA} d_n^{exp}} \left ( 
A^{exp,j}_3 
- 2 p_p^{PWIA} d_p^{exp}
A^{exp,j}_p \right )~,
%\nonumber 
%\\
%& \simeq & 
%{1 \over p_n^{PWIA} d_n^{PWIA}} \left ( A^{FSI,i}_3 
%- 2 p_p^{PWIA} d_p^{PWIA}
%A^{exp,i}_p \right )~.
\label{extr_fico}
\end{eqnarray}
Indeed, $p_{n(p)}^{PWIA}$
can be obtained from a realistic wave function with very small
model dependence (see Ref. \cite{pskv} for an analysis
of the dependence of effective polarizations on different realistic 
potentials).  
In Eq. \eqref{extr_fico}, the experimental dilution factors are 
\begin{eqnarray}
d_{p(n)}^{exp}(x_{Bj},Q^2,z)=
\displaystyle\frac{\sigma^{p(n)} (x_{Bj} ,Q^2,z)}
{ \sigma^{n} (x_{Bj} ,Q^2,z)+2
\sigma^{p}(x_{Bj} ,Q^2,z)},
\label{dilfe}
\end{eqnarray}
where no dependence on the FSI model is { present, differently from   Eq. (\ref{dilf})}.
In Fig. \ref{figlast} one sees
that the uncertainty in the extraction procedure based on 
Eq. (\ref{extr_fico}) is not much bigger than
the one occurred by using Eq. (\ref{extr_final}).
In Fig. \ref{figlast}, Eq. (\ref{extr_fico}) has been actually
evaluated  using 
$A^{FSI,j}_3$ instead of $A^{exp,j}_3$, and using, instead of 
$d_{p(n)}^{exp}$, the dilution factors evaluated with
the parameterizations of unpolarized
parton distributions \cite{gluu} and fragmentation functions
\cite{kret} already described in the previous section. 
Therefore Fig. \ref{figlast} shows that, 
for a safe extraction procedure through Eq. (\ref{extr_fico}), 
the evaluation of distorted effective polarizations and
dilution factors, which appear in Eq. (\ref{extr_final})
and { are depending } on the adopted FSI model,
is actually not required.

Summarizing, the comparisons shown in Figs. \ref{fig8} and 
\ref{figlast} illustrates
two methods for the
successful extraction of the neutron single spin asymmetries
using transversely polarized $^3$He targets at JLab, and they represent
the most relevant outcomes
of the present investigation.

One could argue that the very nice results
obtained within our FSI model, are actually expected to hold in
any description of final state interactions which is
(i) factorized and (ii) basically spin-independent, i.e.,
producing a similar effect in spin-dependent and spin-independent
cross sections. This last feature is very likely to be realized
for any FSI occurring
in processes where the relative energy of the interacting systems
is high, as it is the case in the present study.

\begin{table}
\begin{tabular}{|ccccccccccccccc|}\hline
  ${\cal{E}} $, &&      $x_{Bj}$&     &   $\nu$  &       & $P_\pi$  
&& $d_{n}(x_{Bj},z)$ && $ p_{n} d_{n}$ && $d_{p}(x_{Bj},z)$&& 
$ p_{p} d_{p}$\\
GeV        &\,\,\, &        &\,\,\,&  GeV      &\,\,\,&  GeV/c         
&\,\,\,&              &\,\,\,&     &\,\,\,&     &\,\,\,&           
\\ \hline
     8.8     &&  0.21         &&    7.55      && 3.40       && 0.304         
&&   0.266      && 0.348         &&   -$8.4 10^{-3}$
         \\ \hline
     8.8     &&   0.29        &&    7.15      && 3.19       
&& 0.286         &&   0.251     && 0.357         &&   -$8.5 10^{-3}$  \\ 
\hline
     8.8     &&   0.48        &&    6.36      && 2.77       &&0.257          
&&   0.225     && 0.372         &&   -$8.9 10^{-3}$  
\\ \hline\hline
     11      &&    0.21       &&    9.68      && 4.29       &&0.302          
&&   0.265     && 0.349         &&    -$8.3 10^{-3}$  
\\ \hline
     11      &&   0.29        &&    9.28      &&  4.11      &&  0.285        
&&   0.250      &&  0.357        &&    -$8.5 10^{-3}$  
\\ \hline
\end{tabular}
\caption{
The { PWIA values }of the dilution factors $d_{n(p)}(x_{Bj},z)$ and 
their product with the corresponding effective polarizations,
in  PWIA, for the kinematical conditions of the planned experiments at JLab, 
with scattering angle $\theta_e=30^o$ and detected pion angle  
$\theta_\pi=14 ^o$. The effective polarizations are evaluated with extrema
of integrations depending upon the kinematics. 
 At $Q^2=5.73$ (GeV/c)$^2$, {i.e. the central $Q^2$ value for an energy beam} 
${\cal{E}}=8.8$ GeV, 
one obtains
$p_{n}$= 0.876, $p_{p}$= -0.024 (cf Eq. \eqref{effpol}), very close 
to the corresponding asymptotic values 0.878 and -0.024  
(i.e. in the Bjorken limit). 
}
\end{table}

\vskip 3mm
\begin{table}
 \begin{tabular}{|ccccccccccccccc|}\hline
  ${\cal{E}} $, &&      $x_{Bj}$&     &   $\nu$  &       & $P_\pi$  
&& $d_{n}(x_{Bj},z)$ && $ p_{n} d_{n}$ && 
$d_{p}(x_{Bj},z)$&& $ p_{p} d_{p}$\\
  GeV        &\,\,\, &        &\,\,\,&  GeV      &\,\,\,&  GeV/c         &\,\,\,&              &\,\,\,&     &\,\,\,&     &\,\,\,&           \\ \hline
     8.8     &&  0.21       &&     7.55      && 3.40           &&  0.353         &&  0.267      && 0.405         &&   -$1.1 \cdot 10^{-2}$          \\ \hline
     8.8     &&  0.29       &&     7.15      && 3.19           && 0.332         &&  0.251     && 0.415        &&   -$1.1 \cdot 10^{-2}$  \\ \hline
     8.8     &&  0.48       &&     6.36      && 2.77           &&0.298           &&  0.225      && 0.432        &&   -$1.2 \cdot 10^{-2}$  \\ \hline\hline
     11     &&   0.21       &&     9.68      && 4.29           &&  0.351         &&  0.266      && 0.405         &&    -$1.0 \cdot 10^{-2}$  \\ \hline
     11     &&   0.29       &&     9.28      &&  4.11          &&  0.331       &&  0.250       &&  0.415      &&    -$1.1 \cdot 10^{-2}$  \\ \hline
\end{tabular}
\caption{The same as in Table I, but taking into account FSI 
within GEA framework.
For all the presented kinematical conditions, 
one gets {\em distorted polarizations}, 
evaluated by using  Eq. \eqref{effpol}, 
with distorted distributions.  They amount to 
$p_{n} \simeq 0.756$, 
$p_{p} \simeq -0.0265$ for 
$Q^2=5.73$ (GeV/c)$^2$, {i.e. the  central $Q^2$ value for an energy beam} 
${\cal{E}}=8.8$ GeV.
In these conditions one gets, for the quantities
$<N_n>$ and $<N_p>$, Eq. (\ref{neff}), the values 0.85
and 0.87, respectively.
}
\end{table}
\begin{figure} [h]
\includegraphics{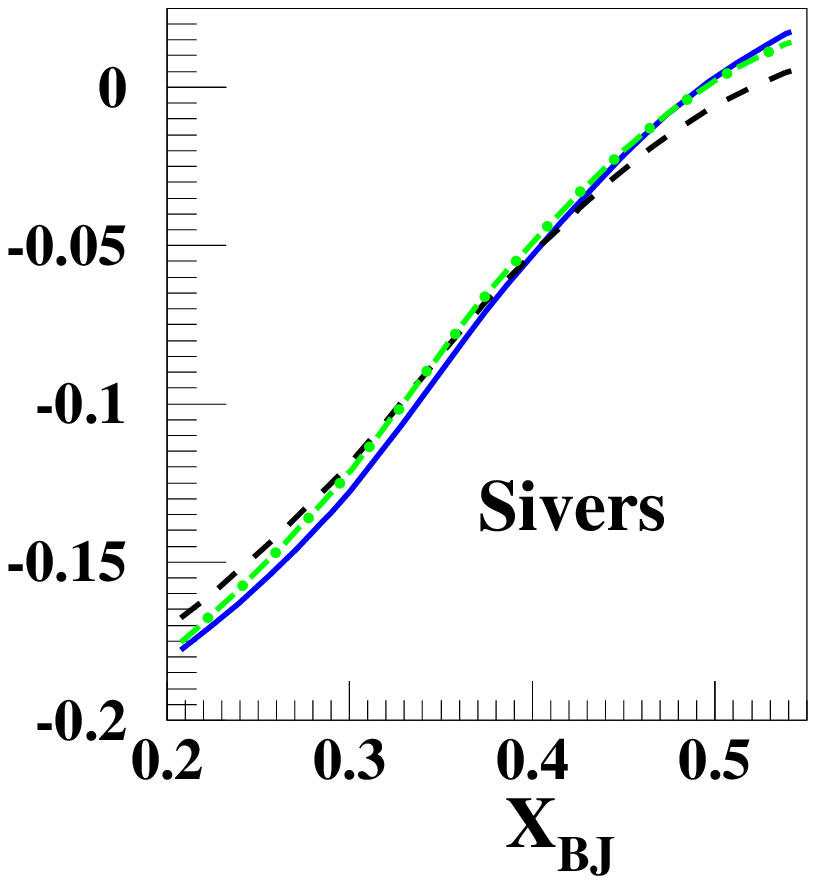}
\includegraphics{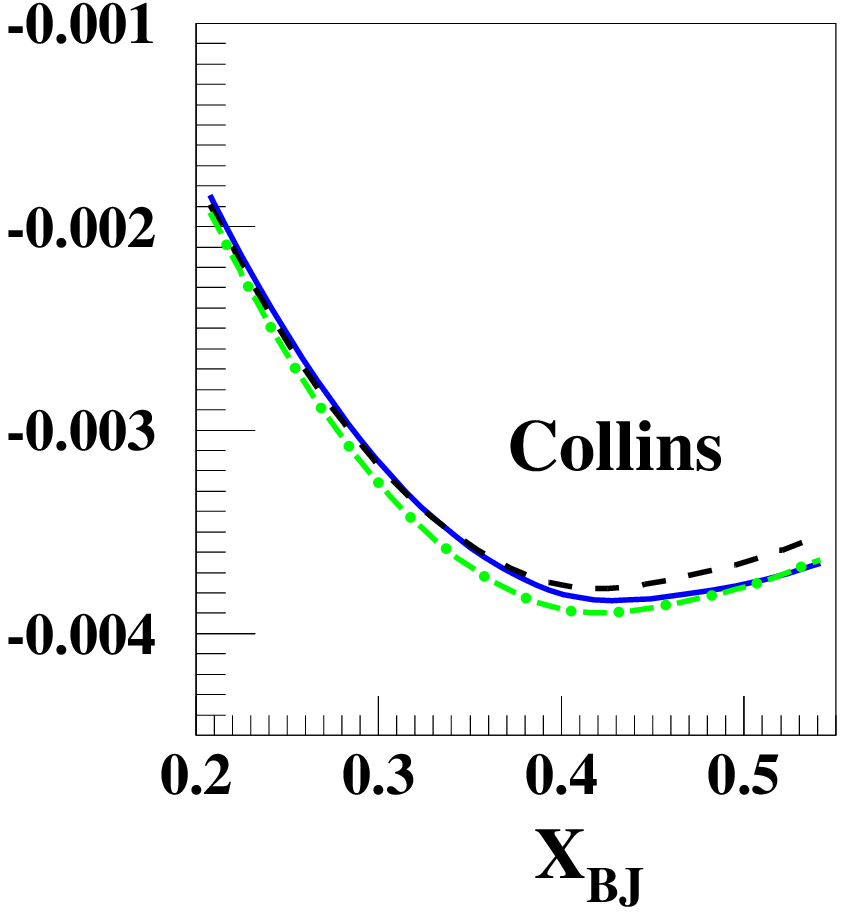}
\vskip 8.cm
\caption{(color online)
The neutron Sivers (left panel)
and Collins (right panel) asymmetries, 
for the JLab kinematics at an initial electron energy of
$\cal E$=8.8 GeV.
Full line: the model for the neutron asymmetry used in the calculation;
dot-dashed line: the neutron asymmetry
extracted from the full calculation of $A_3^j$
with FSI taken into account, using the
extraction formula Eq. (\ref{extr_final});
dashed line: the result obtained using Eq. (\ref{extr_fico}) to extract
the neutron asymmetries from the same calculation. 
Calculations have been performed at $Q^2 = 5.73$ (GeV/c)$^2$,
{i.e. the central $Q^2$ value for an energy beam}   $\cal E$=8.8 GeV 
(see text).}
\label{figlast}
\end{figure}
\section{Conclusions}

Measurements of the Sivers and Collins asymmetries
for both proton and  deuteron have shown 
{{a strong flavor dependence, motivating independent
further investigations using different targets to
safely access the same quantities for the neutron.}}
As for any polarized neutron observable, $^3$He
is the natural
target, due to its specific spin structure. 
Two experiments, aimed at measuring azimuthal asymmetries
in the production of $\pi^\pm$  from transversely
polarized $^3$He, were performed at JLab. From the gathered 
$^3$He data \cite{prljlab},
the Collins and Sivers neutron asymmetries were
extracted 
using a procedure proposed in Ref. \cite{mio}. 
{However, such an extraction procedure
was not considering some relevant nuclear effects,
properly evaluated in the present paper, which 
strengthens {\it a posteriori} the method used
in Ref. \cite{prljlab} to obtain the neutron information.
In particular, the extraction procedure proposed
in Ref. \cite{mio} and used in Ref. \cite{prljlab}} was
able to take care of (i)  the spin structure of $^3$He 
and  (ii) the momentum and energy distributions
of bound nucleons, through  a realistic spin-dependent 
spectral function evaluated by
using nuclear wave functions obtained from the AV18 interaction, 
in plane wave impulse approximation.
The results of Ref. \cite{mio} were obtained in the Bjorken limit, 
namely without considering possible effects of the kinematics
of JLab, dominated by finite values of the energy and momentum transfers, 
and, more important, without FSI effects.
The problem whether or not the extraction procedure based on PWIA calculations
can be extended to a scenario where final state interactions between 
the debris, originated from the struck nucleon,
and the interacting spectator system
are allowed to play a role,
as it likely happens
in the actual JLab kinematics,
has been thoroughly analyzed in the present paper. 
We were able to quantitatively
show that  
the extraction procedure is basically
independent of FSI, evaluated within the generalized 
eikonal approximation.
In particular, in order to perform the needed  full evaluation of 
the FSI effects, we have extended the calculation of a realistic 
distorted spin-dependent spectral function,
introduced in a previous paper of ours \cite{nostro}, where
it was taken into account the two-body break up channel only.
Actually, we have performed a highly non trivial (from the numerical point 
of view) computation of the
contribution to the distorted spin-dependent spectral function from 
the three-body break-up channel, essential to obtain reliable cross sections 
and in turn { to robustly extract valuable} neutron information. 
Once, such a refined
spectral function became available, we  have exploited our results  
for calculating both Sivers and Collins {single spin asymmetries}.
FSI effects have been found to produce sizable effects
in both the unpolarized and polarized cross sections. Differently,   
the SSAs have resulted slightly affected by
FSI, since they are ratio of  cross sections, and therefore the FSI effects
cancel to a large extent.
As a result, the very same extraction procedure proven to be
successful in PWIA can be used also in a scenario where
FSI effects are relevant. This means that all the complexities related 
to Fermi motion, binding and FSI effects 
can be summarized in the nucleon effective polarizations, 
quantities known from accurate few-body calculations
in a rather model independent way. This scheme is valid
in a wide range of FSI models, every time that
FSI are basically spin-independent, as  expected to
happen at high energies (i.e. in the case of JLab or  the
planned Electron Ion Collider) and lead to convolution formulas
for the  nuclear cross sections, namely a folding of  
cross sections off bound nucleons and
distorted spin-dependent spectral functions.

The importance of these results for both the planning and the analysis
of experiments with transversely polarized $^3$He target is clear.
Further studies of the same issue will involve the
implementation of GEA in the relativistic nuclear overlaps,
defined in \cite{tobe}, so that a light-front, distorted, spin
dependent spectral function can be evaluated and relativistic
effects can be taken into account in a consistent framework.

\section*{Acknowledgments}

Calculations were in part performed on the ZEFIRO facility 
of INFN, Commissione 4, Pisa, Italy.
L.P.K. thanks INFN, Perugia, for partial financial support during his stay
in Perugia in 2015 and 2016, and the Dipartimento di Fisica e Geologia
of Perugia University for warm hospitality.

\appendix
\section{Overlaps for the distorted spectral function}
\label{overls}
The overlaps 
\be
{ \tilde {\cal O}}_{\lambda\lambda'}^{N \,M \, M'}
(\epsilon^*_{23},{\bf p}_{mis})=
\nonu =
\la \lambda, \tau^N,
\phi_{\epsilon_{23}^*}(\br) 
e^{-i{\bf p}_{mis}\brho}
{\cal G}(\br,\brho)
| 
\Psi_3^{M}(\br,\brho)\ra
\la \Psi_3^{M'}(\br',\brho')| \lambda', \tau^N, {\cal G}(\br',\brho')
\phi_{\epsilon_{23}^*}(\br') 
e^{-i{\bf p}_{mis}\brho'}\ra~,
\label{aa1}
\ee
corresponding to Eq. (\ref{overfsig}) with
the index $f_{23}$ removed for simplicity,
are built in terms of two- and three-body
wave functions.

In particular, 
when
the energy of the pair
is $\epsilon_{23}^* = t^2 / m$, 
%with ${\bf{t}}$ the relative momentum,
the two-body wave function reads: 

\be
\phi_{s_{23}\sigma_{23}T_{23}\tau_{23}}^{\bf{t}}( {\bf{r}} )
=
4\pi\sum_{l m l_f J_f M_f} \left 
\langle lms_{23}\sigma_{23}|J_fM_f \right \rangle 
Y^*_{lm}(\hat{{\bf t}})~i^{l}
\psi_{ll_f s_{23}}^{J_f}(|{\bf t}|, |{\bf r}|)
Y^{l_f s_{23}}_{J_fM_f}({\bf\hat{r}}) |T_{23} \tau_{23} \rangle,
\ee

with the tensor spherical harmonics defined as
\be
Y^{l_f s_{23}}_{J_f M_f}({\bf\hat{r}})=
\sum_{m_f \sigma_{23}'}
\langle l_f m_f s_{23} \sigma_{23}'|J_f M_f \rangle
Y_{l_f m_f}({\bf\hat{r}}) \chi_{s_{23} \sigma_{23}'}
~.
\ee

When the pair is in the deuteron state,
with binding energy $E_D$,
the two-body wave function reads: 

\be
\phi_{M_D}( E_D, {\bf{r}} )
=
u_0(r) 
Y^{0 1 }_{1 M_D}({\bf\hat{r}})
+
u_2(r) 
Y^{2 1 }_{1 M_D}({\bf\hat{r}})~.
\ee

The three body wave function in \cite{pisa} is defined according
to the following scheme

\begin{eqnarray} 
\langle \sigma_1,\sigma_2,\sigma_3; 
T_{23}, \tau_{23}, \tau; 
{\bfgr \rho},{\bf r}|^3{\rm He};
{1 \over 2} M 
{1 \over 2} T_z
\rangle 
& = &
\langle T_{23} \tau_{23} {1 \over 2} \tau |{1 \over 2} T_z \rangle
\sum_{L_\rho M_\rho}
\sum_{X M_X} \sum_{j_{23}m_{23}}\la XM_X L_\rho M_\rho|{ 1 \over 2}M \ra
~ \la
j_{23}m_{23}{1 \over 2}\sigma_1|XM_X \ra
\nonumber
\\
& \times &
\sum_{s_{23}\sigma_{{23}}}
\sum_{l_{23} \mu_{23} } ~\la {1 \over 2}\sigma_2 {1 \over 2}
\sigma_3| s_{23}\sigma_{{23}} \ra  \la l_{23} \mu_{23} s_{23} \sigma_{{23}} 
|j_{23} m_{23} \ra
\nonumber 
\\
& \times &
{ Y}_{ l_{23} \mu_{23} } (\hat {\bf r})~ { Y}_{L_\rho M_\rho } 
(\hat {\bfgr \rho})
~\phi^{j_{23}l_{23} s_{23}}_{L_\rho X} ( |{\bf r}|,|{\bfgr \rho}|)~.
\label{hwf}
\end{eqnarray} 

The antisymmetrization of the wave function requires
$l_{23}+s_{23}+T_{23}$, where $T_{23}$ is the isospin
of the pair $23$, to be odd. In addition, 
$l_{23}+L_\rho$ has to be even, due to the parity
of $^3$He.

Using these wave functions, 
one has, in the 3bbu channel:

\begin{eqnarray}
\sum\limits_{\sigma_{23} T_{23} \tau_{23}} \int d \hat {\bf t} 
\, { \tilde {\cal O}}_{\lambda\lambda'}^{N \, M \, M'}
(\epsilon^*_{23},{\bf p}_{mis})
& = & 
\int d \hat {\bf{t}}
\sum\limits_{\sigma_{23} \tilde \sigma_{23}} 
%\sum\limits_{T_{23} \tau_{23}}
\sum\limits_{ \{ \alpha, \tilde\alpha\} }
\sum\limits_{ M_f 
M_X \tilde M_X 
m_{23} \tilde m_{23}
}
\nonumber \\
& \times &
\la XM_X L_\rho M_\rho|\frac12 M\ra \la \tilde X\tilde M_X \tilde L_\rho 
\tilde M_\rho|\frac12 M'\ra
\nonumber \\
& \times &
\la j_{23}m_{23} \frac12 \lambda| X M_X\ra \la \tilde j_{23}\tilde m_{23} 
\frac12 \lambda'| \tilde X \tilde M_X\ra
\nonumber\\ 
& \times &
\la l_{23}\mu_{23}s_{23}\sigma_{23} |j_{23}m_{23}\ra
\la \tilde l_{23}\tilde\mu_{23}s_{23} \tilde\sigma_{23}|
\tilde j_{23}\tilde m_{23}\ra
\nonumber \\
& \times &
\la  l_{f} m_{f} s_{23}  \sigma_{23} |J_f M_{f}\ra
\la \tilde l_{f}\tilde m_{f} s_{23}  \tilde\sigma_{23}|J_f M_{f}\ra 
\nonumber
\\ & \times &
O_{\{\alpha\} s_{23}}^{(FSI)}
({\epsilon^*_{23},\bf p}_{mis})~
O_{\{\tilde\alpha\} s_{23}}^{(FSI)}({\epsilon^*_{23},\bf p}_{mis})~,
\label{lp}
\end{eqnarray}

where 
$\{\alpha\} = \{ L_\rho, M_\rho, X, j_{23}, l_f, m_f, 
l_{23}, \mu_{23}, l, J_f \}$
and 

\be
O_{\{\alpha\} s_{23} }^{(FSI)}({\epsilon^*_{23},\bf p}_{mis})= 
4 \pi \int d\bs \rho \int d{\bf r}~ 
{\rm e}^{i{\bf p}_{mis}\bs \rho}
{\cal G} ({\bf r},\bs\rho){\psi_{l l_f s_{23} }^{J_f \, *}(
|{\bf t}|, |{\bf r}|
)}~
{\rm Y}_{l_f m_f}^*(\hat {\bf r})
\nonu \times ~{\rm Y}_{L_\rho M_\rho}(\hat {\bs \rho})
{\rm Y}_{l_{23} \mu_{23}}(\hat {\bf r}) ~
\phi^{j_{23}l_{23} s_{23}}_{L_\rho X} ( |{\bf r}|,|{\bfgr \rho}|)~.
\label{overlapsFSI}
\ee

When the active nucleon $N$ is a proton $p$, 
besides the $3bbu$ channel, 
one can have also the
2bbu channel, for which the overlap becomes

\begin{eqnarray} 
&&
\sum\limits_{M_D}
{\cal O}^{N=p\, M \, M'}_{\lambda\lambda'}
(
E_D,
{\bf p}_{mis}
)
= 
%\nonumber \\ &&
%\left\la \hat S_{Gl}(1,2,3)\left\{
%\Psi_{{\bf P}_{D}},\lambda,{\bf p}_N\right\}|   S_A, \Phi_A\right\ra
%\left\la  \Phi_A,  S_A| \hat S_{Gl}(1,2,3)\left\{\Psi_{{\bf P}_D},\lambda',{\bf p}_N \right\}\right\ra=
\sum_{
M_D M_X \tilde M_X 
m_{23} \tilde m_{23}
\sigma_{23} \tilde \sigma_{23}
}
\nonumber\\ &&
\sum\limits_{\{ \beta, \tilde\beta\}}
\la XM_X L_\rho M_\rho|\frac12 M \ra \la \tilde X\tilde M_X \tilde 
L_\rho \tilde M_\rho|\frac12 M' \ra
\la j_{23}m_{23} \frac12 \lambda| X M_X\ra \la \tilde j_{23}\tilde m_{23} \frac12 \lambda'| \tilde X \tilde M_X\ra
\nonumber\\ &&
\la l_{23}\mu_{23}1 \sigma_{23} |j_{23}m_{23}\ra
\la \tilde l_{23}\tilde\mu_{23} 1 \tilde \sigma_{23}
|\tilde j_{23}\tilde m_{23}\ra
\la  L_{D} m_{L} 1  \sigma_{23} |1 M_{D}\ra
\la \tilde L_{D}\tilde m_{L} 1  \tilde \sigma_{23} | 1 M_{D}\ra
\nonumber\\ &&
O_{\beta}^{(FSI)}
(
E_D,
{\bf p}_{mis}
)~
O_{\tilde\beta}^{(FSI)}(
E_D,
{\bf p}_{mis}
)~,
\label{lp}
\end{eqnarray}
where
\begin{eqnarray}
O_{\beta}^{(FSI)}
(
E_D,
{\bf p}_{mis}
)
& = & \int d\bs \rho \int d{\bf r}
\, {\rm e}^{i{\bf p}_{mis}\bs \rho}
{\cal{G}} ({\bf r},\bs\rho)
%{\Psi_{L_D}(|{\bf r}|)\over |{\bf r}|}~
u_{L_D} (|{\bf r}|)~
{\rm Y}^*_{L_D m_L}(\hat {\bf r})
\nonumber \\ & \times & ~{\rm Y}_{L_\rho M_\rho}(\hat {\bs \rho})
{\rm Y}_{l_{23} \mu_{23}}(\hat {\bf r}) ~
\phi^{j_{23} L_D 1}_{L_\rho X} ( |{\bf r}|,|{\bfgr \rho}|)~,
\label{overlapsFSI}
\end{eqnarray}
and 
$\{\beta\} = \{ L_\rho, M_\rho, X, j_{23},  
l_{23}, 
\mu_{23}, 
L_D=0,2, m_L
\}$.

\section{Properties of the Glauber distorted Spectral Function}
\label{GSF}
Let us consider a reference frame with the $z$-axis along the momentum 
transfer ${\bf q}$. If in such a reference frame a nucleus with $J_A=1/2$ 
has a polarization ${\bf S}_A$, one can
expand the nucleus state by using  pure states polarized with respect to the
quantization axis $\hat q \equiv \hat e_z$, i.e.
$\left |{1\over 2},\pm{1 \over 2}\right\rangle_{\hat q}$. In this case,  a generic state with $J_A=1/2$ and polarization
  directed along some direction 
%  in the reference frame with $\hat q \equiv \hat e_z$ 
  is written as follows
\be
%\left |{1\over 2},S_A\right\rangle_{\hat q}\equiv
\left |{1\over 2},{1 \over 2}\right\rangle_{\hat S_A} = \cos {\beta\over 2}
\left |{1\over 2},{1 \over 2}\right\rangle_{\hat q}+
\sin {\beta\over 2}
\left |{1\over 2},-{1 \over 2}\right\rangle_{\hat q}\label{one}~,
\label{beta}
\ee
where $\cos \beta = {\hat S_A}\cdot {\hat q}$ and 
$\left |{1\over 2},{1 \over 2}\right\rangle_{\hat S_A}$ 
is a pure state polarized
with respect to the quantization axis $\hat S_A$ (see Eq. (43)). In Eq. (19) of \cite{cda1} one can find a general expression of the PWIA  spectral function,
\be
{\bm {P}}_{{\cal M}}({\bf p}, E)={1 \over 2} \left\{ B_0\left[|{\bf p}|,E,({\bf  S}_A \cdot \hat {\bf p})^2\right] +
 {\bm \sigma} \cdot {\bm {\cal F}_{{\cal M}}({\bf p}, E)}\right\}~,
%=\nonu= B_0(|{\bf p}|,E) + \cos \alpha  ~{\cal F}_z + \sin \alpha ~{\cal F}_\perp=
\label{dv}
\ee
where { ${\bf p}=-{\bf p}_{mis}$ is the nucleon three-momentum inside the target,} the index ${\cal M}$ refers to the third component with respect
to the quantization axis $\hat S_A$ and ${\bm {\cal F}}_{{\cal M}}({\bf p}, E)$ is a pseudovector depending upon the vector $\hat {\bf p}$ and the peudovector ${\bf  S}_A$
\be
{\bm {\cal F}}_{{\cal M}}({\bf p}, E) = {\bf  S}_A B_{1,{\cal M}}\left[|{\bf p}|,E,({\bf  S}_A \cdot \hat {\bf p})^2\right] + \hat {\bf p} ~
({\bf  S}_A \cdot \hat {\bf p})~ B_{2,{\cal M}}\left[|{\bf p}|,E,({\bf  S}_A \cdot \hat {\bf p})^2\right] \quad .
\label{tre}
\ee
{{In the case were the FSI is considered through a Glauber operator
at high momentum transfer, 
there is a further  dependence  of the spectral function upon the vector 
${\bf q}$ and Eqs. (\ref{dv}) and (\ref{tre}) are to be replaced by
\be
{\bm {\cal P}}_{{\cal M}}({\bf p}, E, {\bf q})={1 \over 2} \left\{ B_0\left[|{\bf p}|,E,({\bf  S}_A \cdot \hat {\bf p})^2,|{\bf q}|,({\bf  S}_A \cdot \hat {\bf q})^2,\hat {\bf p} \cdot \hat {\bf q}\right] +
 {\bm \sigma} \cdot {\bm {\cal F}_{{\cal M}}({\bf p}, E,{\bf q})}\right\}
%=\nonu= B_0(|{\bf p}|,E) + \cos \alpha  ~{\cal F}_z + \sin \alpha ~{\cal F}_\perp=
\label{dvG}
\ee
\be
\hspace{0mm} {\bm {\cal F}}_{{\cal M}}({\bf p}, E,{\bf q}) = {\bf  S}_A
 B_{1,{\cal M}}\left[|{\bf p}|,E,({\bf  S}_A \cdot \hat {\bf p})^2,
|{\bf q}|,({\bf  S}_A \cdot \hat {\bf q})^2,\hat {\bf p} \cdot 
\hat {\bf q}\right] 
 \nonu
 + ~
\hat {\bf p} ~({\bf  S}_A \cdot \hat {\bf p})~
B_{2,{\cal M}}\left[|{\bf p}|,E,({\bf  S}_A \cdot \hat {\bf p})^2,
|{\bf q}|,({\bf  S}_A \cdot \hat {\bf q})^2,\hat {\bf p} 
\cdot \hat {\bf q}\right]  +
\nonu
\hspace{0mm} + ~ \hat {\bf p} ~({\bf  S}_A \cdot \hat {\bf q})~ 
B_{3,{\cal M}}\left[|{\bf p}|,E,({\bf  S}_A \cdot \hat {\bf p})^2,
|{\bf q}|,({\bf  S}_A \cdot \hat {\bf q})^2,\hat {\bf p} 
\cdot \hat {\bf q}\right] 
\nonu
+ ~
 \hat {\bf q} ~({\bf  S}_A \cdot \hat {\bf p})~
  B_{4,{\cal M}}\left[|{\bf p}|,E,({\bf  S}_A \cdot \hat {\bf p})^2,
|{\bf q}|,({\bf  S}_A \cdot \hat {\bf q})^2,\hat {\bf p} 
\cdot \hat {\bf q}\right] 
  \nonu
  + ~
 \hat {\bf q} ~({\bf  S}_A \cdot \hat {\bf q})~ 
 B_{5,{\cal M}}\left[|{\bf p}|,E,({\bf  S}_A \cdot \hat {\bf p})^2,|{\bf q}|,
({\bf  S}_A \cdot \hat {\bf q})^2,\hat {\bf p} \cdot \hat {\bf q}\right] 
 \nonu
 + ~
 \hat {\bf p} {\rm x} \hat {\bf q}~ 
 B_{6,{\cal M}}\left[|{\bf p}|,E,({\bf  S}_A \cdot \hat {\bf p})^2,
|{\bf q}|,({\bf  S}_A \cdot \hat {\bf q})^2,\hat {\bf p} \cdot \hat 
{\bf q}\right] \quad .
\label{dv1}
\ee
}}
The above expressions for the spectral function, put in evidence the dependence
upon ${\bf  S}_A$, as well as the dependence of the scalar functions  $B_i$ 
($i=1,...,6$) by the possible scalars
$|{\bf p}|,E,({\bf  S}_A \cdot \hat {\bf p})^2,|{\bf q}|,({\bf  S}_A \cdot \hat {\bf q})^2,\hat {\bf p} \cdot \hat {\bf q}$.
{{If ${\bf  S}_A$ is orthogonal to the $z$ axis, ${\bm {\cal F}}_{{\cal M}}({\bf p}, E,{\bf q}) $ reduces to
\be
\hspace{0mm} {\bm {\cal F}}_{{\cal M}}({\bf p}, E,{\bf q}) = {\bf  S}_A
 B_{1,{\cal M}}\left[|{\bf p}|,E,({\bf  S}_A \cdot \hat {\bf p})^2,|{\bf q}|,
\hat {\bf p} \cdot \hat {\bf q}\right] 
 ~
 + ~
\hat {\bf p} ~({\bf  S}_A \cdot \hat {\bf p})~
 B_{2,{\cal M}}\left[|{\bf p}|,E,({\bf  S}_A \cdot \hat {\bf p})^2,|{\bf q}|,
\hat {\bf p} \cdot \hat {\bf q}\right]  +
\nonu
+ ~
\hat {\bf q} ~({\bf  S}_A \cdot \hat {\bf p})~
 B_{4,{\cal M}}\left[|{\bf p}|,E,({\bf  S}_A \cdot \hat {\bf p})^2,
|{\bf q}|,\hat {\bf p} \cdot \hat {\bf q}\right] 
 ~
 + ~
 \hat {\bf p} {\rm x} \hat {\bf q}~ 
 B_{6,{\cal M}}\left[|{\bf p}|,E,({\bf  S}_A \cdot \hat {\bf p})^2,
|{\bf q}|,\hat {\bf p} \cdot \hat {\bf q}\right] \quad .
\label{dv2}
\ee
}}
From Eq. (\ref{dvG}) one has 
\be
 B_0\left[|{\bf p}|,E,({\bf  S}_A \cdot \hat {\bf p})^2,|{\bf q}|,\hat {\bf p} 
\cdot \hat {{\bf q}}\right] = Tr \left [  {\bm {\cal P}}_{{\cal M}}
({\bf p}, E,{\bf q})\right ]
 \label{tr}
\\ &&
{\bm {\cal F}}_{{\cal M}}({\bf p}, E,{\bf q}) = Tr \left [   
{\bm {\cal P}}_{{\cal M}}({\bf p}, E,{\bf q}) ~  {\bm \sigma}  \right ]
\quad .
\label{quattro}
\ee

{ Let us now express the distorted spectral function with a polarization axis along
${\bf S}_A$ (cf Eqs. \eqref{spectr} and \eqref{overfsi}) in terms of the components
given in  Eq. \eqref{spectrg}, that correspond to a polarization axis along $\hat q$ by 
using Eq. (\ref{beta}). 
Since we are interested in a transversely-polarized target, i.e.
${\bf S}_A\equiv \{1,0,0\}$, one has to consider  $\beta = 90^o$, and}
the components of the spectral functions are
\be
\hspace{-7mm}{{{\cal P}}}_{{{\cal M} = \frac12 , \sigma \sigma'}}({\bf p}, E,{\bf q})=
\frac12 \left\{{{{\cal P}}}^{\frac12\frac12}_{{\sigma \sigma'}}({\bf p}, E,{\bf q})+
{{{\cal P}}}^{-\frac12-\frac12}_{{\sigma \sigma'}}({\bf p}, E,{\bf q})+
\left[
{{{\cal P}}}^{\frac12-\frac12}_{{\sigma \sigma'}}({\bf p}, E,{\bf q})+{{{\cal P}}}^{-\frac12\frac12}_{{\sigma \sigma'}}({\bf p}, E,{\bf q})\right]  \right\}.
\label{sf}
\ee

If the nucleus is polarized along $-{\bf  S}_A$, 
the state of the nucleus can be written as follows
\be
%\left |{1\over 2},-S_A\right\rangle_{\hat q}\equiv
\left |{1\over 2},-\frac12\right\rangle_{\hat S_A} = -\sin {\beta \over 2}
\left |{1\over 2},{1 \over 2}\right\rangle_{\hat q}+
\cos {\beta\over 2}
\left |{1\over 2},-{1 \over 2}\right\rangle_{\hat q} \quad 
\label{oneprime}
\ee
and  for 
the spectral function becomes
\be
\hspace{-8mm}{{{\cal P}}}_{{{\cal M} = -\frac12 , \sigma \sigma'}}({\bf p}, E,{\bf q})=
\frac12 \left\{{{{\cal P}}}^{\frac12\frac12}_{{\sigma \sigma'}}({\bf p}, E,{\bf q})+
{{{\cal P}}}^{-\frac12-\frac12}_{{\sigma \sigma'}}({\bf p}, E,{\bf q}) -
 \left[
{{{\cal P}}}^{\frac12-\frac12}_{{\sigma \sigma'}}({\bf p}, E,{\bf q})+{{{\cal P}}}^{-\frac12\frac12}_{{\sigma \sigma'}}({\bf p}, E,{\bf q})\right]  \right\} \quad .
\label{sf2}
\ee
To obtain the real and the imaginary parts of the quantity 
$ \left [ {{\cal P}}^{N \, \frac12 -\frac12}_{\frac12 -\frac12}(E,{\bf p}_{mis})\  + ~
{{\cal P}}^{N \, -\frac12 \frac12}_{\frac12 -\frac12}(E,{\bf p}_{mis})\ \right ]$,
needed to evaluate the single spin asymmetries (see Eq. (\ref{transv})),
let us  first consider the $x$ and the $y$ components of 
${\bm {\cal F}}_{ \frac12 }({\bf p}, E,{\bf q})$ with ${\bf S}_A = {\bf S}_3$ 
along the $x$ axis.

 {
 From Eq. (\ref{dv2}) one has 
 \be
  {\cal F}^{\hat x}_{\frac12 x }({\bf p}, E,{\bf q}) =
   B_{1,\frac12} \left[|{\bf p}|,E,({\bf  S}_A \cdot \hat {\bf p})^2,|{\bf q}|,\hat {\bf p} \cdot \hat {\bf q}\right]
   + \sin^2\theta ~ \cos^2\phi  ~ 
   B_{2,\frac12} \left[|{\bf p}|,E,({\bf  S}_A \cdot \hat {\bf p})^2,|{\bf q}|,\hat {\bf p} \cdot \hat {\bf q}\right]
   \nonu
   +  \sin \theta \sin \phi 
   ~ B_{6,\frac12}\left[|{\bf p}|,E,({\bf  S}_A \cdot \hat {\bf p})^2,|{\bf q}|,\hat {\bf p} \cdot \hat {\bf q}\right]
   \label{x1}
   \\&&
    \nonu
  {\cal F}^{\hat x}_{\frac12 y }({\bf p}, E,{\bf q}) = \sin^2\theta ~ \cos\phi ~ \sin\phi ~ 
  B_{2,\frac12} \left[|{\bf p}|,E,({\bf  S}_A \cdot \hat {\bf p})^2,|{\bf q}|,\hat {\bf p} \cdot \hat {\bf q}\right]
  \nonu
   - \sin \theta \cos \phi ~
   B_{6,\frac12} \left[|{\bf p}|,E,({\bf  S}_A \cdot \hat {\bf p})^2,|{\bf q}|,\hat {\bf p} \cdot \hat {\bf q}\right] \quad ,
 \label{y1}
 \ee
 where the angles $\theta$ and $\phi$ define the direction of the nucleon momentum 
 $\bm p$.
 { From Eq.} (\ref{quattro}) and Eq. (\ref{sf}) one obtains
\be
 {\cal F}^{\hat x}_{\frac12 x }({\bf p}, E,{\bf q}) = Tr \left [  
 {\bm {\cal P}}_{{\cal M}=\frac12}({\bf p}, E,{\bf q}) ~  {\sigma_x}  \right ]
 \nonu
 =  \Re \left [  {{{\cal P}}}^{\frac12\frac12}_{{\frac12-\frac12}}({\bf p}, E,{\bf q}) +
 {{{\cal P}}}^{-\frac12-\frac12}_{{\frac12-\frac12}}({\bf p}, E,{\bf q}) +
 {{{\cal P}}}^{\frac12-\frac12}_{{\frac12-\frac12}}({\bf p}, E,{\bf q}) +
 {{{\cal P}}}^{-\frac12\frac12}_{{\frac12-\frac12}}({\bf p}, E,{\bf q})
  \right]
   \label{x2}
 \\&&
 {\cal F}^{\hat x}_{\frac12 y }({\bf p}, E,{\bf q}) = Tr \left [  {\bm {\cal P}}_{{\cal M}=\frac12}({\bf p}, E,{\bf q}) ~  {\sigma_y}  \right ] =
\nonu
  - \Im \left [  {{{\cal P}}}^{\frac12\frac12}_{{\frac12-\frac12}}({\bf p}, E,{\bf q}) +
 {{{\cal P}}}^{-\frac12-\frac12}_{{\frac12-\frac12}}({\bf p}, E,{\bf q}) +
 {{{\cal P}}}^{\frac12-\frac12}_{{\frac12-\frac12}}({\bf p}, E,{\bf q}) +
 {{{\cal P}}}^{-\frac{1}{2}\frac12}_{{\frac12-\frac12}}({\bf p}, E,{\bf q})
  \right] .
 \label{y2}
 \ee
 
 Then  let us consider the $x$ and the $y$ components of 
 ${\bm {\cal F}}_{ \frac12 }({\bf p}, E,{\bf q})$ with ${\bf S}_3$ opposite to the $x$ axis.
 From Eq. (\ref{dv2}) one has 
 \be
  {\cal F}^{-\hat x}_{\frac12 x }({\bf p}, E,{\bf q}) = 
  - B_{1,\frac12}\left[|{\bf p}|,E,({\bf  S}_A \cdot \hat {\bf p})^2,|{\bf q}|,\hat {\bf p} \cdot \hat {\bf q}\right ]- \sin^2\theta ~ \cos^2\phi  ~ 
  B_{2,\frac12} \left[|{\bf p}|,E,({\bf  S}_A \cdot \hat {\bf p})^2,|{\bf q}|,\hat {\bf p} \cdot \hat {\bf q}\right]
  \nonu +  \sin \theta \sin \phi ~ B_{6,\frac12}\left[|{\bf p}|,E,({\bf  S}_A \cdot \hat {\bf p})^2,|{\bf q}|,\hat {\bf p} \cdot \hat {\bf q}\right]
   \label{x3}
   \\&&
   \nonu
  {\cal F}^{-\hat x}_{\frac12 y }({\bf p}, E,,{\bf q}) = - \sin^2\theta ~ \cos\phi ~ \sin\phi ~ 
  B_{2,\frac12}\left[|{\bf p}|,E,({\bf  S}_A \cdot \hat {\bf p})^2,|{\bf q}|,\hat {\bf p} \cdot \hat {\bf q}\right ]
  \nonu
  - \sin \theta \cos \phi ~ 
  B_{6,\frac12}\left[|{\bf p}|,E,({\bf  S}_A \cdot \hat {\bf p})^2,|{\bf q}|,\hat {\bf p} \cdot \hat {\bf q}\right ]\quad ,
 \label{y3}
 \ee
  while from Eq. (\ref{quattro}) and Eq. (\ref{sf}) one obtains
\be
 {\cal F}^{-\hat x}_{\frac12 x }({\bf p}, E,{\bf q}) = Tr \left [  
 {\bm {\cal P}}_{{\cal M}=-\frac12}({\bf p}, E,{\bf q}) ~  {\sigma_x}  \right ]
 \nonu
 =  \Re \left [  {{{\cal P}}}^{\frac12\frac12}_{{\frac12-\frac12}}({\bf p}, E,{\bf q}) +
 {{{\cal P}}}^{-\frac12-\frac12}_{{\frac12-\frac12}}({\bf p}, E,{\bf q}) -
 {{{\cal P}}}^{\frac12-\frac12}_{{\frac12-\frac12}}({\bf p}, E,{\bf q}) -
 {{{\cal P}}}^{-\frac12\frac12}_{{\frac12-\frac12}}({\bf p}, E,{\bf q})
  \right]
   \label{x4}
 \\&&
 {\cal F}^{-\hat x}_{\frac12 y }({\bf p}, E,{\bf q}) = Tr \left [  
{\bm {\cal P}}_{{\cal M}=-\frac12}({\bf p}, E,{\bf q}) ~  {\sigma_y}  \right ] =
\nonu
  - \Im \left [  {{{\cal P}}}^{\frac12\frac12}_{{\frac12-\frac12}}({\bf p}, E,{\bf q}) +
 {{{\cal P}}}^{-\frac12-\frac12}_{{\frac12-\frac12}}({\bf p}, E,{\bf q}) -
 {{{\cal P}}}^{\frac12-\frac12}_{{\frac12-\frac12}}({\bf p}, E,{\bf q}) -
 {{{\cal P}}}^{-\frac{1}{2}\frac12}_{{\frac12-\frac12}}({\bf p}, E),{\bf q}
  \right] .
 \label{y4}
 \ee
The difference of Eqs. (\ref{x1}) and (\ref{x3}) is equal to the difference of Eqs. (\ref{x2}) and (\ref{x4})
\be
2  B_{1,\frac12}\left[|{\bf p}|,E,({\bf  S}_A \cdot \hat {\bf p})^2,|{\bf q}|,\hat {\bf p} \cdot \hat {\bf q}\right ] + 2  \sin^2\theta ~ \cos^2\phi 
~ B_{2,\frac12}\left[|{\bf p}|,E,({\bf  S}_A \cdot \hat {\bf p})^2,|{\bf q}|,\hat {\bf p} \cdot \hat {\bf q}\right ] 
\nonu
= 
2 \Rea \left [  
 {{{\cal P}}}^{\frac12-\frac12}_{{\frac12-\frac12}}({\bf p}, E,{\bf q}) +
 {{{\cal P}}}^{-\frac12\frac12}_{{\frac12-\frac12}}({\bf p}, E,{\bf q})
  \right]
\label{x}
\ee
and the difference of Eqs. (\ref{y1}) and (\ref{y3}) is equal to the difference of Eqs. (\ref{y2}) and (\ref{y4})
\be
2  \sin^2\theta ~ \cos\phi ~ \sin\phi ~ B_{2,\frac12}\left[|{\bf p}|,E,({\bf  S}_A \cdot \hat {\bf p})^2,|{\bf q}|,\hat {\bf p} \cdot \hat {\bf q}\right ] = 
%\nonu
- 2 \Ima \left [  
 {{{\cal P}}}^{\frac12-\frac12}_{{\frac12-\frac12}}({\bf p}, E,{\bf q}) +
 {{{\cal P}}}^{-\frac{1}{2}\frac12}_{{\frac12-\frac12}}({\bf p}, E,{\bf q})
  \right] .
\label{y}
\ee
Let us stress that the scalar functions $B_1$ and  $B_2$ do depend on the variable 
$\phi$ only through $({\bf  S}_A \cdot \hat {\bf p})^2 = (\sin \theta \cos \phi)^2$, since 
$\hat {\bf p} \cdot \hat {\bf q}= \cos \theta $.

In the nucleon tensor operators  ${\hat w}_{\mu\nu}^{sN}$ that give rise to the Collins 
 and  the Sivers effect, the nucleon momentum can appear directly or through the
  nucleon spin operator. Therefore  terms of zero order in 
${\bm p}_{\perp}/m_N$ can appear, as well as terms of the first, second and third
 order ($\perp$ means orthogonal to the $\hat q = \hat z$ axis) { \cite{barone}}. 
 Once multiplied by the spectral function and integrated over the nucleon momentum, 
 the terms of the second and third order can be discarded, since the spectral function
  decreases rapidly as a function of the nucleon momentum (see, e.g., Fig. 2).

In the imaginary part, $ \Ima \Bigl[w_{\mu\nu}^{sN \frac12-\frac12}\Bigr]$, 
the terms of zero order and of the first order in ${\bm p}_{\perp}/m_N$, 
once multiplied by the left hand side of Eq. (\ref{y}) and integrated over 
$\phi$ , do not give contribution to the hadronic tensor, since one has to 
integrate quantities like $(\cos \phi ~ \sin \phi)$,  
$(\cos^2 \phi ~ \sin \phi)$ or  $(\cos \phi ~ \sin^2 \phi)$ times a function 
of $ \cos^2 \phi$. Then the product of the imaginary quantities in 
Eq. (\ref{pr}) does not give contribution 
to the cross section.

An analogous analysis can be performed on the real part 
$ \Rea \Bigl[w_{\mu\nu}^{sN \frac12-\frac12}\Bigr]$ of the nucleon tensor.
 In this case the terms of zero order in ${\bm p}_{\perp}$ give a non-zero contribution, 
 while  the first order terms yield zero, once the integration over $\phi$ is performed.
% It is important to notice that this analysis is independent of the specific tensor structure of the %Collins effect and therefore applies to the Sivers effect as well.

Let us finally notice that since the transverse components of { ${\bf p} $ 
can be disregarded, as discussed above}, the expressions for $\Delta \sigma_{Col}^N$ and for $\Delta \sigma_{Siv}^N$ of Eqs. (66) and (67) of our paper, that were obtained in a reference frame where $\bm p_\perp = 0 $ (see, e.g., Eqs. 
(6.5.18) and (6.5.17) of Ref. \cite{barone}), can be safely used.
}

\end{document}